\documentclass[%
preprint,
showpacs,preprintnumbers,
showkeys,
 amsmath,amssymb,
prd
]{revtex4-1}

\usepackage[T1]{fontenc} 
\usepackage{subcaption}
\usepackage{amssymb}
\usepackage{hyperref}
\usepackage{bm}
\usepackage{verbatim}
\usepackage[normalem]{ulem}
\usepackage{setspace}
\usepackage{float}

\usepackage[compat=1.1.0]{tikz-feynhand}
\usepackage{tikz-feynman}
\tikzfeynmanset{compat=1.1.0}
\usepackage{feynmp}
\usepackage{tikzsymbols}
\usepackage{array}
\usepackage{pifont} 
\usepackage{bbold}
\usepackage{pifont}
\usepackage{array}
\newcolumntype{L}[1]{>{\raggedright\let\newline\\\arraybackslash\hspace{0pt}}m{#1}}
\newcolumntype{C}[1]{>{\centering\let\newline\\\arraybackslash\hspace{0pt}}m{#1}}
\newcolumntype{R}[1]{>{\raggedleft\let\newline\\\arraybackslash\hspace{0pt}}m{#1}}
\newcommand{\nqbd}{$0\nu 4\beta$}

\usepackage{natbib}
\setcitestyle{sort&compress,numbers}
\begin{document} 
\preprint{arXiv:1903.xxxx}
\preprint{Prepared for submission to Phys.~Rev.~D}
\title{Radiative Dirac Neutrino Mass with Dark Matter and it's implication to $0\nu 4\beta$ in the $U(1)_{B-L}$ extension of the Standard Model}
\author{Arnab Dasgupta}
\email{arnabdasgupta28@gmail.com}
 \altaffiliation{School of Liberal Arts, Seoul-Tech, Seoul 139-743, Korea}
\author{Sin Kyu Kang}
\email{skkang@seoultech.ac.kr}
 \altaffiliation{School of Liberal Arts, Seoul-Tech, Seoul 139-743, Korea}
  \author{Oleg Popov}
  \email{opopo001@ucr.edu}
\altaffiliation{Institute of Convergence Fundamental Studies, \\ Seoul National University of Science and Technology, \\Seoul 139-743, Korea}
\date{\today}

\begin{abstract}
The Standard Model gauge symmetry is extended by $U(1)_{B-L}$ which when spontaneously broken leads to residual $\mathbb{Z}_4$ symmetry. $U(1)_{B-L}$ gauge symmetry made anomaly free by introducing exotic SM singlets with corresponding $U(1)_{B-L}$ charges of $13$, $-14$, and $15$. $\mathbb{Z}_4$ symmetry ensures the Dirac nature of neutrinos, simultaneously stabilizing dark matter. Dirac neutrino mass is generated through scotogenic scenario. Dark matter, direct detection, cosmological constraints, and collider constraints analysis is performed. $\mathbb{Z}_4$ symmetry predicts the exact absence of neutrinoless double beta decay ($0\nu 2\beta$) and gives a prediction for an enhanced neutrinoless quadruple beta decay ($0\nu 4\beta$) via which this model can be tested. Model allows for Majorana dark matter as well as for long-lived dark matter candidates.
\end{abstract}
\maketitle
%
\twocolumngrid
\begin{singlespace}
\tableofcontents
\end{singlespace}
\onecolumngrid
\section{Introduction}
\label{sec:intro}
The Standard Model (SM) of strong and electroweak interactions has proven to be very successful so far with the last remaining piece experimentally discovered on July, 4'th 2012~\cite{Aad:2012tfa,Chatrchyan:2012xdj}.Nevertheless there are experimental observations that require new physics beyond Standard Model (BSM). One of these problems is the experimental observation of neutrino oscillations~\cite{PhysRevLett.77.1683,PhysRevC.80.015807,Fukuda:2002pe,PhysRevLett.90.021802,PhysRevLett.81.1562,PhysRevD.74.072003,PhysRevLett.111.211803,PhysRevLett.108.171803,PhysRevLett.108.191802,PhysRevD.86.052008} back in 1990's. Theoretical explanation of neutrino masses requires addition of new particles BSM. The most minimalistic and simplest realizations of this are the seesaw mechanism of type I~\cite{Mohapatra:1979ia,Minkowski:1977sc,Yanagida:1979ab,GellMann:1979grs} which adds a fermion singlet to SM. Next would be seesaw of type II~\cite{Schechter:1980gr,Magg:1980ut,Cheng:1980qt,Mohapatra:1980yp} which extends the SM by a scalar triplet. Last of this kind of realizations is seesaw of type III in which SM is extended by a fermionic electroweak triplet. All these tree level realizations of naturally small neutrino masses require either a small couplings or heavy new physics in order to explain the smallness of neutrino masses and they are lead to unique dimension-five effective operator 
\begin{equation}
    \frac{f_{ij}}{2\Lambda}\left(\nu_i \phi^0-l_i \phi^+\right)\left(\nu_j \phi^0-l_j \phi^+\right)+\text{h.c.},
\end{equation}
known as Weinberg operator~\cite{Weinberg:1979sa}. 
In order to avoid the requirement of heavy new physics or small couplings, for instance neutrino masses can be generated radiatively at one-loop order. Examples of this realizations include~\cite{Zee:1980ai} the Zee model from 1980, the canonical scotogenic model~\cite{Ma:2006km} (\emph{scotos} from Greek meaning darkness) from 2006, and radiative inverse seesaw model~\cite{Fraser:2014yha}. 
Since neutrinos are neutral and colorless they can be of Dirac or Majorana type. Currently there is no experimental evidence toward any direction. But if neutrinos are Dirac in nature there must be a symmetry (conserved quantity) responsible for the absence of Majorana mass of neutrinos. This issue was systematically studied in~\cite{Ma:2016mwh}. The symmetry for the Dirac nature of neutrinos can be the lepton number already present in SM as an accidental symmetry. Experiments such as COURE, GERDA $0\nu 2\beta$, NEMO 3, The MAJORANA Neutrinoless Double-beta Decay Experiment, \emph{etc.} looking for neutrinoless double and quadruple beta decay can solve this problem in the near future. On the other hand, if $(0\nu 2\beta)$ experiment sees no positive results this could hint in the direction of Dirac neutrinos. But there exist even more exotic scenarios like was explained in~\cite{Hirsch:2017col}. In the case of absence of positive results from neutrinoless double beta decay and confirmation of neutrinoless quadruple beta decay, one needs to find a theoretical explanation for this kind of experimental observation.

In our work we present a UV complete model where lepton number is gauged in $U(1)_{B-L}$ symmetry. Spontaneously breaking $U(1)_{B-L}$ to residual $\mathbb{Z}_4$ discrete symmetry allows for Dirac neutrinos which obtain their masses radiatively via scotogenic scenario. Model naturally predicts neutrinoless quadruple beta decay whereas neutrinoless double beta decay is exactly absent. Furthermore, model allows for stable dark matter, fermionic or bosonic, and leptogenesis. Similar works done on $U(1)_{B-L}$ extension of SM are~\cite{Benavides:2015afa, Liu:2016oph, Geng:2015qha, Boucenna:2014pga, Lim:2005aa, Witten:2000dt, Kim:1999yu, CentellesChulia:2019gic, Ma:2016nnn,Basso:2008iv}.

The paper is organized as follows: in Sec.~\ref{sec:model} the model is introduced and the cancellation of chiral anomalies is explained; Sec.~\ref{sec:mnu} demonstrates how radiative Dirac neutrino masses are generated; Secs.~\ref{sec:fermion} and~\ref{sec:scalar} give the fermion and scalar mass spectrum, respectively; Sec.~\ref{sec:DM} discusses dark matter candidates; in Sec.~\ref{sec:nqbd} we go over the neutrinoless quadruple beta decay prediction;  Sec.~\ref{sec:const} presents the results and discusses relevant constraints for our model; and Sec.~\ref{sec:conclusion} concludes.
\section{Model}
\label{sec:model}
SM gauge symmetry is extended to $SU(3)_c\times SU(2)_L\times U(1)_Y\times U(1)_{B-L}$ and the field content of the model is shown in Tab.~\ref{tab:particles}. All fields are given as left-chiral fields and "c" in superscript denotes the charge conjugation. The Yukawa part of the Lagrangian of the model is given below
\begin{align}
    \label{eq:lag}
    \mathcal{L}_{\text{Yuk}}&=\mathcal{L}_{\text{Yuk}}^{\text{SM}}-\bar{N}_{aR} Y_L^{ab} L_{bi}\eta_j \epsilon^{ij}-\bar{\nu}_{aR} Y_R^{ab} N_{bL}\chi\nonumber \\
    &-\bar{N}_{aR} Y_{ND}^{ab} N_{bL} S_4-N_{aR} Y_{NM}^{ab} N_{bR} S_4\nonumber \\
    &-\Psi_{1aL} Y_{12}^{ab} \Psi_{2bL}\chi-\Psi_{2aL} Y_{23}^{ab}\Psi_{3bL}\chi^*-\Psi_{1aL} Y_{13}^{ab}\Psi_{3bL}\frac{\left(S_4^*\right)^7}{\Lambda^6}-\Psi_{2aL} Y_{22}^{ab} \Psi_{2bL} \frac{S_4^7}{\Lambda^6}+\text{H.c.}
\end{align}
The model is constructed as follows: $\nu_R$ is introduced as the Dirac partner for Left-handed neutrinos, $N_{L,R}$ fermions, $\eta$, and $\chi$ scalars are introduced to complete the loop for radiative neutrino mass generation, i.e. scotogenic scenario, $\Psi_{i}$ are introduced for anomaly cancellation, and lastly $S_4$ is needed for spontaneous symmetry breaking (SSB) of $U(1)_{B-L}$ to residual $\mathbb{Z}_4$ discrete symmetry in the leptonic sector. Here, the residual $\mathbb{Z}_4$ symmetry is given by $e^{\imath (B-L)2\pi/4}=w^{(B-L)}$, where $w=e^{\imath 2\pi/4}$ with $w^4=1$.
\begin{table}[h]
    \centering
    \begin{tabular}{cccccc}
    \hline\hline
        Field & SU(3)$_c$ & SU(2)$_L$ & U(1)$_Y$ & U(1)$_{B-L}$ & Flavor \\ \hline
        Q & {\bf 3} & {\bf 2} & $\frac{1}{6}$ & $\frac{1}{3}$ & 3 \\
        u$^c$ & $\pmb{\bar{3}}$ & {\bf 1} & $-\frac{2}{3}$ & $-\frac{1}{3}$ & 3 \\
        d$^c$ & $\pmb{\bar{3}}$ & {\bf 1} & $\frac{1}{3}$ & $-\frac{1}{3}$ & 3 \\
        L & {\bf 1} & {\bf 2} & $-\frac{1}{2}$ & $-1$ & 3 \\
        e$^c$ & {\bf 1} & {\bf 1} & $1$ & 1 & 3 \\
        $\nu^c$ & {\bf 1} & {\bf 1} & $0$ & 5 & 3 \\
        N & {\bf 1} & {\bf 1} & $0$ & $-6$ & 3 \\
        N$^c$ & {\bf 1} & {\bf 1} & $0$ & 2 & 3 \\
        $\Psi_{\text{\tiny{1}}}$ & {\bf 1} & {\bf 1} & $0$ & $13$ & 3 \\
        $\Psi_{\text{\tiny{2}}}$ & {\bf 1} & {\bf 1} & $0$ & $-14$ & $2\times 3$ \\ 
        $\Psi_{\text{\tiny{3}}}$ & {\bf 1} & {\bf 1} & $0$ & $15$ & 3 \\ \hline
        H & {\bf 1} & {\bf 2} & $\frac{1}{2}$ & $0$ & $1$ \\
        $\eta$ & {\bf 1} & {\bf 2} & $\frac{1}{2}$ & $-1$ & $1$ \\
        $\chi$ & {\bf 1} & {\bf 1} & $0$ & $1$ & $1$ \\
        S & {\bf 1} & {\bf 1} & $0$ & $2$ & $1$ \\
        S$_4$ & {\bf 1} & {\bf 1} & $0$ & $4$ & $1$ \\
    \hline\hline
    \end{tabular}
    \caption{Model particle content. All fields are given as Left-handed chiral fields. Bold-faced numbers represent non-Abelian group irreducible representations.}
    \label{tab:particles}
\end{table}
$H$ serves the role of Standard Model (SM) Higgs field that couples and gives masses to SM quarks and charged leptons. $S_4$ is introduced to spontaneously break $U(1)_{B-L}$ gauge symmetry. Since $SU(2)_L$ (same for $SU(3)_c$) group is orthogonal to $U(1)_{B-L}$, $SU(2)_L$ irreducible representation's (irrep's) components must transform identically under $U(1)_{B-L}$. Therefore, for quarks, i.e. $u_{L,R}$, $d_{L,R}$, $U(1)_{(B-L)}$ gauge symmetry is broken to global $U(1)_{(B-L)}$ symmetry. Whereas for $\eta^+, \eta^0,e_{L,R}$, and $\nu_{L,R}$, $U(1)_{(B-L)}$ gauge symmetry is broken to $\mathbb{Z}_4$ symmetry under which they transform as $w^*$. All field transformations under residual $\mathbb{Z}_4$ and $U(1)_{B-L}$ global symmetries are summarized in Tab.~\ref{tab:z4}. Fermions that transform as $w^2$ are of Majorana type, i.e. $N_{L,R}$ and $\Psi_{2L}$ in this case and fermions that transform as $w$ and $w^*$ under $\mathbb{Z}_4$ residual symmetry arrange themselves into Dirac pairs, $\nu_{L,R}, \Psi_{1,3 L}$ are among them. When electroweak symmetry is broken by $H$ the $\eta^{0}$ and $\chi^{\star}$ scalars mix through $H$, which is needed for the neutrino mass generation. Furthermore, when $U(1)_{B-L}$ symmetry is broken by $S_4$ vacuum expectation value (VEV), the mass eigenstates of $(\eta^0,\chi^{\star})$(call them $\xi_{1,2}$) obtain an effective operator $\xi_i^4$+H.c., which is invariant under $Z_4$ and generates the neutrinoless quadruple beta decay. Neutrinoless double beta decay is forbidden by $\mathbb{Z}_4$ symmetry, therefore neutrinoless quadruple beta decay will be dominant. More on this in Secs.~\ref{sec:nqbd} and~\ref{sec:scalar}.
\begin{table}[]
\centering
    \caption{Global $U(1)_{(B-L)}$ and $\mathbb{Z}_4$ transformations of fields. $a=1,2,3$ is the flavor index and $i=1,2$.}
    \label{tab:z4}
    \begin{tabular}{cc}
    \hline\hline
        $\mathbb{Z}_4$ & Fields \\ \hline
        $\mathbb{1}$ & $H$, $S_4$ \\
        $w^*$ & $\nu^a$, $e^a$, $\Psi_{1L}^{a*}$, $\Psi_{3R}^a$, $\eta^0$, $\eta^{+}$, $\chi^*$ \\
        $w^2$ & $N_{L,R}^a$, $\Psi_{i2L}^{a}$, $S$\\ \hline
        Global $U(1)_{(B-L)}$ & Fields \\ \hline
        $\frac{1}{3}$ & $u^a$, $d^a$ \\
    \hline\hline
    \end{tabular}
\end{table}
\subsection*{Chiral anomalies }
\label{app:ano}
Model is chiral anomaly free and cancellation of anomalies per family is shown in Tab.~\ref{tab:ano}. $U(1)_Y$ gravitational anomaly is cancelled like in SM and $U(1)_{B-L}$ gravitational anomaly is cancelled as follows:
\begin{align}
    \left.\sum Q_{B-L}\right|_{\text{Grav.}}&=3\left(2\times\frac{1}{3}-\frac{1}{3}-\frac{1}{3}\right)-2\times 1+1+5-6+2+13-2\times 14+15=0
\end{align}
\begin{table}[!h]
    \centering
    \caption{Chiral anomaly cancellation.}
    \label{tab:ano}
        \includegraphics[width=0.8\textwidth]{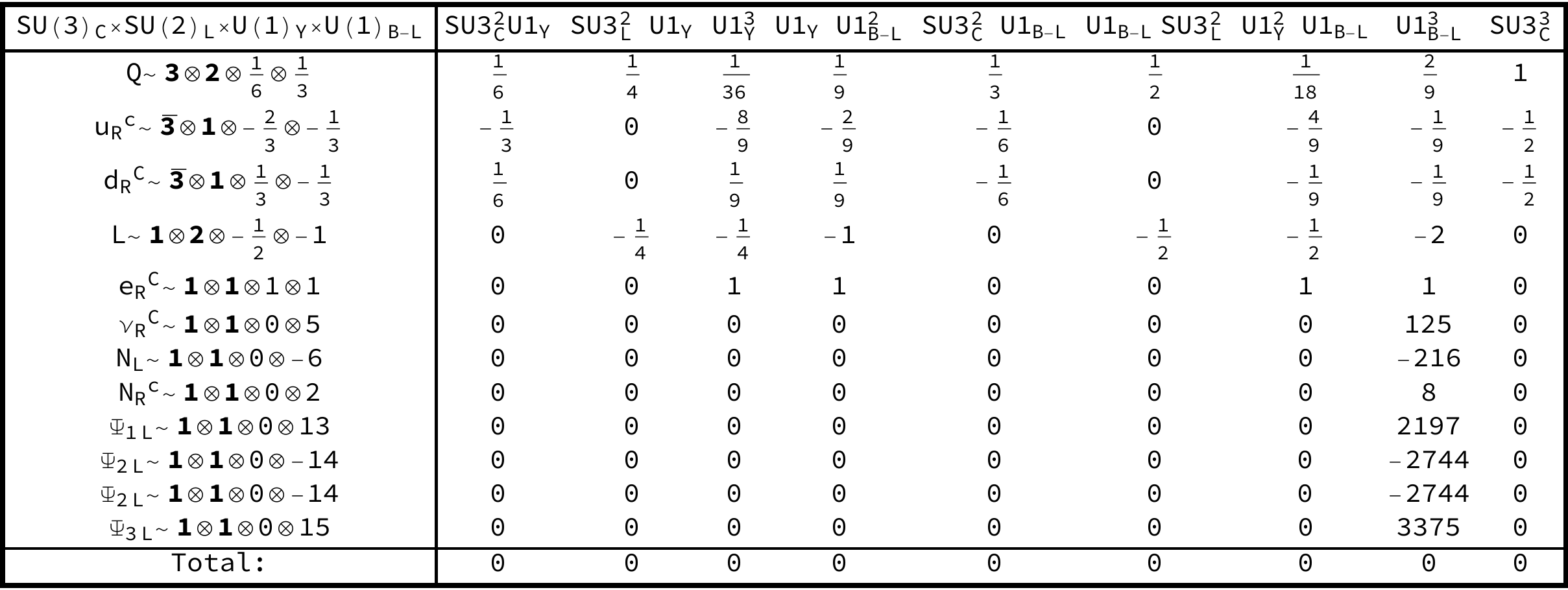}\footnotemark
\end{table}
\footnotetext{Generated using \href{http://renatofonseca.net/susyno.php}{Susyno}~\cite{Fonseca:2011sy}}
\subsection*{Abelian kinetic mixing}
\label{sec:akm}
Model Lagrangian must be augmented with renormalizable Abelian kinetic mixing(KM) $\epsilon_0$ counter-term, since the one-loop corrections has singular contribution, same as in~\cite{Rizzo:2018vlb}
\begin{align}
    \label{eq:gauge_lag}
    \mathcal{L}_{Gauge}&=-\frac{1}{4}B_{\mu\nu}B^{\mu\nu}-\frac{1}{4}B_{\mu\nu}^{\prime}B^{\prime \mu\nu}+\frac{\varepsilon_0}{2c_w}B_{\mu\nu}B^{\prime \mu\nu},
\end{align}
where $B$ and $B^{\prime}$ are strength tensors for hypercharge($U(1)_Y$) and $B-L$($U(1)_{B-L}$) gauge groups. respectively. $\varepsilon_0$ represents the bare Abelian kinetic mixing counter-term which must be included to renormalize the divergent one-loop corrections~\cite{Holdom:1985ag,Holdom:1986eq}
\begin{align}
    \label{eq:ferm_gkm}
    \varepsilon_f&=\frac{c_w}{12\pi^2}g^{\prime}g_{bl}\left[\sum_i Q_{Y_i}Q_{BL_i}\left(-\epsilon^{-1}-\ln \mu^2\right)+\sum_i Q_{Y_i}Q_{BL_i}\ln m_{f_i}^2\right] , \\
    \varepsilon_s&=\frac{c_w}{12\pi^2}g^{\prime}g_{bl}\left[\sum_i \frac{-1}{2}Q_{Y_i}Q_{BL_i}\left(-\epsilon^{-1}-\ln \mu^2\right)+\sum_i \frac{-1}{2} Q_{Y_i}Q_{BL_i} \ln m_{s_i}^2\right].
\end{align}
Fermions that contribute Abelian KM are $Q,u^c,d^c,L,e^c$ and scalar that contributes is $\eta$. Their contributions to the divergence are given as
\begin{align}
    F:&\sum_i Q_{Y_i}Q_{BL_i}=3\times 2\frac{1}{6}\frac{1}{3} + 3\times 1\frac{-2}{3}\frac{-1}{3} + 3\times 1\frac{1}{3}\frac{-1}{3} + 1\times 2\frac{-1}{2}(-1) + 1=\frac{8}{3}/\text{flavor}\\
    S:&\sum_i \frac{-1}{2}Q_{Y_i}Q_{BL_i}=-\frac{1}{2}\times 1\times 2\frac{1}{2}(-1)=\frac{1}{2}\\
    \text{Total}:&3\times \frac{8}{3}+\frac{1}{2}=\frac{17}{2}.
\end{align}
In order to regularize the divergence $\varepsilon_0$ must be given by
\begin{align}
    \label{eq:eps0}
    \varepsilon_0&= (-)\frac{c_w}{12\pi^2}g^{\prime}g_{bl}\frac{17}{2}\left(-\epsilon^{-1}-\ln \mu^2\right) + \varepsilon_0^{\text{finite}},
\end{align}
where the tree level finite piece is denoted by $\varepsilon_0^{\text{finite}}$ and the one-loop corrected finite contribution to Abelian KM is given by
\begin{align}
    \label{eq:finite_km}
    \varepsilon^{\prime}&= \varepsilon_0^{\text{finite}} + \frac{c_w}{12\pi^2}g^{\prime}g_{bl}\left[\underset{F_i}{\sum} Q_{Y_i}Q_{BL_i}\ln m_{f_i}^2 + \underset{s_i}{\sum} \frac{-1}{2} Q_{Y_i}Q_{BL_i} \ln m_{s_i}^2\right].
\end{align}
\section{Neutrino masses}
\label{sec:mnu}
\begin{figure}[ht]
\centering
\begin{tikzpicture}
\begin{feynman}
\vertex (i1);
\vertex [right=2cm of i1] (a);
\vertex [right=2cm of a] (b);
\vertex [right=2cm of b] (c);
\vertex [right=2cm of c] (f1);
\vertex [below=1cm of b] (bb) {$\left\langle S_4\right\rangle$};
\vertex [above=2cm of b] (tb1);
\vertex [above=1cm of tb1] (tb2) {$\left\langle H\right\rangle$};

\diagram* {
i1 -- [fermion, edge label'=$\nu_{L}$] (a) -- [fermion, edge label'=$N_R$] (b) -- [fermion, edge label'=$N_L$] (c) -- [fermion, edge label'=$\nu_{R}$] (f1),
b -- [anti charged scalar, insertion=0.9] (bb),
a -- [anti charged scalar, quarter left, edge label=\(\eta\)] (tb1),
tb1 -- [charged scalar, quarter left, edge label=$\chi$] (c),
tb1 -- [anti charged scalar, insertion=0.9] (tb2),
};
\end{feynman}
\end{tikzpicture}
\caption{Dirac radiative neutrino mass with $\mathbb{Z}_4$ as scotogenic symmetry.}
\label{fig:mnu}
\end{figure}
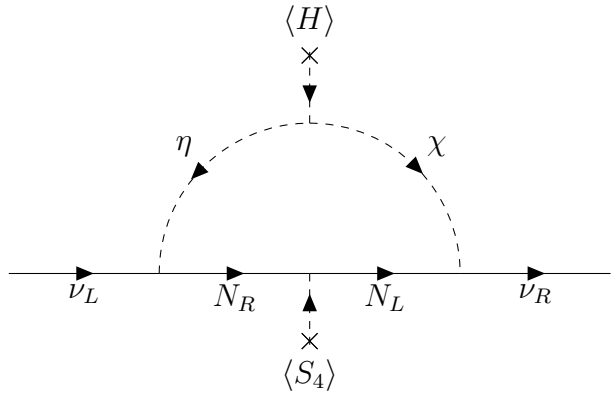

Neutrino tree level mass is forbidden by $U(1)_{B-L}$ symmetry. This is the $S$ symmetry from Ref.~\cite{Ma:2016mwh} and the neutrino mass is generated via first scenario of one-loop radiative case from Ref.~\cite{Ma:2016mwh}. Neutrino masses are obtained via a diagram shown in Fig.~\ref{fig:mnu}. Neutrinos transform as $w^*$ under residual $\mathbb{Z}_4$ symmetry, therefore $\mathbb{Z}_4$ guarantees the Dirac nature of neutrinos in our model. Interesting feature of this model is that Dirac neutrino masses are generated through the Majorana dark sector $N_{1,2}$ fermions which transform as $w^2$ under $\mathbb{Z}_4$ symmetry and are allowed to have Majorana masses. Other interesting property is that the $\mathbb{Z}_4$ residual symmetry which originated from gauged $U(1)_{(B-L)}$ symmetry is responsible both for Diracness of the neutrinos as well as for the stability of dark matter in our model. It is actually the $\mathbb{Z}_4$ plus the Lorentz symmetry that stabilizes the dark matter. 
Neutrino radiative mass is given as 
\small{
\begin{align}
    m_{\nu}^{ab}&=\frac{s_N c_N s_{\xi} c_{\xi}}{16\pi^2} Y_{L}^{ac}\left\{m_{N_1}\left[F\left(\frac{m_{\xi_1}^2}{m_{N_1}^2}\right)-F\left(\frac{m_{\xi_2}^2}{m_{N_1}^2}\right)\right]+m_{N_2}\left[F\left(\frac{m_{\xi_2}^2}{m_{N_2}^2}\right)-F\left(\frac{m_{\xi_1}^2}{m_{N_2}^2}\right)\right]\right\}_{cd} Y_R^{db},
\end{align}
}
where $F(x)$ is defined as 
\begin{align}
F(x)&=\frac{x}{1-x}\text{ln}x,
\end{align}
and mixing angles of $\left(\eta,\chi\right)$ and $\left(N_L,N_R^c\right)$ are given in Eqs.~\ref{eq:etachimix} and~\ref{eq:nmix}, respectively.
\section{Fermion sector}
\label{sec:fermion}
The SM fermions generate their masses in a usual way. Since neutrinos transform as $w^*$ under residual $\mathbb{Z}_4$ symmetry, their masses are of Dirac type and were given in Sec.~\ref{sec:mnu}. $N_{L,R}$ transform as $w^2$ under $\mathbb{Z}_4$ therefore they obtain Majorana masses through the seesaw-I texture matrix form
\begin{align}
    \mathcal{L}_{N}&=\left(\bar{N}_L^c,\bar{N}_R\right)\left(\begin{matrix}0 & Y_{ND} v_4 \\ Y_{ND} v_4 & Y_{NM}^{\dagger} v_4\end{matrix}\right)\left(\begin{matrix}N_L \\ N_R^c\end{matrix}\right)+\text{H.c.}
\end{align}
In general Yukawas here can be complex but the Majorana phases of $N_{L,R}$ can be used to remove this phases, so they are not physical. If the $(1,1)$ component of the $\left(\bar{N}_L^c,\bar{N}_R\right)$ mass matrix was non-zero this would not be the case. See App.~\ref{ap:mN} for more details on this.
Eigenvalues and eigenvectors are given by
\begin{align}
\label{eq:mN12}
    m_{N_{1,2}}&=\frac{v_4}{2}\left(\left|Y_{NM}\right|\pm\sqrt{\left|Y_{NM}\right|^2+4\left|Y_{ND}\right|^2}\right),\\
    \left(\begin{matrix}N_1 \\ N_2\end{matrix}\right)&=\left(\begin{matrix}\text{cos}\theta & \text{sin}\theta \\ -\text{sin}\theta & \text{cos}\theta\end{matrix}\right)\left(\begin{matrix}N_L \\ N_R^c\end{matrix}\right),\\
\label{eq:nmix}
    \text{tan}(2\theta_N)&=-2\frac{\left|Y_{ND}\right|}{\left|Y_{NM}\right|}.
\end{align}
In order for $\Psi_{1,2,3}$ to get their masses, for instance, a SM singlet scalar with $U(1)_{B-L}\sim 28$ (say $S_{28}$) can be introduced. When $S_{28}$ obtains a non-zero VEV $\left\langle(S_{28}\right\rangle\sim v_{28}$ which is allowed by $\mathbb{Z}_4$ symmetry, $\Psi_{i}$ masses would be generated through Lagrangian terms $\Psi_{1L}^a Y_{13}^{ab}\Psi_{3L}^b S_{28}^*$ and $\Psi_{2L}^a Y_{22}^{ab} \Psi_{2L}^b S_{28}$. But in our case, we generate $\Psi_{1,2,3}$ masses through effective dimension-ten Lagrangian terms given in Eq.~\ref{eq:lag}. $\Lambda$ scale is associated with mass scale involved in generation of $m_{\Psi_{i}}$. Since $\Psi_{1}^*,\Psi_{3}\sim w$ and $\Psi_{2i}\sim w^2$ under $\mathbb{Z}_4$, $\Psi_1$ and $\Psi_3$ form a Dirac fermion and $\Psi_{2i}$ form two Majorana fermions. Their masses are given by $m_{\Psi_{1,3}}^{ab}=Y_{13}^{ab}\frac{\left\langle S_{8}\right\rangle^7}{\Lambda^6}$ and $m_{\Psi_{22}}^{\alpha\beta} = Y_{22}^{\alpha\beta} \frac{\left\langle S_{8}\right\rangle^7}{\Lambda^6}$ with $a,b=1-3$ and $\alpha,\beta=1-6$, respectively. For the number estimates: if $Y_{13,22}\sim O(1)$ and $m_{13,22}\sim O(1\text{TeV})$ we get $\log(\Lambda)\approx (7\log v_4 -3)/6$ relation between $v_4$ and $\Lambda$ scales. If $v_4\approx 10^5(10^{11})$ GeV then $\Lambda\approx 2\cdot 10^5(10^{12})$ GeV, respectively.
\\
Remark regarding the mixing of $N_{L,R}$ and $\nu_{L,R}$ with $\Psi_{2i}$ and $\Psi_{1,3}$, respectively. Since there is no symmetry distinguishing $N_{R,L}$ from $\Psi_{2}$ and similarly for $\nu_{L,R}$ and $\Psi_{1,3}$, these will mix via dimension-6, 7, and 8 effective operators given by
\begin{align}
    \Bar{N}_R Y_{\scriptstyle{R2}}\Psi_2 \frac{S_4^3}{\Lambda^2} + N_L Y_{\scriptstyle{L2}} \Psi_2 \frac{S_4^5}{\Lambda^4} + \text{H.c.},\\
    \Bar{\Psi}_3 Y_{\scriptstyle{3R}} \nu_R \frac{S_4^5}{\Lambda^4} + \Psi_1 Y_{\scriptstyle{1L}} LH \frac{S_4^{\star 3}}{\Lambda^3} + \text{H.c.}
\end{align}
On the other hand if $S_{28}$ is included, as was explained above, then $\Psi_{1,2,3}$ do generate their masses. But since scalars with charge $\sim 20$ and $\sim 12$ under $U(1)_{B-L}$ are not included, there will be no mixing between $\Psi$ sector and $(N,\nu)$ sector. Which means there is an inherent $\mathbb{Z}_2$ symmetry induced under which $\Psi$ sector is odd and all other particles are even(trivial). If this is the case, then lightest of $\Psi$ eigenstates and $\mathbb{Z}_4$ LSP can be DM candidates which would give a multi-component DM scenario.
\section{Scalar sector}
\label{sec:scalar}
Most general scalar potential is given as
\begin{align}
\label{eq:potential}
    V\left(H,\eta,\chi,S,S_4\right)=&-m_{H}^2H^{\dagger}H+\frac{1}{2}\lambda_{H}\left(H^{\dagger}H\right)^2+m_{\eta}^2\eta^{\dagger}\eta+\frac{1}{2}\lambda_{\eta}\left(\eta^{\dagger}\eta\right)^2\nonumber \\
    &+m_{\chi}^2\chi^{*}\chi+\frac{1}{2}\lambda_{\chi}\left(\chi^{*}\chi\right)^2+m_{S}^2S^{*}S+\frac{1}{2}\lambda_{S}\left(S^{*}S\right)^2\nonumber \\
    &-m_{S_4}^2S_4^{*}S_4+\frac{1}{2}\lambda_{S_4}\left(S_4^{*}S_4\right)^2+\lambda_{H\eta}\left(H^{\dagger}H\right)\left(\eta^{\dagger}\eta\right)\nonumber \\
    &+\lambda_{H\eta}^{\prime}\left(H^{\dagger}\eta\right)\left(\eta^{\dagger}H\right)+\lambda_{H\chi}\left(H^{\dagger}H\right)\left(\chi^{*}\chi\right)\nonumber \\
    &+\lambda_{HS}\left(H^{\dagger}H\right)\left(S^{*}S\right)+\lambda_{HS_4}\left(H^{\dagger}H\right)\left(S_4^{*}S_4\right)\nonumber \\
    &+\lambda_{\eta\chi}\left(\eta^{\dagger}\eta\right)\left(\chi^{*}\chi\right)+\lambda_{\eta S}\left(\eta^{\dagger}\eta\right)\left(S^{*}S\right)\nonumber \\
    &+\lambda_{\eta S_4}\left(\eta^{\dagger}\eta\right)\left(S_4^{*}S_4\right)+\lambda_{\chi S}\left(\chi^{*}\chi\right)\left(S^{*}S\right)+\lambda_{\chi S_4}\left(\chi^{*}\chi\right)\left(S_4^{*}S_4\right)\nonumber \\
    &+\lambda_{S S_4}\left(S^{*}S\right)\left(S_4^{*}S_4\right)+\left[\mu_{H}\left(H^{\dagger}\eta\right)\chi+\mu_{S}\chi^2 S^*+\mu_{S_4}S^2 S_4^*\right.\nonumber \\
    &\left.+\lambda_{1}\left(H^{\dagger}\eta\right)\chi^*S+\lambda_2\chi^2SS_4^*+\text{H.c.}\right]
\end{align}
Potential minimization conditions are
\begin{align}
    &\lambda_H v^2+\lambda_{HS_4}v_4^2-2m^2_H=0,\\
    &\lambda_{S_4} v_4^2+\lambda_{HS_4}v^2-2m^2_{S_4}=0.
\end{align}
Due to $\mathbb{Z}_4$ residual symmetry mass eigenstates can be divided into three groups: trivially transforming under $\mathbb{Z}_4$ (singlet representation), transforming as $w$ or $w^*$ under $\mathbb{Z}_4$ (complex irreducible representation), and transforming as $w^2$ under $\mathbb{Z}_4$ (real irreducible representation).
Scalars transforming trivially under $\mathbb{Z}_4$ are those that obtain none-zero VEVs $S_4^0$, $H^0$ and $H^0$'s charged multiplet partner $H^{\pm}$. Their mass matrices are given as
\begin{align}
    m^2&=\left(\begin{matrix}\lambda_H v^2 & \lambda_{H S_4}v v_4 \\ \lambda_{H S_4}v v_4 & \lambda_{S_4}v_4^2\end{matrix}\right),
\end{align}
for the $\left(Re[H^0],Re[S_4]\right)$ basis. Here $\left\langle H^0\right\rangle=\frac{v}{\sqrt{2}}$ and $\left\langle S_4\right\rangle=\frac{v_4}{\sqrt{2}}$. Corresponding mass eigenvalues are 
\begin{align}
    m_{1,2}^2&=\frac{\left(\lambda_H v^2+\lambda_{S_4}v_4^2\right)\pm\sqrt{\left(\lambda_H v^2-\lambda_{S_4}v_4^2\right)^2+4\lambda_{H S_4}^2 v^2 v_4^2}}{2},
\end{align}
and mixing angle is given by
\begin{align}
    \text{tan}\left(2\theta_{H^0 S_4}\right)&=\frac{2\lambda_{H S_4}v v_4}{\lambda_H v^2-\lambda_{S_4}v_4^2},\\
    \text{with }\left(\begin{matrix}s_h \\ s_H\end{matrix}\right)&=\left(\begin{matrix}\text{cos}\theta_{H^0 S_4} & \text{sin}\theta_{H^0 S_4} \\ -\text{sin}\theta_{H^0 S_4} & \text{cos}\theta_{H^0 S_4}\end{matrix}\right)\left(\begin{matrix}Re[H^0] \\ Re[S_4]\end{matrix}\right).
\end{align}
$Im[H^0]$ and $Im[S_4]$ correspond to would-be Nambu-Goldstone bosons of $Z$ Standard Model gauge boson corresponding to weak neutral current and $Z^{\prime}$ corresponding to spontaneously broken $U(1)_{B-L}$, hence they get eaten-up and have zero mass matrix. Similarly for $H^{\pm}$, it is a would-be Nambu-Goldstone boson and corresponds to SM $W^{\pm}$. Here $H^{\pm}$ does not mix with $\eta^{\pm}$ due to $\mathbb{Z}_4$ residual discrete symmetry, the former transforms trivially and the later transforms as $w^*$ under $\mathbb{Z}_4$ residual discrete symmetry. The mass of $\eta^{\pm}$ is given as $m^2_{\eta^{\pm}} = m_{\eta}^2 + \lambda_{H\eta}v^2/2 + \lambda_{\eta S_4}v_4^2/2$. 
Scalars transforming as $w$ or $w^*$ under $\mathbb{Z}_4$ also mix, their corresponding mass matrix is given by
\begin{align}
    m^2&=\left(\begin{matrix}m^2_{\eta}+\left(\lambda_{H\eta}+\lambda_{H\eta 2}\right) \frac{v^2}{2}+\lambda_{\eta S_4} \frac{v_4^2}{2} & \pm\mu_H\frac{v}{\sqrt{2}} \\ \pm\mu_H\frac{v}{\sqrt{2}} & m^2_{\chi}+\lambda_{H\chi} \frac{v^2}{2}+\lambda_{\chi S_4} \frac{v_4^2}{2}\end{matrix}\right),
\end{align}
with plus sign corresponding to the $\left(Re[\eta],Re[\chi]\right)$ basis and minus sign corresponding to the $\left(Im[\eta],Im[\chi]\right)$ basis. Corresponding mass eigenvalues are the same for both scalar and pseudo-scalar parts, since they transform as complex $w$ representation under $\mathbb{Z}_4$, and are given as 
\begin{align}
\label{eq:mxi}
    m_{1,2}^2&=\frac{1}{2}\left[m^2_{\eta}+m^2_{\chi}+\left(\lambda_{H\eta}+\lambda_{H\eta 2}+\lambda_{H\chi}\right) \frac{v^2}{2}+\left(\lambda_{\eta S_4}+\lambda_{\chi S_4}\right) \frac{v_4^2}{2}\right]\nonumber \\
    &\pm\frac{1}{2}\sqrt{\left[m^2_{\eta}-m^2_{\chi}+\left(\lambda_{H\eta}+\lambda_{H\eta 2}-\lambda_{H\chi}\right) \frac{v^2}{2}+\left(\lambda_{\eta S_4}-\lambda_{\chi S_4}\right) \frac{v_4^2}{2}\right]^2+4\mu_H^2\frac{v^2}{2}},
\end{align}
and mixing angles are given by
\begin{align}
\label{eq:etachimix}
    \text{tan}\left(2\theta_{\xi R/I}\right)&=\frac{\pm 2\mu_H\frac{v}{\sqrt{2}}}{m^2_{\eta}-m^2_{\chi}+\left(\lambda_{H\eta}+\lambda_{H\eta 2}-\lambda_{H\chi}\right) \frac{v^2}{2}+\left(\lambda_{\eta S_4}-\lambda_{\chi S_4}\right) \frac{v_4^2}{2}},\\
    \text{with }\left(\begin{matrix}\xi_1 \\ \xi_2\end{matrix}\right)&=\left(\begin{matrix}\text{cos}\theta_{\xi} & \text{sin}\theta_{\xi} \\ -\text{sin}\theta_{\xi} & \text{cos}\theta_{\xi}\end{matrix}\right)\left(\begin{matrix}\eta^0 \\ \chi^{\star}\end{matrix}\right).
\end{align}
Lastly, scalars transforming as $w^2$ under $\mathbb{Z}_4$ consist only of $S$. Corresponding scalar and pseuso-scalar mass eigenvalues are 
\begin{align}
    m^2_{\text{Re}[S]}&=m_S^2+\lambda_{H S} \frac{v^2}{2}+\lambda_{S S_4} \frac{v_4^2}{2}+\sqrt{2}\mu_{S_4}v_4,\\
    m^2_{\text{Im}[S]}&=m_S^2+\lambda_{H S} \frac{v^2}{2}+\lambda_{S S_4} \frac{v_4^2}{2}-\sqrt{2}\mu_{S_4}v_4.
\end{align}
Here the mass splitting is due to $\mu_{S_4}$ term in Eq.~\ref{eq:potential} which is allowed by $\mathbb{Z}_4$ residual symmetry, since $S\sim w^2$. 
\section{Dark Matter}
\label{sec:DM}
Dark matter is stabilized by same $\mathbb{Z}_4$ symmetry that ensures the Dirac nature of the neutrinos. Since our model has beyond Standard Model (BSM) fields that transform as $w$ and $w^2$ under $\mathbb{Z}_4$, Dirac as well as Majorana type DM candidates are possible. For Dirac type DM, either $\left(\Psi_1\Psi_3\right)$ Dirac fermion or lighter of $\xi_{1,3}$ mass eigenstates is possible. Whereas if DM of Majorana type, lighter mass eigenstate of $N_{L,R}$ or lighter mass eigenstate of $\Psi_{2 i}$ is a viable candidate. 

Now in order to calculate the relic abundance of the particle dark matter which was in thermal equilibrium, we would need to calculate the Boltzmann equation
\begin{align}
    \frac{d Y_{DM}}{d z} = \frac{-2s(z)}{H(z)z}\langle \sigma v\rangle\left(Y^2_{DM} - (Y^{eq}_{DM})^2\right)
\end{align}
where $Y_{DM} = n_{DM}/s$, $n_{DM}$ is the number density of the dark matter and $s$ is the entropy density. $H$ is the Hubble expansion, $z=M_{DM}/T$ where $T$ is the background temperature and $\langle \sigma v\rangle$ is the thermally averaged cross-section of the dark matter annihilation process given as 
\begin{align}
    \langle \sigma v \rangle &= \frac{1}{8M^4_{DM}T K^2_2(M_{DM}/T)}\int^{\infty}_{4M^2_{DM}}\sigma(s-4M^2_{DM})\sqrt{s}K_1(\sqrt{s}/T).
\end{align}
 We can write the partial wave expansion $\langle \sigma v \rangle = a + bv^2$. Now, the solution of the above Boltzmann equation in terms of this expansion can be given as 
\begin{align}
\Omega h^2 &\sim \frac{1.04\times 10^9 x_f}{M_{pl}\sqrt{g_*}(a+3b/x_f)}
\end{align}
where $M_{pl} = 2.4\times 10^{18}$ GeV and $g_*$ is the number of relativistic 
degrees of freedom at the time of freeze-out. The freeze-out tempertaure can be 
calculated by the following expression
\begin{align}
    x_f &= \ln \left[\frac{0.038g_{DM}M_{pl}M_{DM}\langle \sigma v \rangle}{g^{1/2}_* x^{1/2}_f}\right]
\end{align}
which in turn derived from the equality condition of rate of expansion of the Universe $H\approx g^{1/2}_* T^2/M_{pl}$. 

\begin{figure}[!h]
    \centering
    \includegraphics[width=0.6\textwidth]{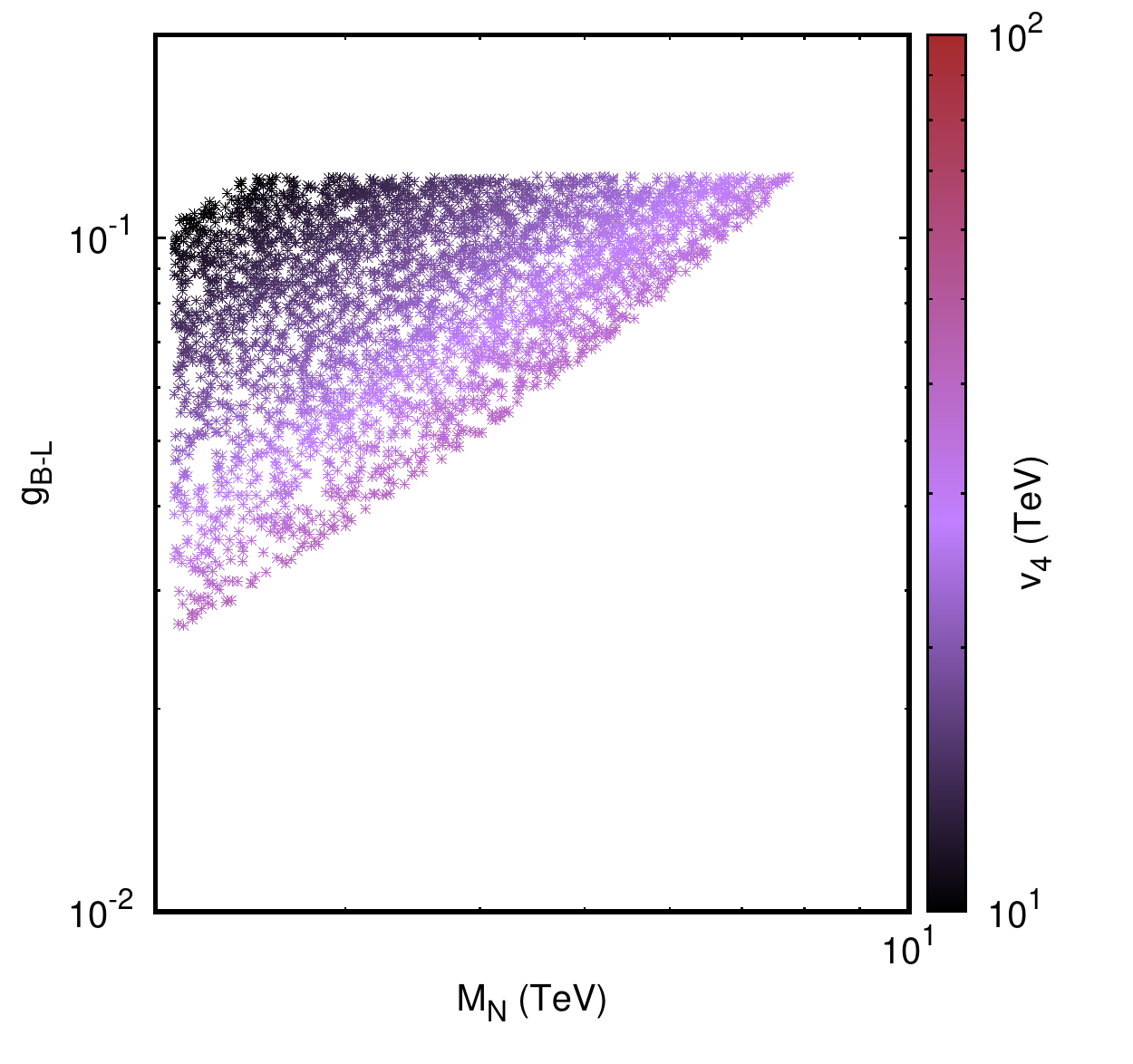}
    \caption{In the above figure we have shown the scatter plot of the $U(1)_{B-L}$ gauge coupling $g_{B-L}$ versus the mass ($\left|m_{N}\right|\approx v_4 \frac{Y_D^2}{Y_M}$ for $\frac{Y_D}{Y_M}\ll 1$) of the Dark Matter ($N$ in this case) while varying the VEV($v_4$) of scalar $S_4$.}
    \label{fig:NDM}
\end{figure}

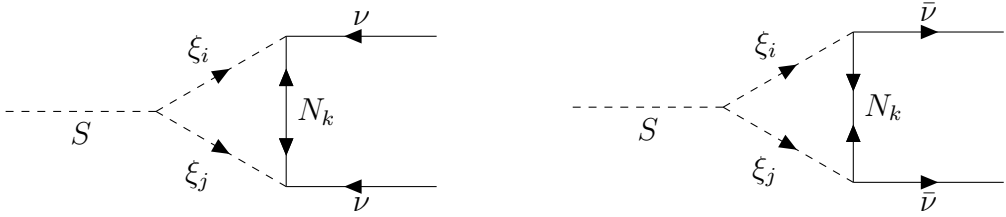
\begin{figure}
    \centering
\begin{subfigure}[b]{0.45\textwidth}
\begin{tikzpicture}
\begin{feynman}
\vertex (i1);
\vertex [right=2cm of i1] (a);
\vertex [right=1.73cm of a] (a2);
\vertex [above=1cm of a2] (b);
\vertex [below=1cm of a2] (c);
\vertex [right=2cm of b] (f1);
\vertex [right=2cm of c] (f2);

\diagram* {
i1 -- [scalar, edge label'=$S$] (a) -- [charged scalar, edge label=$\xi_i$] (b) -- [anti fermion, edge label=$\nu$] (f1),
a -- [charged scalar, edge label'=$\xi_j$] (c) -- [anti fermion, edge label'=$\nu$] (f2),
b -- [anti majorana, edge label=$N_k$] (c),
};
\end{feynman}
\end{tikzpicture}
\end{subfigure}
\begin{subfigure}[b]{0.45\textwidth}
\begin{tikzpicture}
\begin{feynman}
\vertex (i1);
\vertex [right=2cm of i1] (a);
\vertex [right=1.73cm of a] (a2);
\vertex [above=1cm of a2] (b);
\vertex [below=1cm of a2] (c);
\vertex [right=2cm of b] (f1);
\vertex [right=2cm of c] (f2);

\diagram* {
i1 -- [scalar, edge label'=$S$] (a) -- [charged scalar, edge label=$\xi_i$] (b) -- [fermion, edge label=$\Bar{\nu}$] (f1),
a -- [charged scalar, edge label'=$\xi_j$] (c) -- [fermion, edge label'=$\Bar{\nu}$] (f2),
b -- [majorana, edge label=$N_k$] (c),
};
\end{feynman}
\end{tikzpicture}
\end{subfigure}
    \caption{$S$ decay to neutrino pair diagram.}
    \label{fig:sdecay}
\end{figure}

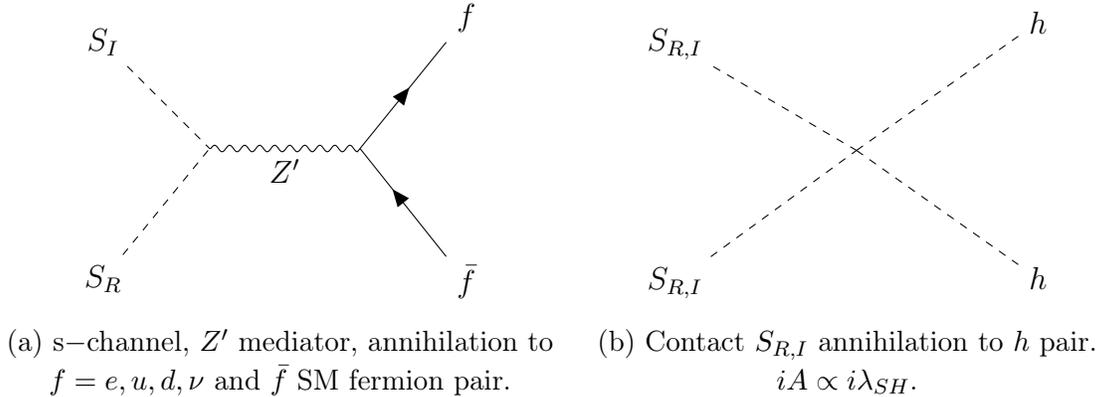
\begin{figure}[!h]
    \centering
    \begin{subfigure}[b]{0.45\textwidth}
\begin{tikzpicture}
\begin{feynman}
\vertex (i1) {$S_I$};
\vertex [below=1.414cm of i1] (im);
\vertex [below=1.414cm of im] (i2) {$S_R$};
\vertex [right=1.414cm of im] (a);
\vertex [right=2cm of a] (b);
\vertex [right=1.414cm of b] (fm);
\vertex [above=1.414cm of fm] (f1) {$f$};
\vertex [below=1.414cm of fm] (f2) {$\bar{f}$};

\diagram* {
(i1) -- [scalar] (a) -- [boson, edge label'=$Z^{\prime}$] (b) -- [fermion] (f1),
(i2) -- [scalar] (a),
(b) -- [anti fermion] (f2),
};
\end{feynman}
\end{tikzpicture}
    \caption{s$-$channel, $Z^{\prime}$ mediator, annihilation to $f=e,u,d,\nu$ and $\Bar{f}$ SM fermion pair.}
    \label{fig:ssigma_vf_s}
    \end{subfigure}
    \begin{subfigure}[b]{0.45\textwidth}
\begin{tikzpicture}
\begin{feynman}
\vertex (i1) {$S_{R,I}$};
\vertex [below=1.414cm of i1] (im);
\vertex [below=1.414cm of im] (i2) {$S_{R,I}$};
\vertex [right=2.414cm of im] (o);
\vertex [right=2.414cm of o] (fm);
\vertex [above=1.414cm of fm] (f1) {$h$};
\vertex [below=1.414cm of fm] (f2) {$h$};

\diagram* {
(i1) -- [scalar] (o) -- [scalar] (f1),
(i2) -- [scalar] (o) -- [scalar] (f2),
};
\end{feynman}
\end{tikzpicture}
    \caption{Contact $S_{R,I}$ annihilation to $h$ pair. $i A\propto i\lambda_{SH}$.}
    \label{fig:ssigma_contact}
    \end{subfigure}
    \caption[$S_{R,I}$ annihilation diagrams.]{$S_{R,I}$ annihilation diagrams.}
    \label{fig:ssigma}
\end{figure}

\begin{figure}[!h]
    \centering
    \includegraphics[width=0.8\textwidth]{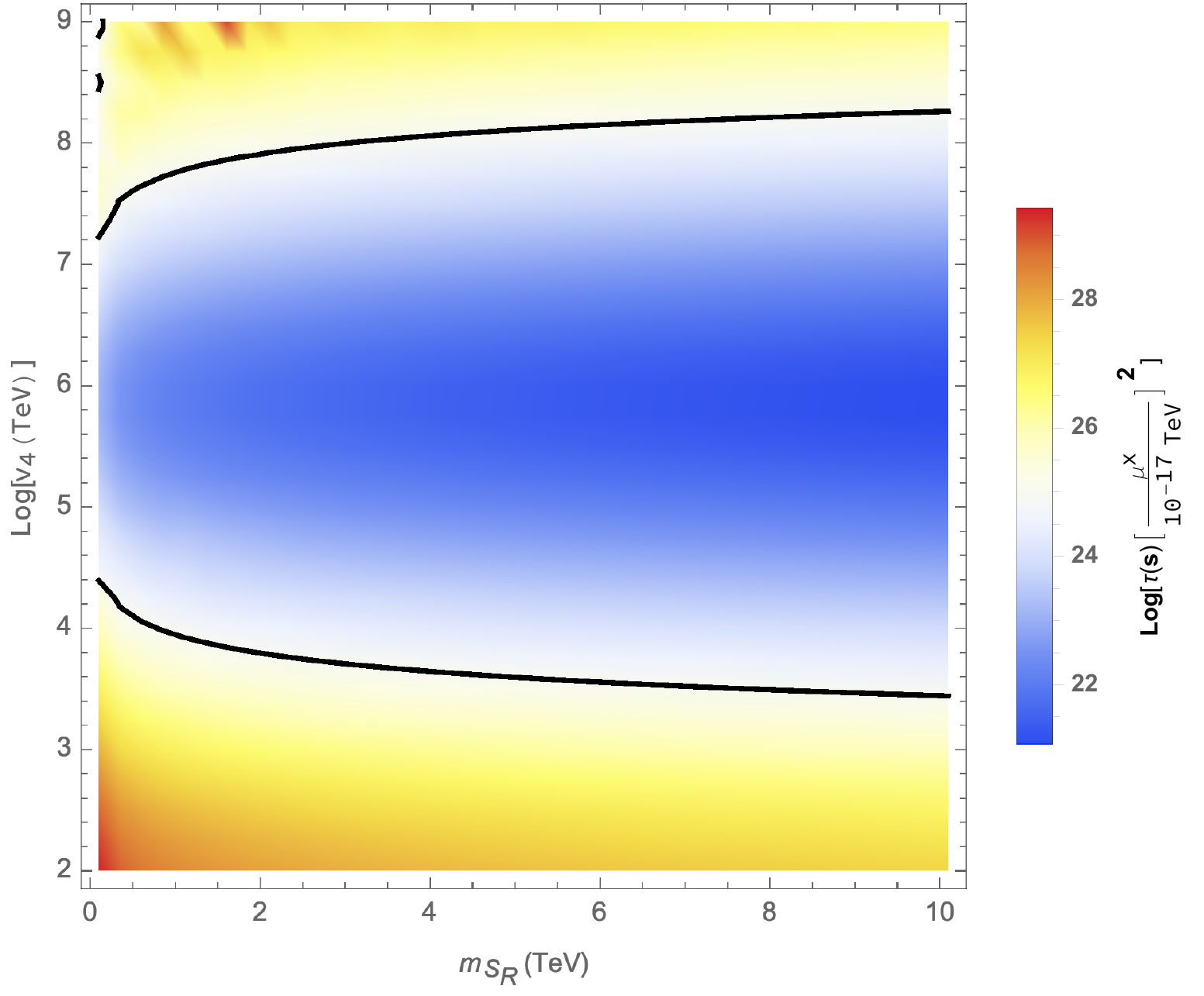}
    \caption{$S_{R,I}$ lifetime dependence on $v_4$, $m_S$, and $\mu^x$ for $\frac{m_{\xi_1}}{m_{\xi_2}}-1=10^{-2}$ and $Y_{M,L}\sim O(1)$, $Y_{R}\sim 10^{-4}$, $\theta_{\xi}=\pi/4$, and $Y_D=10^{-2}$. The solid line indicates the lower limit on the lifetime condition $\tau_s > 10^{25}$s.}
    \label{fig:tau_s}
\end{figure}

Now, since in our case we have additional particles with mass differences close the dark matter, then they can be thermally accessible during the freeze-out. This will eventually give rise to many additional channels through which the dark matter can co-annihilate and give Standard Model ({\bf SM}) particles in the final states. The effective cross-section in this case would be as follows
\begin{align}
    \sigma_eff &= \sum_{i,j}^N \langle \sigma_{ij}v\rangle r_i r_j = \sum_{i,j}^N \langle \sigma_{ij}v\rangle \frac{g_ig_j}{g_{eff}^2}(1+\Delta_i)^{3/2}(1+\Delta_j)^{3/2}e^{-x_f(\Delta_i + \Delta_j)}
\end{align}
where $x_f = M_{DM}/T$, $\Delta_i = M_i/M_{DM}-1$ and 
\begin{align}
    g_{eff} = \sum_{i=1}^N g_i(1+\Delta_i)^{3/2}e^{-x_f\Delta_i} 
\end{align}
And the thermally averaged cross-section is given as 
\begin{align}
    \langle \sigma_{ij} v \rangle &= \frac{x_f}{8m^2_im^2_jK_2((M_i/M_{DM})x_f)K_2((M_j/M_{DM})x_f)} \nonumber \\
    &\times\int_{(M_i+M_j)^2}^\infty \sigma_{ij}(s-2(M^2_i + M^2_j))\sqrt{s}K_1(\sqrt{s}x_f/M_{DM})
\end{align}
One remarkable thing here is that the symmetry that stabilizes DM is the same symmetry that makes neutrinos of Dirac type. The consequence of this is that neutrinos transform non-trivially under DM symmetry, $\mathbb{Z}_4$ in this case. Therefore, any field that transforms as $w^2$ under $\mathbb{Z}_4$ and is in tensor irrep of Poincare symmetry will always decay to pair of neutrinos. On the other hand, fields that transform as $w^2$ and are in spinor irrep of Poincare symmetry will not be able to decay to only neutrinos, therefore the lightest can be DM candidate. 
\\
$\Psi_i$ will not be considered for DM candidate since, as can be seen from eq.~\ref{eq:lag}, they do not participate directly in neutrino mass and \nqbd~ generations and will not lead to interesting phenomenology. Main candidates to consider are $\xi$, $N$, $S$. $\xi$ has a mixing with the neutral component of the $\eta$ doublet, therefore it will have a direct detection channel mediated by $Z$ SM gauge boson and is severely constrained~\cite{Akerib:2016vxi,Akerib:2018lyp,Aprile:2018dbl}. $N$ for which $m_{N}\approx - v_4 \frac{Y_D^2}{Y_M}$ is the best DM candidate, since this $N$ is naturally LSP as required by the smallness of neutrino mass and enhancement of \nqbd. The only neutral $\mathbb{Z}_4$ non-trivial particle that is lighter than $N$ is neutrino, but $N$ decay to neutrinos is forbidden by $\mathbb{Z}_4$ and Poincare symmetry. Decay to the other $\mathbb{Z}_4$ non-trivial particles is forbidden by $U(1)_{em}\times\mathbb{Z}_4$. The annihilation channels for $N$ as a DM candiate are shown in fig.\ref{fig:Nsigma}. And since the dominant channel will be near resonance i.e ($m_{Z^\prime} = 2 m_N$), we have imposed the resonance condition while doing the analysis. The allowed parameter region to satisfy the relic is shown in fig. \ref{fig:NDM}.
\\
\begin{figure}[!h]
    \centering
    \includegraphics[width=0.6\textwidth]{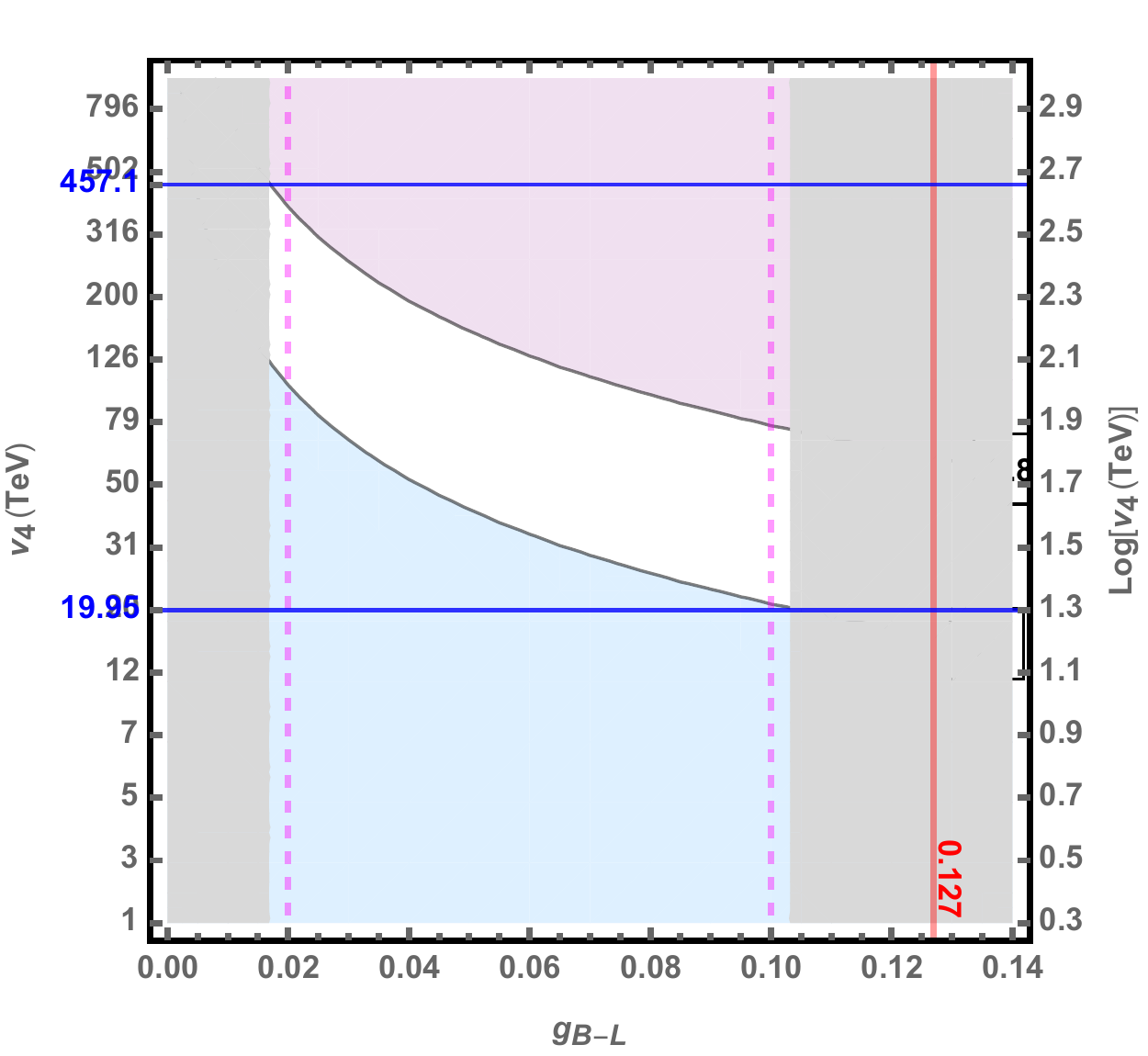}
    \caption{Parameter space of $v_4$, $g_{B-L}$, and $m_{DM}$ constraint by DM relic abundance, $Z^{\prime}$ direct detection, and $m_{Z^{\prime}}=2 m_{N}$ resonance requirement.}
    \label{fig:NDM_const}
\end{figure}
From the plot in Fig.~\ref{fig:NDM} we infer that in order for $N$ to be a plausible dark matter candidate the mass of the lightest $N$ has to be between $2.2 \sim 7.8$ TeV and the coupling $g_{B-L}$ to be between $0.1 \sim 0.02$, Fig.~\ref{fig:NDM_const}.  For the analysis we have implemented the model into {\tt SARAH 4} \cite{Staub:2013tta} and then we took the output to {\tt SPheno 3.1} \cite{Porod:2011nf} to calculate the mass spectrum. Finally for the dark matter analysis we used {\tt MicrOmega 4.3} \cite{Barducci:2016pcb}, using the mass spectrum from {\tt SPheno 3.1}.
\\
Now we focus on $S$ being DM candidate. For $S$ to be a viable DM candidate we assume the following particle mass hierarchy: $m_{\Psi_i},m_{N_j},m_{\xi_k},m_{s4},m_{\eta^+} > m_s > m_h > m_{w},m_z > m_{e,u,d} > m_{\nu}$. Since $S$ is a neutral scalar boson that transforms as $w^2$ under residual $\mathbb{Z}_4$ symmetry, $\mathbb{Z}_4$, $U(1)_{em}$, and Poincare symmetries allow $S$ to decay only to $\nu$'s. Assuming all BSM $\mathbb{Z}_4$ non-singlets are heavier than $S$, the decay of $S$ to neutrino pair is radiative and shown in Fig.~\ref{fig:sdecay}.
The amplitude of the diagram in Fig.~\ref{fig:sdecay} is given by
\begin{align}
    i A &= \left[(s_N,c_N)^2_k Y_L m_k Y_L \left(\begin{matrix} c_{\xi}^2 & -s_{\xi} c_{\xi} \\ -s_{\xi} c_{\xi} & s_{\xi}^2\end{matrix}\right)_{ij} y(k_1)y(k2) \right. \nonumber \\
    &\left.+(c_N,-s_N)^2_k Y_R^{\star}m_k Y_R^{\star} \left(\begin{matrix} s_{\xi}^2 & s_{\xi} c_{\xi} \\ s_{\xi} c_{\xi} & c_{\xi}^2\end{matrix}\right)_{ij} x^{\dagger}(k_1)x^{\dagger}(k_2)\right] \mu_{ij}^x C_0 (0,0,m_s^2,m_j,m_k,m_i),
\end{align}
where $\mu_{ij}^x$ is given in eqs.~\ref{eq:muxr} and~\ref{eq:muxi}, $x,y$ are spinors in 2 component notation, $s(c)_N$ is the mixing angle of $N$ states given in eq.~\ref{eq:nmix}, $s(c)_{\xi}$ is the mixing of $(\eta^0,\chi^{\ast})$ states given in eq.~\ref{eq:etachimix}, and $C_0$ is given in eq.~\ref{eq:c0}.
Then the decay width is given by 
\begin{align}
    &\Gamma(S\rightarrow \nu\nu)=\frac{\sum \left|A\right|^2}{16\pi m_s} \\
    &= \frac{m_s}{16\pi}\left[\left|\underset{ijk}{\sum}\mu_{ij}^{x} C_0(0,0,m_s^2,m_j,m_k,m_i) A_L\right|^2 + \left|\underset{ijk}{\sum}\mu_{ij}^{x} C_0(0,0,m_s^2,m_j,m_k,m_i) A_R\right|^2 \right], \nonumber
\end{align}
where
\begin{align}
    A_L &= (s_N,c_N)^2_k Y_L m_k Y_L \left(\begin{matrix} c_{\xi}^2 & -s_{\xi} c_{\xi} \\ -s_{\xi} c_{\xi} & s_{\xi}^2\end{matrix}\right)_{ij}, \\
    A_R &= (c_N,-s_N)^2_k Y_R^{\star}m_k Y_R^{\star} \left(\begin{matrix} s_{\xi}^2 & s_{\xi} c_{\xi} \\ s_{\xi} c_{\xi} & c_{\xi}^2\end{matrix}\right)_{ij}.
\end{align}
We assume the $S_{R,I}$ mass scale is above EW scale ($v=246$GeV) but below $U(1)_{B-L}$ spontaneous breaking ($v < m_s < v_4$), so at the moment of freeze-out of $S_{R,I}$ EW symmetry is conserved whereas $U(1)_{B-L}$ symmetry is broken to $\mathbb{Z}_4$. Then annihilation of $S$ to SM particles will proceed through the  Feynman diagrams shown in Fig.~\ref{fig:ssigma}.
The inelastic scatterring of ($S_R,(n,p)\rightarrow S_I,(n,p)$) assuming $S_R$ as {\bf DM} via t$-$channel $Z^{\prime}$ mediator can be avoided using same trick as was used in~\cite{Arhrib:2015dez}, namely by making $\left|\Delta m_s^2\right|=2\sqrt{2} |\mu_{S_4}| v_4 > O(10^{2-4}keV^2)$. 
Furthermore, $Z$ SM gauge boson mediated t$-$channel DD is absent since $S_{R,I}$ couples only to $Z^{\prime}$ and  $Z-Z^{\prime}$ mixing appears only at the one-loop order.
\\
$S_{R,I}$'s lifetime, $\tau_s$, dependence on $v_4$, $m_S$, and $\mu^x$ (coupling between $\xi_i\xi_j$ and $s_{R,I}$; eqs.~\ref{eq:muxr} and~\ref{eq:muxi}) is shown in Fig.~\ref{fig:tau_s}.
The age of the Universe is $4.35\times 10^{17}$s, but the bound on decaying dark matter~\cite{Slatyer:2016qyl} is much greater, $\tau > 10^{25}$, from the cosmic microwave background (CMB) constraint.

\begin{figure}[!h]
    \centering
    \begin{subfigure}[b]{0.49\textwidth}
    \includegraphics[width=0.99\textwidth]{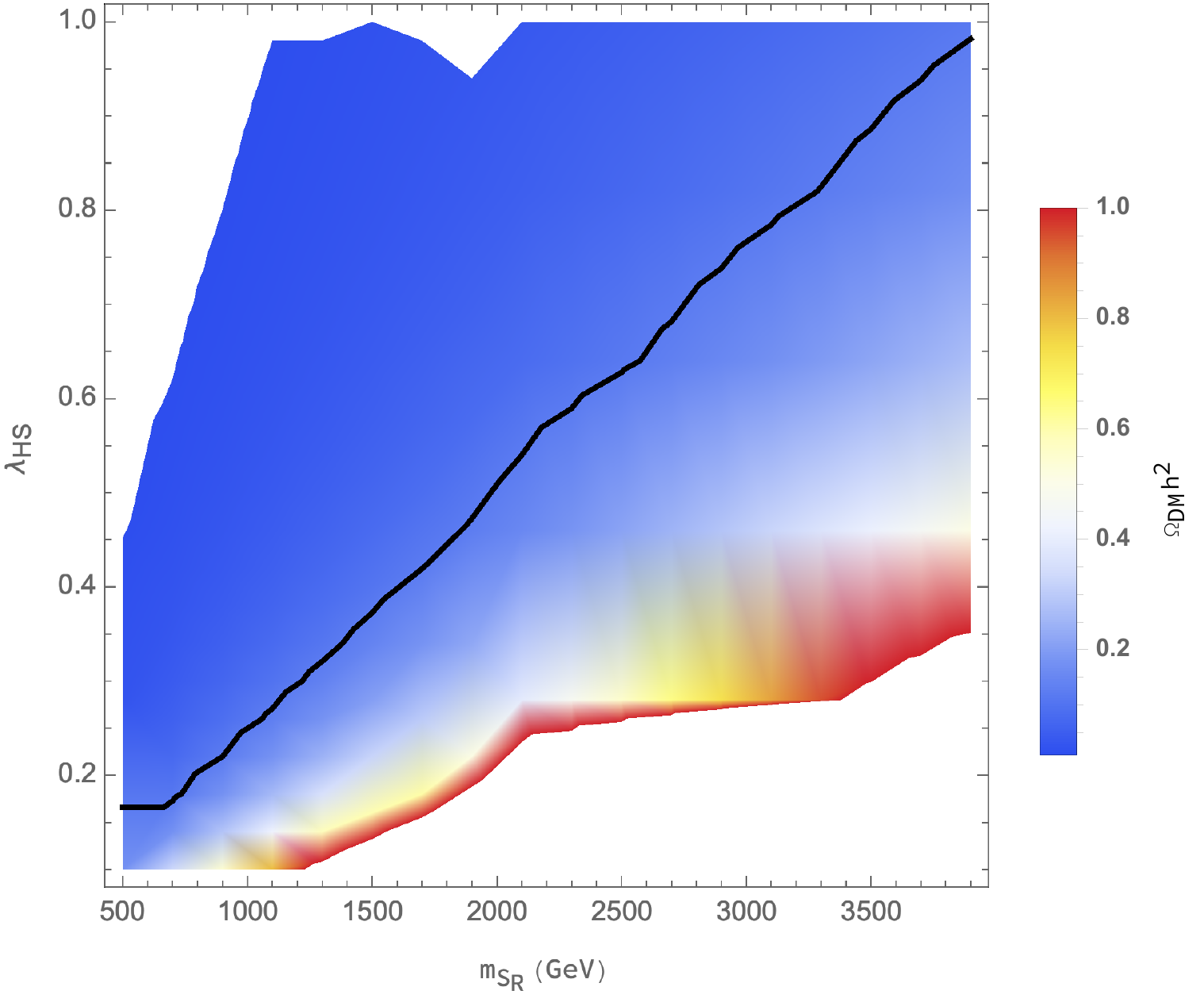}
    \caption{ }
    \label{fig:sOmega}
    \end{subfigure}
    \begin{subfigure}[b]{0.49\textwidth}
    \includegraphics[width=0.99\textwidth]{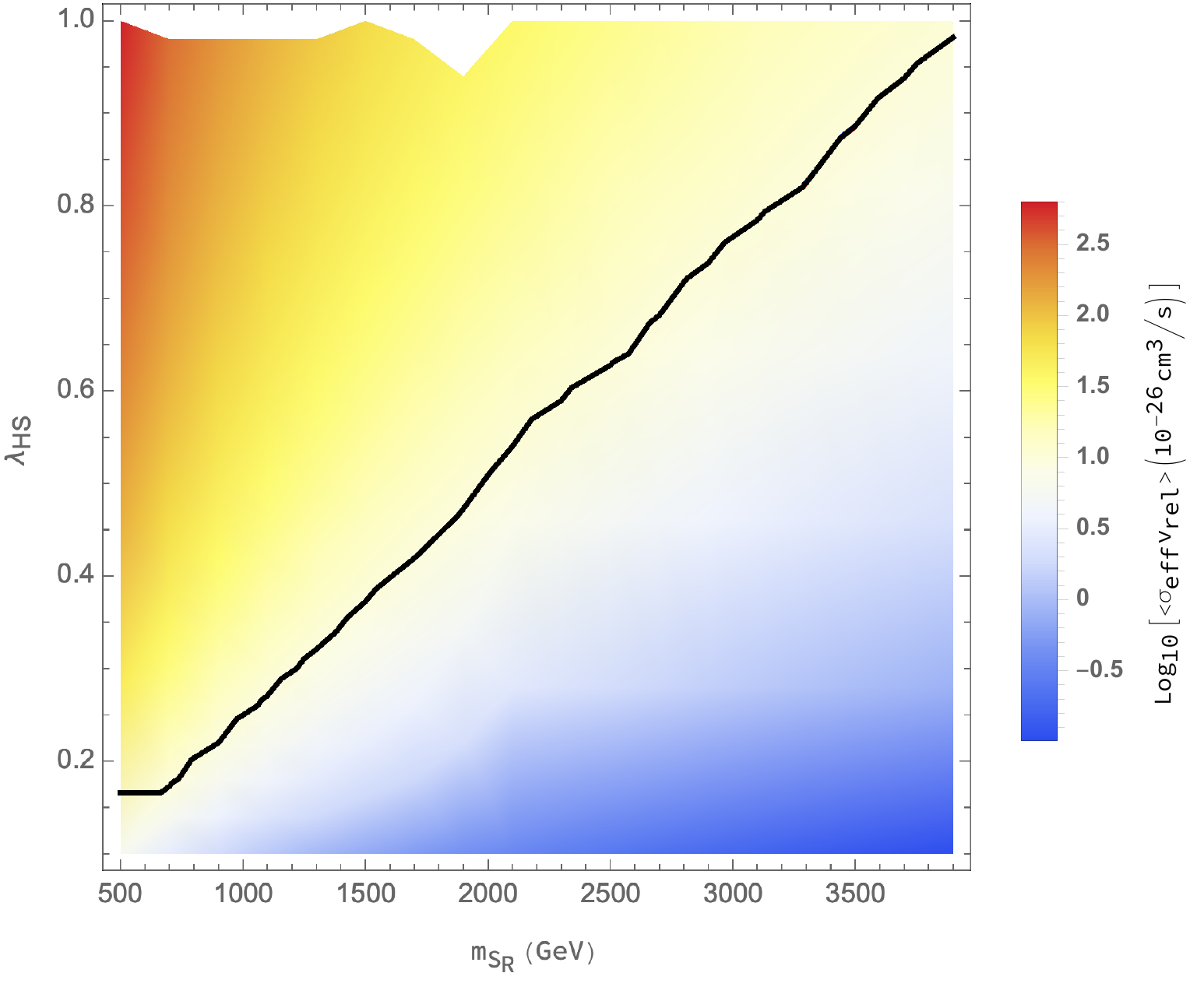}
    \caption{ }
    \label{fig:ssigmaeff}
    \end{subfigure}
    \caption{DM relic abundance $\Omega_{DM}h^2$ and $\left\langle\sigma_{eff}v_{rel}\right\rangle$ cross-section dependence on $\lambda_{HS}$ and $m_S$ for $\frac{m_{S_I}}{m_{S_R}}-1=10^{-2}$, $M_{Z^{\prime}}=4.2$TeV, and $g_{B-L}=0.127$. The solid lines correspond to the relic abundance of $\Omega_{S}h^2=0.120$.}
    \label{fig:srelic}
\end{figure}

\begin{figure}[!h]
    \centering
    \includegraphics[width=0.8\textwidth]{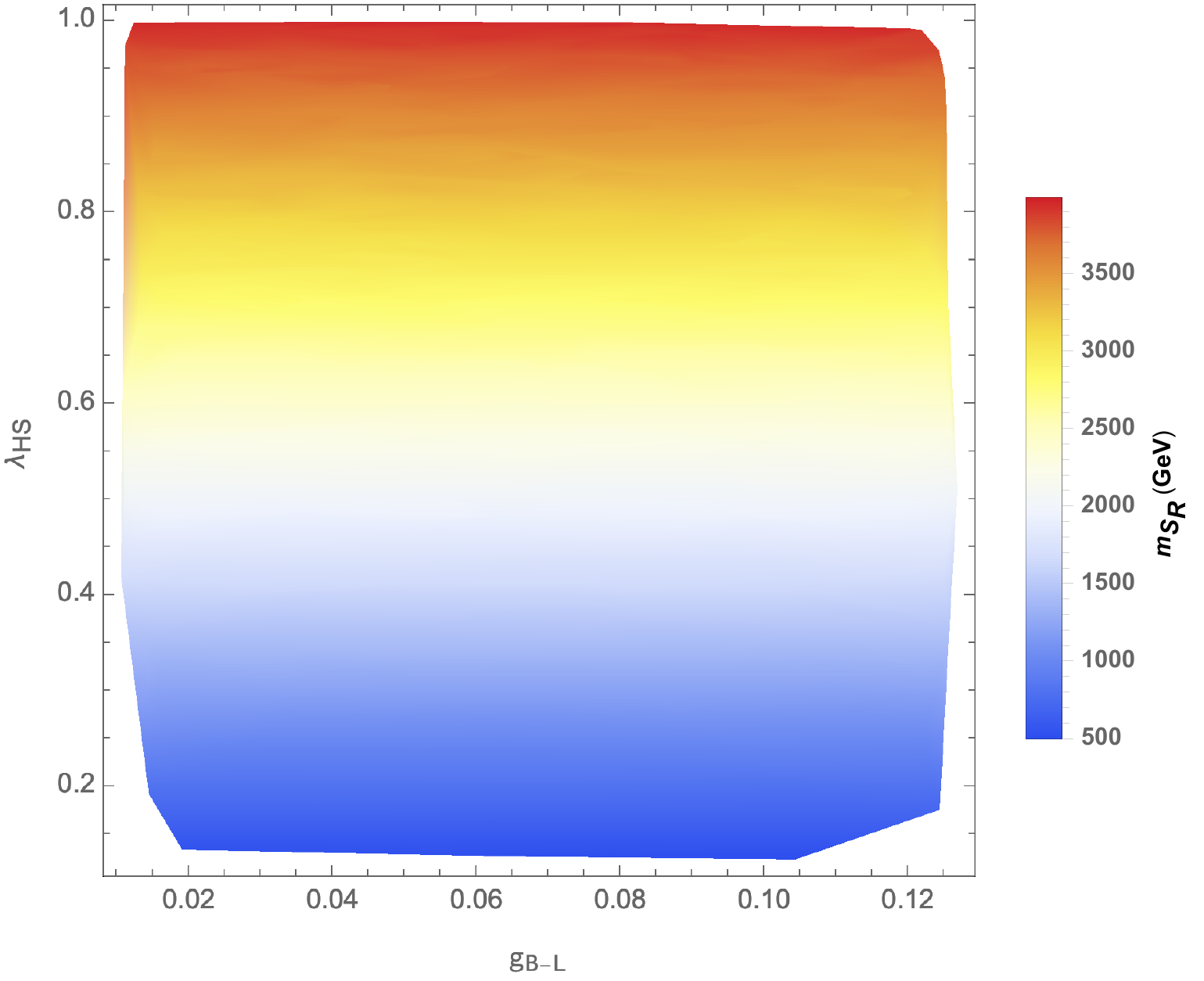}
    \caption{Correlation between $\lambda_{HS}$, $m_S$, and $g_{B-L}$ for $\frac{m_{S_I}}{m_{S_R}}-1=10^{-2}$, $M_{Z^{\prime}}=4.2$TeV fixed that satisfy DM relic abundance of $0.119<\Omega_{S}h^2<0.121$.}
    \label{fig:mslg_plot}
\end{figure}

Important remark regarding Fig.~\ref{fig:tau_s} and $S_{R,I}$ being a viable DM candidate is that to make $S_{R,I}$ long-lived, $\tau_s>10^{25}$s, $\mu^x$ must be tiny ($\approx 10^{-5}$eV). As will be explained in sec.~\ref{sec:nqbd}, in order to have enhanced \nqbd~ $\mu^x\sim \mu_S\approx 10^6$TeV is required. So, for $S_{R,I}$ to be a viable DM candidate means strongly suppressed \nqbd. There are two ways to make $\mu^x$ tiny: either $\left|\sqrt{2}\mu_s + \lambda_2 v_4\right| < 2 \times 10^{-15}$TeV (strong fine-tuning), which will allow for observable \nqbd~ via the other $S$ component ($S_I$) or $\mu_S,\lambda_2 v_4<10^{-15}$TeV in which can \nqbd~ will be strongly suppressed.
\\
We assume that the mass splitting $\Delta m^2_{S}$ between $S_R$ and $S_I$ is small, therefore both $S_R$ and $S_I$ freeze-out simultaneously (with $S_I$ decaying to $S_R$ for $m_{S_R}<m_{S_I}$). Diagrams shown in Fig.~\ref{fig:ssigma} contribute to $\sigma(S_{R,I}S_{R,I}\rightarrow XX)$ $S$ annihilation cross-section in order to get the correct relic abundance for $S_{R,I}$, $\Omega_{S}h^2=0.120$~\cite{Aghanim:2018eyx}. The contact diagram annihilation to Higgs pair is dominant since the $Z^{\prime}$ s$-$channel diagram is suppressed due to large $m_{Z^{\prime}}>4.2$TeV(Sec.~\ref{sec:const}). Even at the resonance, $m_{Z^{\prime}\approx 2 m_S}$, the $Z^{\prime}$ s$-$channel diagram is sub-dominant due to $g_{B-L}<0.127$(Sec.~\ref{sec:const}). Therefore, $S_R$ relic abundance and effective annihilation cross-section for $S_{R,I}$ as a function of DM mass($m_{S_R}$) and coupling $\lambda_{HS}$ with other parameters fixed is plotted in Fig.~\ref{fig:srelic}.
The correlation between $m_{S_R}$, $\lambda_{HS}$, and $g_{B-L}$ is shown in Fig.~\ref{fig:mslg_plot} for the range $0.119<\Omega_{S}h^2<0.121$ and $\frac{m_{S_I}}{m_{S_R}}-1=10^{-2}$, $M_{Z^{\prime}}=4.2$TeV fixed.
As can be seen $S$ is a viable long-lived DM candidate that also allows for correct neutrino masses to be satisfied but will simultaneously lead to highly suppressed \nqbd~ signal. In this case $0\nu 2\beta$ decay is forbidden by Dirac nature of neutrino masses, whereas \nqbd~ signal is highly suppressed. As was shown above, the situation with $N$ is quite different!
%
\section{Neutrinoless Quadruple Beta Decay}
\label{sec:nqbd}
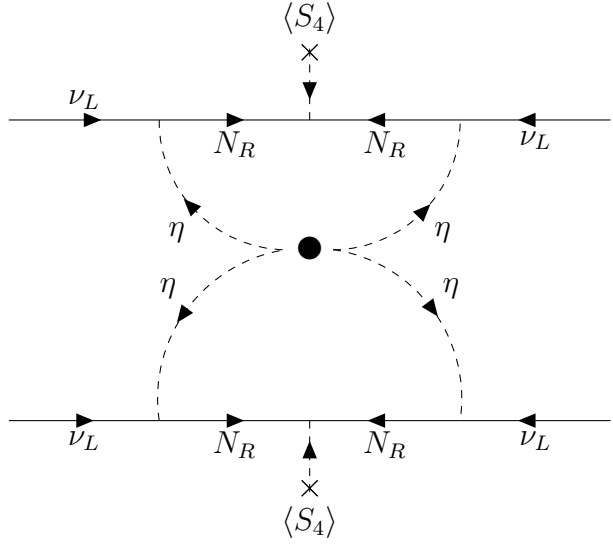
\begin{figure}[ht]
\centering
\begin{tikzpicture}
\begin{feynman}
\vertex (i1);
\vertex [right=2cm of i1] (a);
\vertex [above=4cm of i1] (ii2);
\vertex [right=2cm of a] (b);
\vertex [right=2cm of b] (c);
\vertex [right=2cm of c] (f1);
\vertex [below=1cm of b] (bb) {$\left\langle S_4\right\rangle$};
\vertex [above=2cm of b] (tb1){\ding{108}};
\vertex [right=2cm of ii2] (a2);
\vertex [right=2cm of a2] (b2);
\vertex [right=2cm of b2] (c2);
\vertex [right=2cm of c2] (f2);
\vertex [above=1cm of b2] (bb2) {$\left\langle S_4\right\rangle$};

\diagram* {
i1 -- [fermion, edge label'=$\nu_{L}$] (a) -- [fermion, edge label'=$N_R$] (b) -- [anti fermion, edge label'=$N_R$] (c) -- [anti fermion, edge label'=$\nu_{L}$] (f1),
(a2) -- [fermion, edge label'=$N_R$] (b2) -- [anti fermion, edge label'=$N_R$] (c2) -- [anti fermion, edge label'=$\nu_{L}$] (f2),
b -- [anti charged scalar, insertion=0.9] (bb),
b2 -- [anti charged scalar, insertion=0.9] (bb2),
a -- [anti charged scalar, quarter left, edge label=\(\eta\)] (tb1),
a2 -- [anti charged scalar, quarter right, edge label'=\(\eta\)] (tb1),
tb1 -- [charged scalar, quarter left, edge label=$\eta$] (c),
tb1 -- [charged scalar, quarter right, edge label'=$\eta$] (c2),
(a2) -- [anti fermion, edge label'=$\nu_{L}$] (ii2),
};
\end{feynman}
\end{tikzpicture}
\caption{Diagram contribution to neutrino quadruple beta decay. Blob vertex is given explicitly in Fig.~\ref{fig:eta4}}
\label{fig:nqbd}
\end{figure}

In our construction of the model, by design, due to $\mathbb{Z}_4$ residual symmetry neutrinoless double beta decay ($0\nu 2\beta$) is exactly absent. Therefore the dominant multipole will be neutrinoless quadruple beta decay ($0\nu 4\beta$). Contribution to neutrinoless quadruple beta decay is shown in Fig.~\ref{fig:nqbd}. There will be also a diagram mediated by $\nu_R$ right-chiral neutrinos with $N_R$ replaced by $N_L$ in Fig.~\ref{fig:nqbd}. But due to suppression with neutrino mass at every leg and seesaw suppressed $N_L=\left(\text{cos}\theta N_1-\text{sin}\theta N_2\right)$ Majorana mass (Eq.~\ref{eq:mN12}), contribution mediated by $\nu_R$ can be safely ignored.\\
Reference~\cite{Heeck:2013rpa} is the first paper to study experimental side of $0\nu 4\beta$ with $B-L$ breaking to $\mathbb{Z}_{2n}$ where $n=2$ naturally leading to neutrinoless quadruple beta decay.\\
Neutrinoless quadruple beta decay has been searched for and experimentally studied by NEMO$-$3 collaboration in Refs.~\cite{Arnold:2017bnh,Guzowski:2018neg}. Another study was performed using $^{150}Nd$~\cite{Kidd:2018fbb} nuclei at Kimballton Underground Research Facility setting upper limit for half life-time for $0\nu 4\beta$.

\begin{figure}[!h]
\centering
\begin{subfigure}[b]{0.4\textwidth}
\begin{tikzpicture}
\begin{feynman}
\vertex (i1);
\vertex [above=2cm of i1] (i2);
\vertex [right=1cm of i1] (ii1);
\vertex [right=1cm of i2] (ii2);
\vertex [below=1cm of ii1] (Hib) {$\left\langle H\right\rangle$};
\vertex [above=1cm of ii2] (Hiu) {$\left\langle H\right\rangle$};
\vertex [above=1cm of ii1] (a);
\vertex [right=1cm of a] (b);
\vertex [right=1cm of b] (c);
\vertex [below=1cm of c] (ff1);
\vertex [above=1cm of c] (ff2);
\vertex [right=1cm of ff1] (f1);
\vertex [right=1cm of ff2] (f2);
\vertex [below=1cm of ff1] (Hfb) {$\left\langle H\right\rangle$};
\vertex [above=1cm of ff2] (Hfu) {$\left\langle H\right\rangle$};
\vertex [below=1cm of b] (mb) {$\left\langle S_4\right\rangle$};

\diagram* {
i1 -- [charged scalar, edge label'=$\eta$] (ii1) -- [anti charged scalar, edge label=$\chi$] (a) -- [anti charged scalar, edge label'=$S$] (b) -- [charged scalar, edge label'=$S$] (c) -- [charged scalar, edge label=$\chi$] (ff1) -- [anti charged scalar, edge label'=$\eta$] (f1),
(ii2) -- [anti charged scalar, edge label'=$\chi$] (a),
(ii2) -- [anti charged scalar, edge label'=$\eta$] (i2),
c -- [charged scalar, edge label'=$\chi$] (ff2) -- [anti charged scalar, edge label=$\eta$] (f2),
b -- [anti charged scalar, insertion=0.9] (mb),
(ii1) -- [charged scalar, insertion=0.9] (Hib),
(ii2) -- [charged scalar, insertion=0.9] (Hiu),
(ff1) -- [charged scalar, insertion=0.9] (Hfb),
(ff2) -- [charged scalar, insertion=0.9] (Hfu),
};
\end{feynman}
\end{tikzpicture}
\caption{In the interaction eigenstates $(\eta,\chi)$.}
\label{fig:eta4intb}
\end{subfigure}
\begin{subfigure}[b]{0.4\textwidth}
\centering
\begin{tikzpicture}
\begin{feynman}
\vertex (i1) {$\xi_j$};
\vertex [below=1.414cm of i1] (im);
\vertex [below=1.414cm of im] (i2) {$\xi_i$};
\vertex [right=1.414cm of im] (a);
\vertex [right=2cm of a] (b);
\vertex [right=1.414cm of b] (fm);
\vertex [above=1.414cm of fm] (f1) {$\xi_k$};
\vertex [below=1.414cm of fm] (f2) {$\xi_l$};

\diagram* {
(i1) -- [charged scalar] (a) -- [scalar, edge label'=$S_{R,I}$] (b) -- [anti charged scalar] (f1),
(i2) -- [charged scalar] (a),
(b) -- [anti charged scalar] (f2),
};
\end{feynman}
\end{tikzpicture}
\caption{In the mass eigenstates $(\xi_1,\xi_2)$.}
\label{fig:eta4massb}
\end{subfigure}
\caption{Quartic effective $\eta$ vertex.}
\label{fig:eta4}
\end{figure}
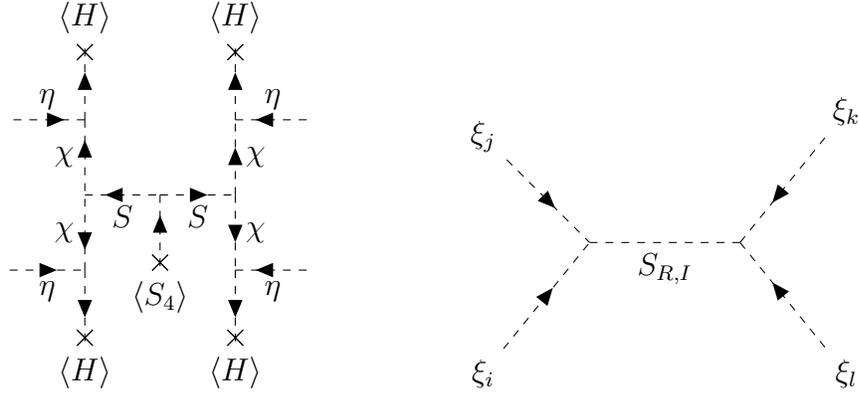
The diagram in Fig.~\ref{fig:eta4} effectively gives the $\mathbb{Z}_4$ invariant vertex
\begin{equation}
    \left(\frac{\mu^R_{ij}\mu^R_{kl}}{m_{s_R}^2}+\frac{\mu^I_{ij}\mu^I_{kl}}{m_{s_I}^2}\right)\xi_i\xi_j\xi_k\xi_l+\text{H.c.}
\end{equation}
The relation between interaction eigenstates $(\eta^0,\chi)$ and mass eigenstates $\xi_i$ is given in Eq.~\ref{eq:etachimix} and $\Delta m_s^2=m_{s_R}^2-m_{s_I}^2=2\sqrt{2}\mu_{S_4}v_4$ is due to $\mu_{S_4}$ term in the scalar potential Eq.~\ref{eq:potential}. The coefficients $\mu_{ij}^{R,I}$ in the basis $(\xi_1,\xi_2)$ are given as
\begin{align}
    \label{eq:muxr}
    \mu_{ij}^R&=\left(\begin{matrix}s^2 & cs \\ cs & c^2\end{matrix}\right)\left(\sqrt{2}\mu_S+\lambda_2\nu_4\right),\\
    \label{eq:muxi}
    \mu_{ij}^I&=\left(\begin{matrix}s^2 & cs \\ cs & c^2\end{matrix}\right)\imath\left(-\sqrt{2}\mu_S+\lambda_2\nu_4\right),
\end{align}
where $s$ and $c$ stand for sin$\theta_{\xi}$ and cos$\theta_{\xi}$, respectively, and $\theta_{\xi}$ was defined in Eq.~\ref{eq:etachimix}. 
$0\nu 4\beta$ can be calculated as two one-loop diagrams. Neutrinoless quadruple beta decay is given by 
\begin{equation}
    \label{eq:quad}
    \frac{Q^{abcd}_{0\nu 4\beta}}{\Lambda^2}\left(\nu_a \nu_b\right)\left(\nu_c\nu_d\right),
\end{equation}
where $Q^{abcd}$ represents quadruple strength, $\Lambda$ is the new physics scale relevant for the neutrinoless quadruple beta decay. $Q^{abcd}/\Lambda^2$ explicitly is given by
\begin{align}
\label{eq:nqbd}
    \frac{Q^{abcd}_{0\nu 4\beta}}{\Lambda^2}&=\frac{-\imath}{\left(16\pi^2\right)^2} c_4 Y_L^{a\alpha}\left(v_s\right)_j\left[s_N^2\frac{F(x_{l1},y_{lj},x_{j1})}{m_{N_1}}+c_N^2\frac{F(x_{l2},y_{lj},x_{j2})}{m_{N_2}}\right]_{\alpha\beta}\left(M_s\right)_{jl}\left(v_s\right)_l Y_L^{\beta b}\times \nonumber \\
    &Y_L^{c\gamma}\left(v_s\right)_i\left[s_N^2\frac{F(x_{k1},y_{ki},x_{i1})}{m_{N_1}}+c_N^2\frac{F(x_{k2},y_{ki},x_{i2})}{m_{N_2}}\right]_{\gamma\delta}\left(M_s\right)_{ik}\left(v_s\right)_k Y_L^{\delta d}\times \nonumber \\
    &\frac{\left(m_{S_I}^2-m_{S_R}^2\right)\left(2\mu_s^2+\lambda_2^2 v_4^2\right)+\left(m_{S_I}^2+m_{S_R}^2\right)2\sqrt{2}\mu_S\lambda_2 v_4}{m_{S_I}^2 m_{S_R}^2},
\end{align}
where the sum over repeated indices is assumed and the Majorana $N$ mass represents the $\Lambda$ scale in Eq.~\ref{eq:quad}. $a,b,c,d,\alpha,\beta,\gamma,\delta$ are flavor indices and take values $1-3$. $v_S$ and $M_S$ are given as
\begin{align}
    v_S&=\left(\begin{matrix}\text{cos}\theta_{\xi} \\ -\text{sin}\theta_{\xi}\end{matrix}\right),\\
    M_S&=\left(\begin{matrix}\text{sin}^2\theta_{\xi} & \text{sin}\theta_{\xi}\text{cos}\theta_{\xi} \\ \text{cos}\theta_{\xi}\text{sin}\theta_{\xi} & \text{cos}^2\theta_{\xi}\end{matrix}\right).
\end{align}
$\theta_{\xi}$ is mixing angle between $\eta$ and $\chi$ scalars and was given in Eq.~\ref{eq:etachimix}. $F(x,y,z)$ is the loop function and is given by
\begin{align}
    F(x,y,z)&=\frac{\text{ln}x}{(1-z)(1-x)}-\frac{z\text{ln}y}{(1-z)(1-y)}.
\end{align}
In Eq.~\ref{eq:nqbd}, $x_{ij}$ and $y_{ij}$ are given by
\begin{align}
    x_{ij}&=\frac{m_{i}^2}{m_{N_j}^2}\\
    y_{ij}&=\frac{m_i^2}{m_j^2},
\end{align}
where $m_i$ is the mass eigenstate of $\xi_i$ given in Eq.~\ref{eq:mxi} and $m_{N_i}$ is the Majorana mass eigenstate of $N_{i}$ given in Eq.~\ref{eq:mN12}. $s_N$ and $c_N$ stand for the sine and cosine of the mixing angle of the $N_{L,R}$ fermions and are given in Eq.~\ref{eq:nmix}. Lastly, $c_4$ is the combinatorics factor and is given as
\begin{align}
    c_4&=\left\{\begin{matrix}4! & i=j=k=l|i,j,k,l\in\{1,2\} \\ 3! & i=j=k\neq l\lor i=j=l\neq k\lor i=k=l\neq j\lor i\neq j=k=l |i,j,k,l\in\{1,2\} \\ 4 & (i=j\land k=l)\lor (i=k\land j=l)\lor (i=l\land j=k)|i,j,k,l\in\{1,2\}\end{matrix}\right. .
\end{align}
Important remark regarding eq.~\ref{eq:nqbd} is the presence of the $\lambda_2 v_4$ cross term in the last line. If $\lambda_2$ was absent (forbidden) in the model then $0\nu 4\beta$ would be proportional to the splitting of $S$ scalar and pseudo-scalar masses, which is controlled by $\mu_{S_4}$ term in eq.~\ref{eq:potential}. Neutrino mass suppression factors like $Y_L$, $\theta_{\xi}$, $\Delta m_{\xi}$, $v_4$ also suppress \nqbd~ but $\mu_S$ freedom can be used to control the enhancement of \nqbd. In the case if $\mu_S\gg v_4\sim O(10^{2-3}\text{TeV})$ the $\mu_S^2$ term will dominate and \nqbd~ will scale as $\mu_S^2$.

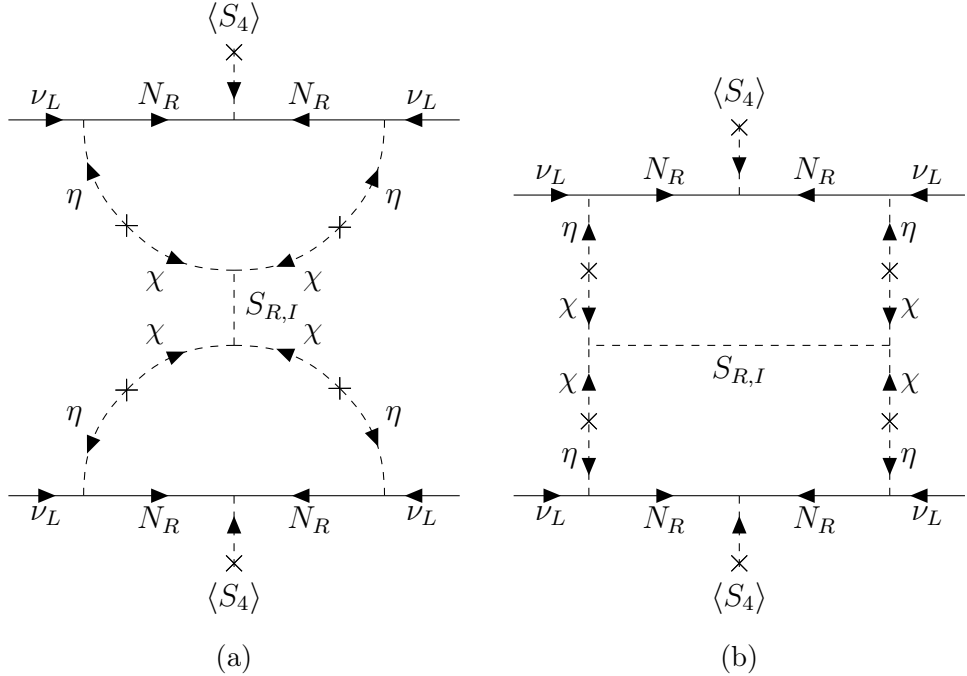
\begin{figure}[ht]
\centering
\begin{subfigure}[b]{0.4\textwidth}
\begin{tikzpicture}
\begin{feynman}
\vertex (i1);
\vertex [right=1cm of i1] (a);
\vertex [above=5cm of i1] (ii2);
\vertex [right=2cm of a] (b);
\vertex [right=2cm of b] (c);
\vertex [right=1cm of c] (f1);
\vertex [below=1cm of b] (bb) {$\left\langle S_4\right\rangle$};
\vertex [above=2cm of b] (tb1);
\vertex [above=1.417cm of b] (xbm);
\vertex [left=1.417cm of xbm] (xbl);
\vertex [right=1.417cm of xbm] (xbr);
\vertex [above=1cm of tb1] (tb2);
\vertex [right=1cm of ii2] (a2);
\vertex [right=2cm of a2] (b2);
\vertex [right=2cm of b2] (c2);
\vertex [right=1cm of c2] (f2);
\vertex [above=1cm of b2] (bb2) {$\left\langle S_4\right\rangle$};
\vertex [below=1.417cm of b2] (xum);
\vertex [left=1.417cm of xum] (xul);
\vertex [right=1.417cm of xum] (xur);

\diagram* {
i1 -- [fermion, edge label'=$\nu_{L}$] (a) -- [fermion, edge label'=$N_R$] (b) -- [anti fermion, edge label'=$N_R$] (c) -- [anti fermion, edge label'=$\nu_{L}$] (f1),
(a2) -- [fermion, edge label=$N_R$] (b2) -- [anti fermion, edge label=$N_R$] (c2) -- [anti fermion, edge label=$\nu_{L}$] (f2),
b -- [anti charged scalar, insertion=0.9] (bb),
b2 -- [anti charged scalar, insertion=0.9] (bb2),
a -- [anti charged scalar, out=90, in=-135, insertion=0.99, edge label=$\eta$] (xbl) -- [charged scalar, out=45, in=180, edge label=$\chi$] (tb1),
tb1 -- [anti charged scalar,  out=0, in=135, insertion=0.99, edge label=$\chi$] (xbr) -- [charged scalar,  out=-45, in=90, edge label=$\eta$] (c),
a2 -- [anti charged scalar, out=-90, in=135, insertion=0.99, edge label'=$\eta$] (xul) -- [charged scalar, out=-45, in=-180, edge label'=$\chi$] (tb2),
tb2 -- [anti charged scalar,  out=-0, in=-135, insertion=0.99, edge label'=$\chi$] (xur) -- [charged scalar,  out=45, in=-90, edge label'=$\eta$] (c2),
(a2) -- [anti fermion, edge label'=$\nu_{L}$] (ii2),
(tb1) -- [scalar, edge label'=$S_{R,I}$] (tb2),
};
\end{feynman}
\end{tikzpicture}
\caption{ }
\label{fig:nqbd2_1}
\end{subfigure}
\begin{subfigure}[b]{0.4\textwidth}
\centering
\begin{tikzpicture}
\begin{feynman}
\vertex (i1);
\vertex [right=1cm of i1] (a);
\vertex [above=4cm of i1] (ii2);
\vertex [right=2cm of a] (b);
\vertex [right=2cm of b] (c);
\vertex [right=1cm of c] (f1);
\vertex [below=1cm of b] (bb) {$\left\langle S_4\right\rangle$};
\vertex [above=1cm of a] (tbab);
\vertex [above=1cm of tbab] (tbam);
\vertex [above=1cm of tbam] (tbau);
\vertex [above=1cm of c] (tbcb);
\vertex [above=1cm of tbcb] (tbcm);
\vertex [above=1cm of tbcm] (tbcu);
\vertex [right=1cm of ii2] (a2);
\vertex [right=2cm of a2] (b2);
\vertex [right=2cm of b2] (c2);
\vertex [right=1cm of c2] (f2);
\vertex [above=1cm of b2] (bb2) {$\left\langle S_4\right\rangle$};

\diagram* {
i1 -- [fermion, edge label'=$\nu_{L}$] (a) -- [fermion, edge label'=$N_R$] (b) -- [anti fermion, edge label'=$N_R$] (c) -- [anti fermion, edge label'=$\nu_{L}$] (f1),
(a2) -- [fermion, edge label=$N_R$] (b2) -- [anti fermion, edge label=$N_R$] (c2) -- [anti fermion, edge label=$\nu_{L}$] (f2),
b -- [anti charged scalar, insertion=0.9] (bb),
b2 -- [anti charged scalar, insertion=0.9] (bb2),
a -- [anti charged scalar, insertion=0.99, edge label=\(\eta\)] (tbab) -- [charged scalar, edge label=\(\chi\)] (tbam) -- [anti charged scalar, insertion=0.99, edge label=\(\chi\)] (tbau) -- [charged scalar, edge label=\(\eta\)] (a2),
c -- [anti charged scalar, insertion=0.99, edge label'=\(\eta\)] (tbcb) -- [charged scalar, edge label'=\(\chi\)] (tbcm) -- [anti charged scalar, insertion=0.99, edge label'=\(\chi\)] (tbcu) -- [charged scalar, edge label'=\(\eta\)] (c2),
(a2) -- [anti fermion, edge label'=$\nu_{L}$] (ii2),
(tbcm) -- [scalar, edge label=$S_{R,I}$] (tbam),
};
\end{feynman}
\end{tikzpicture}
\caption{ }
\label{fig:nqbd2_2}
\end{subfigure}
\caption{2 two-loop diagrams contributing to neutrino quadruple beta decay.}
\label{fig:nqbd2}
\end{figure}

\begin{figure}[!h]
    \centering
    \includegraphics[width=0.8\textwidth]{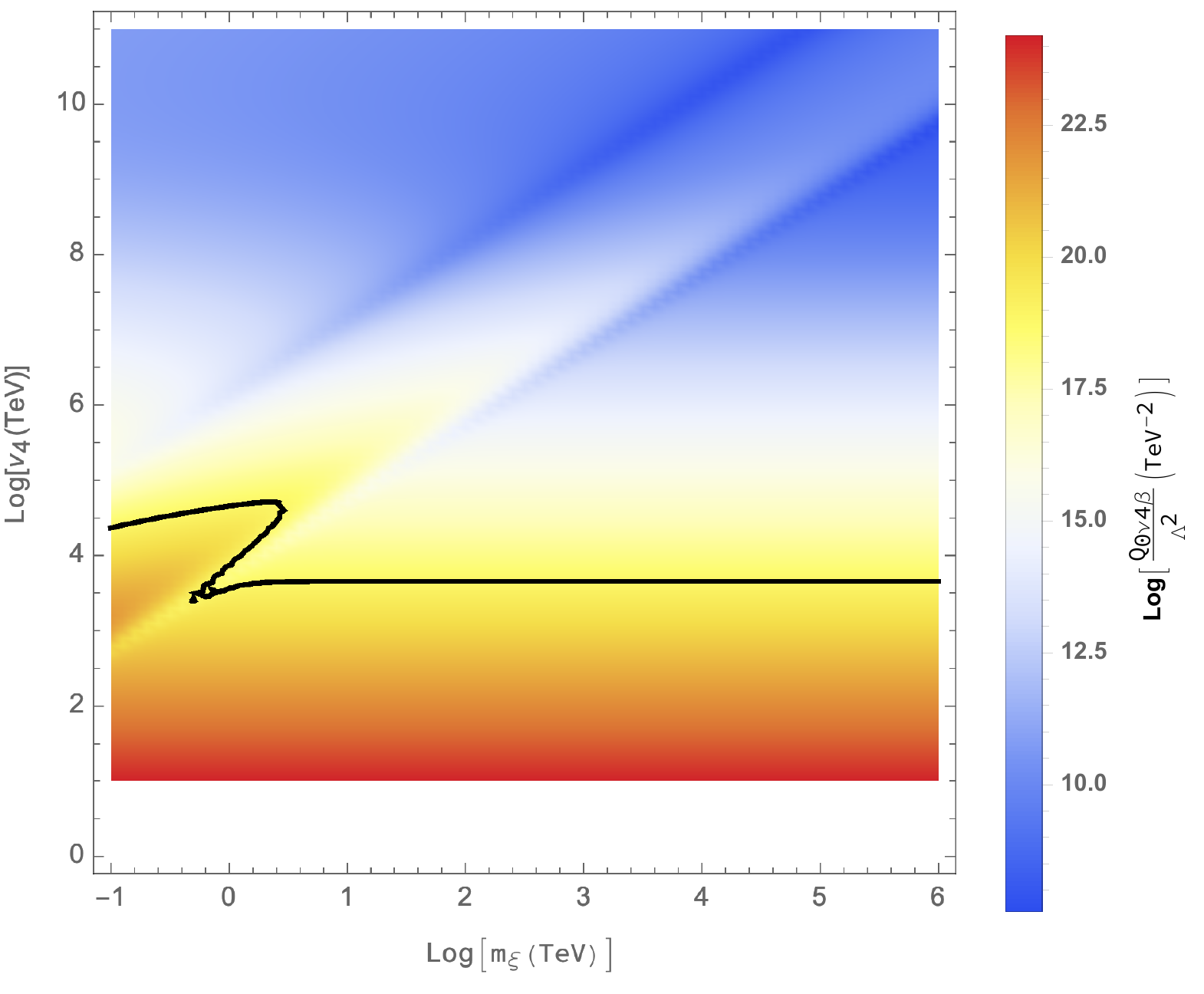}
    \caption{Neutrinoless quadruple beta decay dependence(eq.~\ref{eq:nqbd}) on $m_{\xi}$ and $v_4$ for $\mu_S=10^9$TeV, $\frac{m_{\xi_2}}{m_{\xi_1}}-1=10^{-2}$, $\frac{Y_D}{Y_M}=10^{-3}$, $Y_{M,L}\sim O(1)$, $\theta_{\xi}=\pi/4$, and $m_{S_{R(I)}}=0.8(2)$TeV. For $v_4\ll\mu_S$ $\frac{Q_{0\nu 4\beta}}{\Lambda^2}\propto \mu_S^2$. Solid curve corresponds to current half-life constraint on $\frac{Q_{0\nu 4\beta}}{\Lambda^2}$ from NEMO-3~\cite{Arnold:2017bnh} and Kimballton Underground Research Facility~\cite{Kidd:2018fbb} experiments.}
    \label{fig:nqbdplt}
\end{figure}

\begin{figure}[!h]
    \centering
    \includegraphics[width=0.8\textwidth]{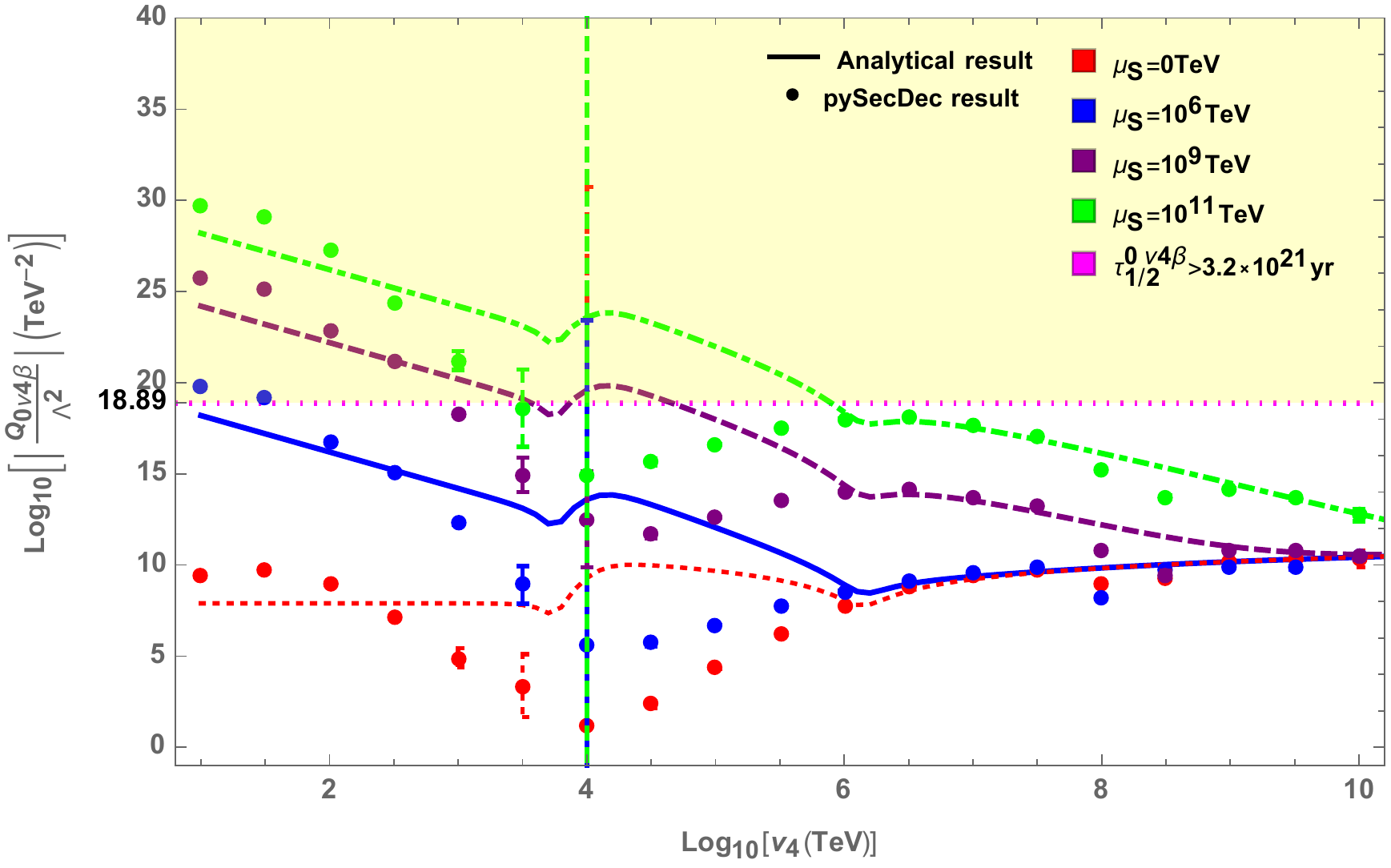}
    \caption{Comparison of numerically obtained result from \emph{pySecDec} with analytically derived formula in eq.~\ref{eq:nqbd} for neutrinoless quadruple beta decay for different values of $v_4$, and $\mu_S$ for $m_{\xi}=1$TeV, $\frac{m_{\xi_2}}{m_{\xi_1}}-1=10^{-2}$, $\frac{Y_D}{Y_M}=10^{-3}$, $Y_{M,L}\sim O(1)$, $\lambda_2\sim O(1)$, $\theta_{\xi}=\pi/4$, and $m_{S_{R(I)}}=0.8(2)$TeV.}
    \label{fig:nqbd2D}
\end{figure}

Below numerical calculation of $\frac{Q_{0\nu 4\beta}}{\Lambda^2}$ is performed using \emph{pySecDec}~\cite{Carter:2010hi} software tool. Diagrams that have dominant contribution to $\frac{Q_{0\nu 4\beta}}{\Lambda^2}$ are the ones with $\nu_L$ legs and are shown in Fig.~\ref{fig:nqbd2}. There are also diagrams with $\nu_L$ replaced by $\nu_R$ but they are suppressed by a factor of $\frac{m_{\nu}}{p_{\nu}}$ for each $\nu_L\rightarrow\nu_R$ leg replacement.
Diagrams in Fig.~\ref{fig:nqbd2} produce loop integrals
\begin{align}
    \label{eq:int1_1}
    i A_1&= i y_a(p_4) y_c (p_3)\left[Y_L^{\prime} m_{N_{m}}Y_L^{\prime}\right]^{ab} \left[Y_L^{\prime} m_{N_{n}}Y_L^{\prime}\right]^{cd} \mu_{ij}^{x}\mu_{kl}^x \int\frac{d^d k}{\left(2\pi\right)^d} \int\frac{d^d l}{\left(2\pi\right)^d} \left[\left(l+p_2\right)^2-m_{s_i}^2\right]^{-1} \\
    &\left[\left(l-p_4\right)^2-m_{s_j}^2\right]^{-1} \left[\left(k+p_3\right)^2-m_{s_k}^2\right]^{-1} \left[\left(k-p_1\right)^2-m_{s_l}^2\right]^{-1} \left[l^2-m_{N_m}^2\right]^{-1} \left[k^2-m_{N_n}^2\right]^{-1} \nonumber \\
    &\left[\left(p_1+p_3\right)^2-m_x^2\right]^{-1} x_b(p_2) x_d(p_1), \nonumber \\
    \label{eq:int2_1}
    i A_2&= i y_a(p_4) y_c (p_3)\left[Y_L^{\prime} m_{N_{m}}Y_L^{\prime}\right]^{ab} \left[Y_L^{\prime} m_{N_{n}}Y_L^{\prime}\right]^{cd} \mu_{ij}^{x}\mu_{kl}^x \int\frac{d^d k}{\left(2\pi\right)^d} \int\frac{d^d l}{\left(2\pi\right)^d} \left[\left(l+p_2\right)^2-m_{s_i}^2\right]^{-1} \\
    &\left[\left(l-p_4\right)^2-m_{s_j}^2\right]^{-1} \left[\left(k+p_3\right)^2-m_{s_k}^2\right]^{-1} \left[\left(k-p_1\right)^2-m_{s_l}^2\right]^{-1} \left[l^2-m_{N_m}^2\right]^{-1} \left[k^2-m_{N_n}^2\right]^{-1} \nonumber \\
    &\left[\left(l-k+p_1+p_2\right)^2-m_x^2\right]^{-1} x_b(p_2) x_d(p_1), \nonumber \\
\end{align}
with $\sum_i p_i=0$ and $Y_L^{\prime}$ given below.
\begin{align}
    \Bar{N}_{aR}Y_L^{ab}L_{bi}\eta_j\epsilon^{ij}+\text{H.c.}&=\left(s_N \Bar{N}_1 + c_N \Bar{N}_2\right)_a Y_L^{ab}\nu_{Lb}\left(c_{\xi}\xi_1 - s_{\xi}\xi_2\right)+\text{H.c} \\
    &=\Bar{N}_{a}^i Y_{L,i}^{\prime ab,j} \nu_{Lb}\xi_j + \text{H.c.},\nonumber
\end{align}
where
\begin{align}
    Y_{L,i}^{\prime ab,j}&=\left(\sin\theta_N, \cos\theta_N\right)_{i} Y_L^{ab}\left(\cos\theta_{\xi}, -\sin\theta_{\xi}\right)^{j}.
\end{align}
After using \emph{pySecDec} python code to calculate these integrals numerically, we compare numerical results with analytically obtained results in eq.~\ref{eq:nqbd} and plot both in Figs.~\ref{fig:nqbdplt} and~\ref{fig:nqbd2D}.
$\frac{Q_{0\nu 4\beta}}{\Lambda^2}$ dependence on $m_{\xi}$, $v_4$, and $\mu_S$ for the analytical result from eq.~\ref{eq:nqbd} is plotted in fig.~\ref{fig:nqbdplt} with the other parameters fixed. As can be seen from eq.~\ref{eq:nqbd}, for $\lambda_2 v_4\ll\mu_S$ $\frac{Q_{0\nu 4\beta}}{\Lambda^2}\propto \mu_S^2$ and the $\mu_S$ can be used to enhance the $\frac{Q_{0\nu 4\beta}}{\Lambda^2}$ for possible detection in the upcoming \nqbd~ experiments.
\\
Current half-life lower limit on $\frac{Q_{0\nu 4\beta}}{\Lambda^2}$ is given as $\tau_{1/2}^{0\nu 4\beta}>3.2\times 10^{21}$years~\cite{Arnold:2017bnh}. The relation between half-life and $\frac{Q_{0\nu 4\beta}}{\Lambda^2}$ is given as
\begin{align}
    \tau_{1/2}^{-1} &= \Gamma = G_{0\nu 4\beta} \left|A_{0\nu 4\beta}\right|^2 = G_{0\nu 4\beta} \left(\frac{G_F^4}{q^4}\frac{Q_{0\nu 4\beta}}{\Lambda^2}\right)^2,
\end{align}
where $G_{0\nu 4\beta}$ is the four particle phase space factor and $A_{0\nu 4\beta}$ is the matrix element for \nqbd~ process. \cite{Arnold:2017bnh} and~\cite{Kidd:2018fbb} use $^{150}$Nd$\rightarrow ^{150}$Gd which has $Q=2.079-2.084$MeV. $q$ can be estimated as $p_{\nu}=|q|\approx 1$fm$^{-1}\approx 100$MeV.
Then $\frac{Q_{0\nu 4\beta}}{\Lambda^2}$ can be estimated from
\begin{align}
    \Gamma_{0\nu 4\beta} &= Q^{11} (\frac{G_F^4}{q^4}\frac{Q_{0\nu 4\beta}}{\Lambda^2})^2 q^{18}\text{~\cite{Heeck:2013rpa}},
\end{align}
where the last factor was inserted for dimensional matching. Using this estimate and half-life lower limit of $\tau_{1/2}^{0\nu 4\beta}>3.2\times 10^{21}$ we get
\begin{align}
    \frac{Q_{0\nu 4\beta}}{\Lambda^2} &\leq \left(\tau_{1/2}Q^{11}q^{10}G_F^8\right)^{-1/2} = 7.8\times 10^{18} \text{TeV}^{-2}.
\end{align}
Fig.~\ref{fig:nqbd2D} shows the comparison of numerical results from \emph{pySecDec} with approximate analytical expression from eq.~\ref{eq:nqbd}. As can be seen from the plot, the coupled loop(Fig.~\ref{fig:nqbd2_2}) is relevant at $v_4<10^6$TeV scales, where it of the order of the decoupled loop(Fig.~\ref{fig:nqbd2_1}) and can interfere destructively ($10^3\text{TeV}<v_4<10^6\text{TeV}$). For $v_4>10^6$TeV coupled loop becomes irrelevant. Important thing to notice is that $\mu_S$ plays crucial role at enhancing \nqbd~ at $v_4<10^6$TeV scales (for $\lambda_2=1$). So, this model predicts a possibility for an enhanced $\frac{Q_{0\nu 4\beta}}{\Lambda^2}$ which can be probed in the future \nqbd~ experiments.
\section{Results and discussion}
\label{sec:const}
\begin{figure}[!h]
    \centering
    \includegraphics[width=0.7\textwidth]{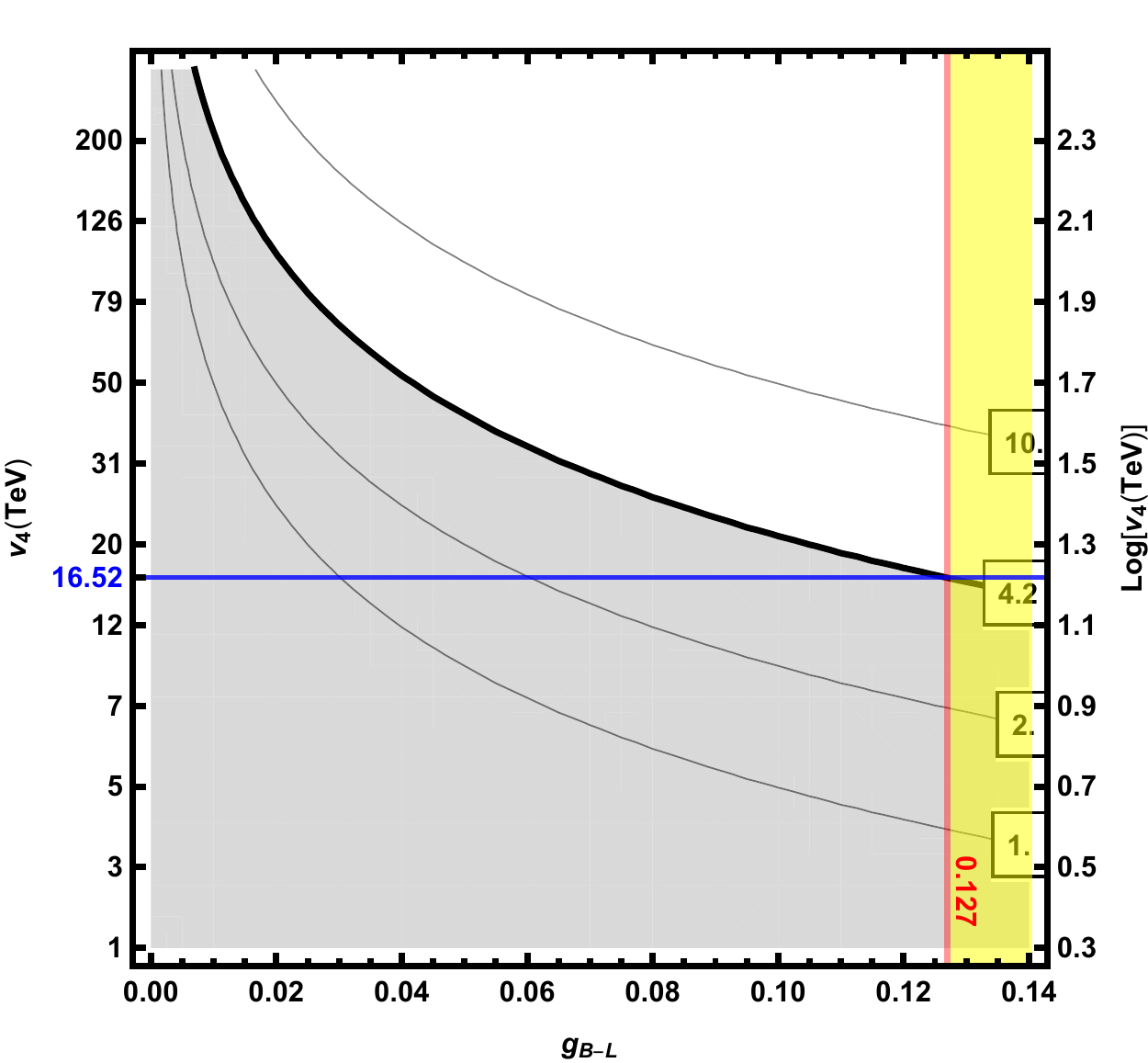}
    \caption{Plot showing correlation between $g_{B-L}$ and $v_4$ with unitarity and collider constraints, $g_{B-L}<0.127$ and $M_Z^{\prime}>4.2$TeV(Tab.~\ref{tab:pheno_bounds}), implemented. Contours represent $M_Z^{\prime}$ in TeV units.}
    \label{fig:gbl_v4}
\end{figure}

\begin{figure}[!h]
    \centering
    \begin{subfigure}[b]{0.8\textwidth}
    \includegraphics[width=1\textwidth]{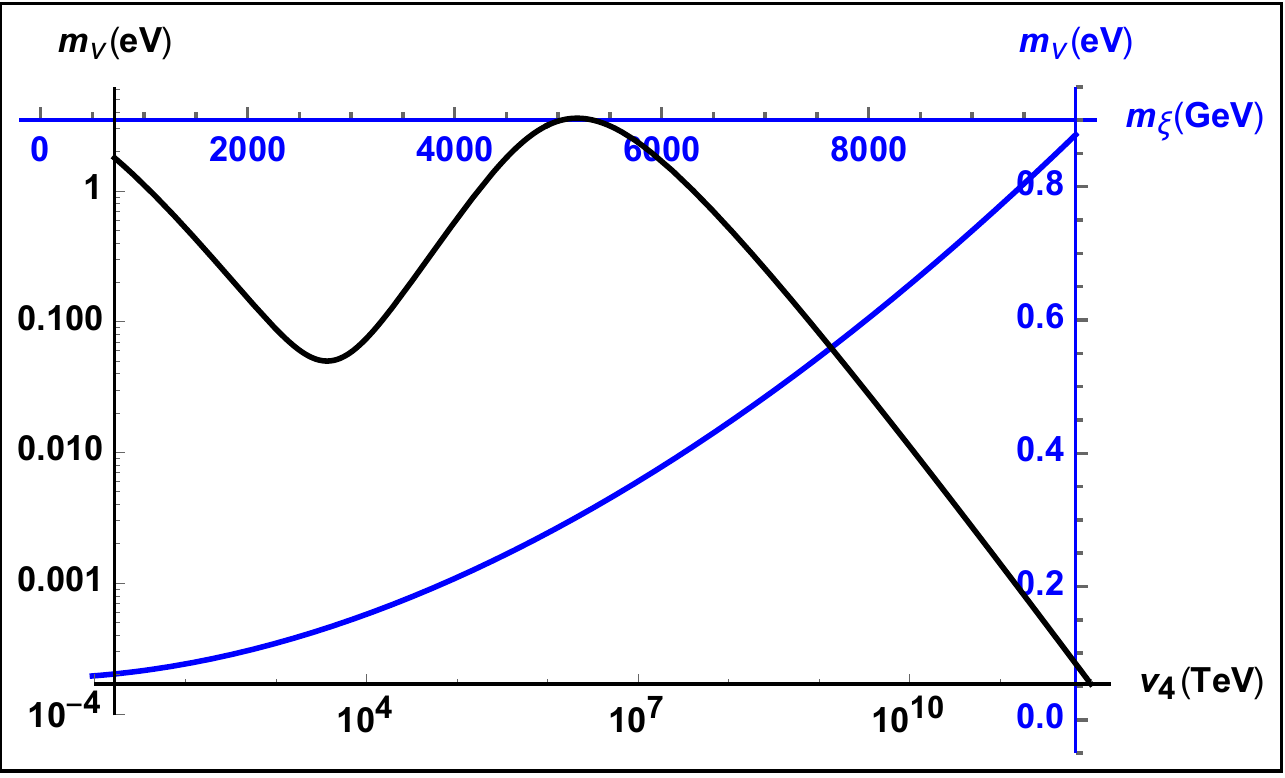}
    \caption{ }
    \label{fig:mnu_dep_1}
    \end{subfigure}
    \\
    \begin{subfigure}[b]{0.8\textwidth}
    \includegraphics[width=1\textwidth, trim={1.5cm 0 0 0}, clip]{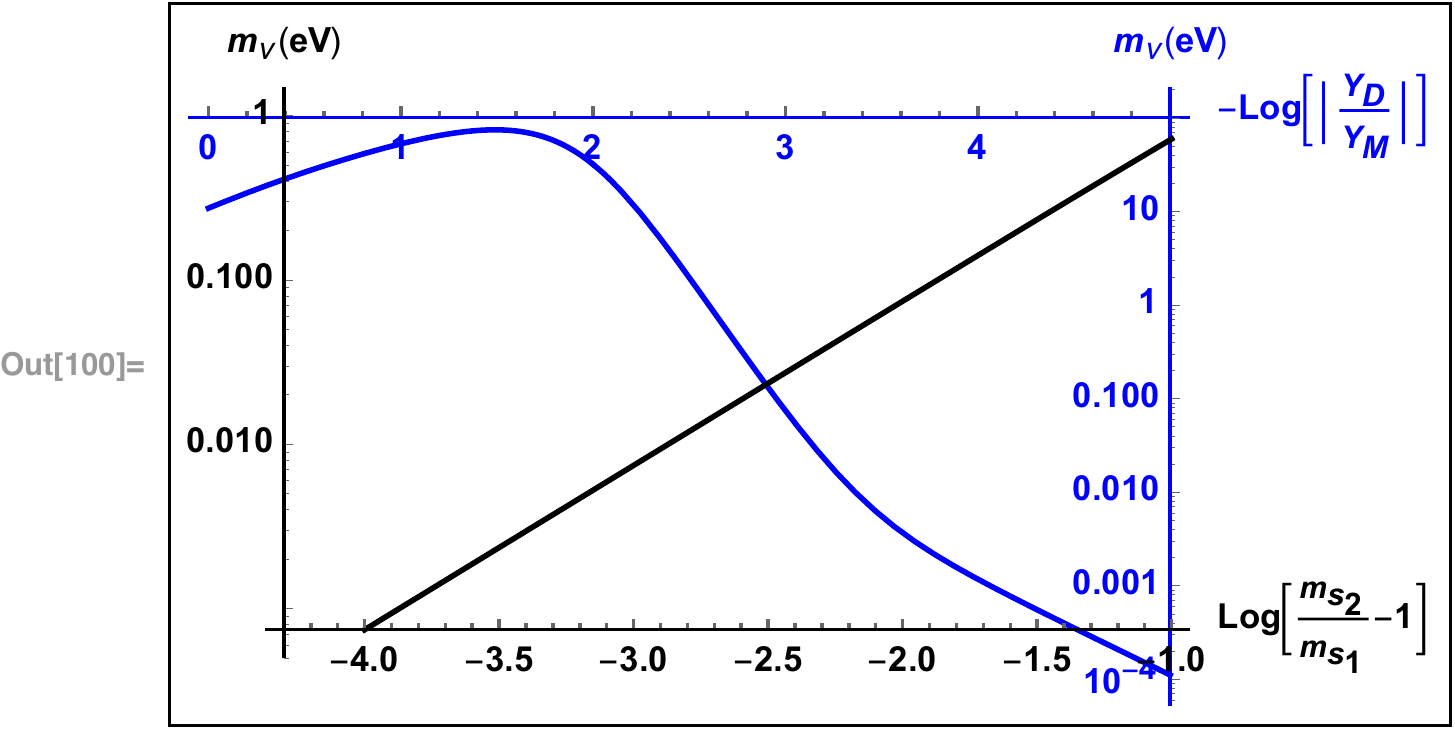}
    \caption{ }
    \label{fig:mnu_dep_2}
    \end{subfigure}
    \caption{Neutrino mass scale dependence on $v_4$, $m_{\xi}$, $\frac{Y_D}{Y_M}$, and $\frac{m_{\xi_1}}{m_{\xi_2}}$. $Y_{M,L}\sim O(1)$, $Y_{R}\sim 10^{-4}$, $\theta_{\xi}=\pi/4$ and the parameters that are not being scanned are fixed to $v_4\sim 10^7$, $m_{\xi}\sim 1$TeV, $\frac{Y_D}{Y_M}\sim 10^{-3}$, and $\frac{m_{\xi_1}}{m_{\xi_2}}-1=10^{-2}$.}
    \label{fig:mnu_dep}
\end{figure}

\begin{figure}[!h]
    \centering
    \begin{subfigure}[b]{0.8\textwidth}
    \includegraphics[width=1\textwidth, trim={1.5cm 0 0 0}, clip]{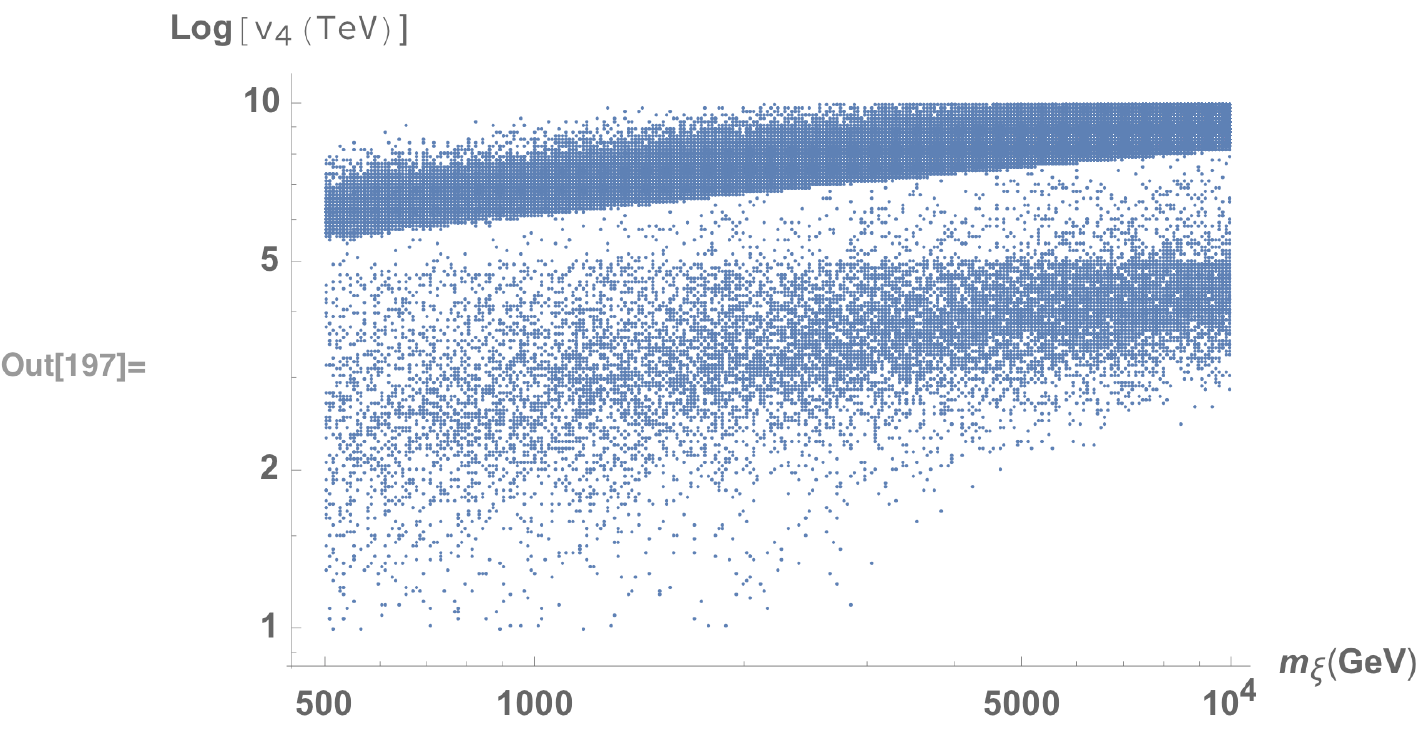}
    \caption{ }
    \label{fig:mnu_corr_1}
    \end{subfigure}
    \\
    \begin{subfigure}[b]{0.8\textwidth}
    \includegraphics[width=1\textwidth, trim={1.5cm 0 0 0}, clip]{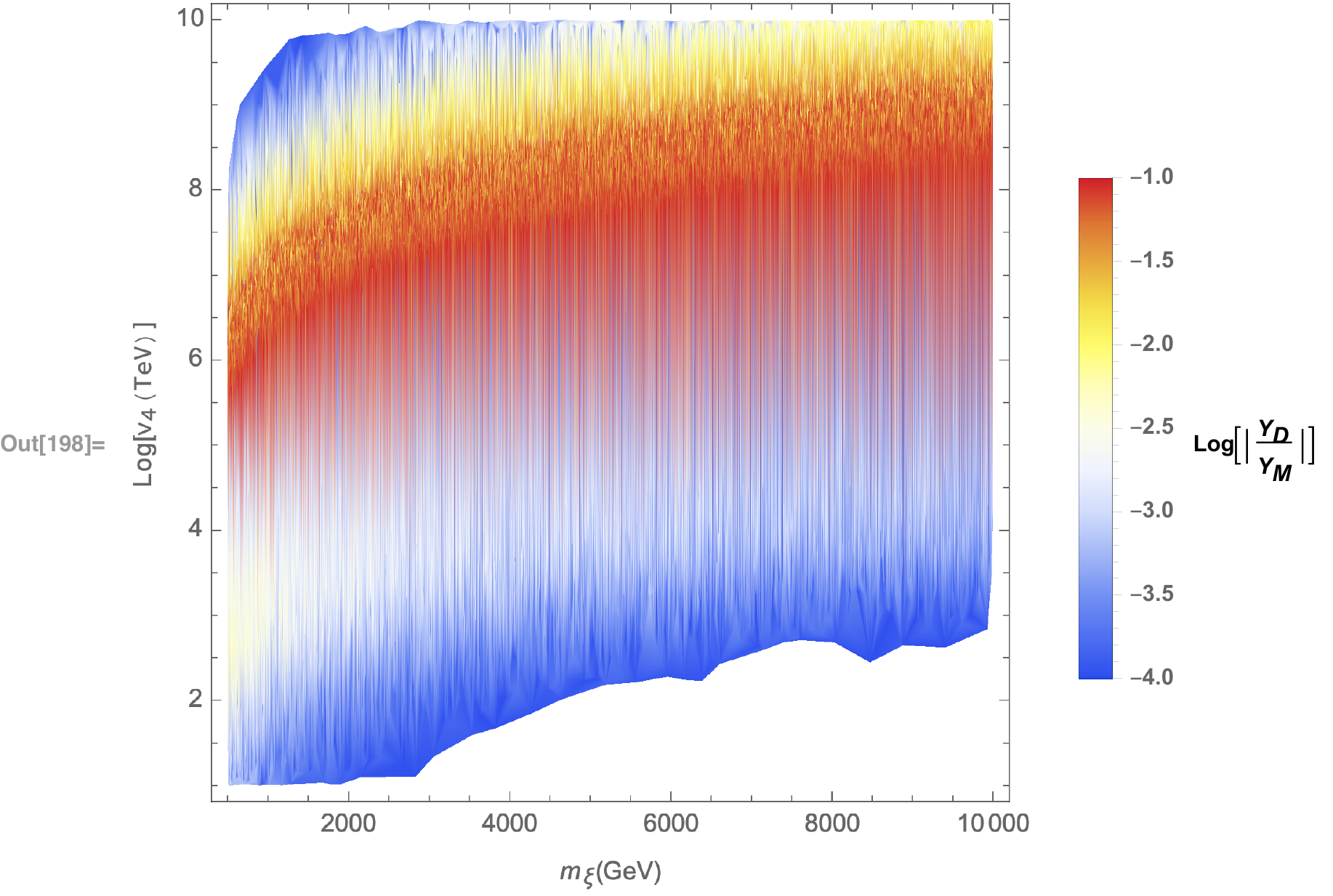}
    \caption{ }
    \label{fig:mnu_corr_2}
    \end{subfigure}
    \caption{Correlation between $v_4$, $m_{\xi}$, and $\frac{Y_D}{Y_M}$ where $\frac{m_{\xi_1}}{m_{\xi_2}}-1=10^{-2}$ and $Y_{M,L}\sim O(1)$, $Y_{R}\sim 10^{-4}$, $\theta_{\xi}=\pi/4$.}
    \label{fig:mnu_corr}
\end{figure}

\begin{table}[ht]
    \centering
    \begin{tabular}{C{5cm} C{5cm} C{5cm}}
    \hline\hline
        Limit & Condition & Reference \\ \hline\hline
        $g_{\scriptstyle{B-L}}<0.25$ & $\forall M_{\scriptstyle{Z^{\prime}}}$ & \cite{Aaboud:2019zxd} \\ \hline
        $M_{\scriptstyle{Z^{\prime}}}>4.2$TeV & $L=36.1$fb$^{-1}$;$\sqrt{s}=13$TeV & \cite{Aaboud:2017buh} \\ \hline
        $M_{\scriptstyle{Z^{\prime}}}>4.4$TeV & $\sigma\times Br(pp\rightarrow Z^{\prime}\rightarrow e\mu)\land Br(Z^{\prime}\rightarrow e\mu)=0.10$, Model independent & CMS CR-2018-371~\cite{Radburn-smith:2649415} \\ \hline
        $g_{\scriptstyle{BL}}$<0.236(0.127) & $Q_{\scriptstyle{BL}}^{Max}=15(28)$ & Perturbativity bound \\ \hline\hline
        $\epsilon <8\times 10^{-4}$ & $\alpha_{BL}=\alpha_{em}\land m_{\scriptstyle{h^{\prime}}}<8\text{GeV}/c^2\land m_{\scriptstyle{A^{\prime}}}<1\text{GeV}/c^2$ & \cite{TheBelle:2015mwa} \\ \hline
        $\epsilon<1.3-3\times 10^{-3}$ & $m_{\scriptstyle{A^{\prime}}}\sim O(10-100\text{MeV}/c^2)$ & \cite{Giovannella:2011nh} \\ \hline
        $\epsilon <10^{-4}-10^{-3}$ & $20\text{MeV}<m_{\scriptstyle{A^{\prime}}}<10.2\text{GeV}$ & \cite{Echenard:2016tiu,ArkaniHamed:2008qn} \\ \hline
        $\epsilon < 10^{-3}$ & $m_{\scriptstyle{A^{\prime}}}\approx O(\text{GeV})$ & \cite{Lees:2017lec,Banerjee:2016tad} \\
    \hline\hline
    \end{tabular}
    \caption{Some relevant phenomenological bounds.}
    \label{tab:pheno_bounds}
\end{table}
Concerning collider constraints on the model: as can be seen from Tab.~\ref{tab:pheno_bounds} $g_{B-L}$ has a upper bound of $0.25$ from collider searches and upper bound of $0.127$ from perturbativity constraints (due to large $Q_{B-L}$ charges in the model). This together with a lower bound on $M_{Z^{\prime}}$ give a lower bound on $U(1)_{B-L}$ breaking scale $v_4$. Constraints on $v_4$ are shown in Fig.~\ref{fig:gbl_v4}.
Neutrino mass scale dependency on $v_4$, $m_{\xi}$, $\frac{Y_D}{Y_M}$, and $\frac{m_{\xi_1}}{m_{\xi_2}}$ is shown in Fig.~\ref{fig:mnu_dep}.
The correlation between $v_4$ scale, $m_{\xi}$, and $\frac{Y_D}{Y_M}$ ratio for a fixed $\frac{m_{\xi_1}}{m_{\xi_2}}-1=10^{-2}$ mass splitting that produce neutrino mass of the order $O(1-0.1\text{eV})$ is shown in Fig.~\ref{fig:mnu_corr}.
Neutrino mass can be made small by the following ways: loop suppression, small $Y_{L,R,D}$ Yukawas, large $v_4$ $U(1)_{B-L}$ breaking scale, and small mass splitting between $\xi_i$ mass eigenstates. In order to suppress neutrino mass but keep \nqbd~ large the following parameter choices were made: $Y_R$, $Y_D\ll 1$ to suppress $m_{\nu}$; $Y_L\sim O(1)$, $\theta_{\xi}=\pi/4$(max. $\xi$ mixing) since \nqbd~ has quartic dependence on them; $Y_M\sim O(1)$ since \nqbd~ depends quadratically on it; $\mu_S\gg v_4$ since it is used to enhance the \nqbd.

Detailed study of phenomenology of the $U(1)_{B-L}$ model is done in~\cite{Basso:2008iv}. Now, in our model the right handed neutrinos can have a strongly hierarchical neutrino Yukawa structure. Which can create leptonic asymmetry through the decays right handed neutrino as shown in \cite{Racker:2013lua,Hugle:2018qbw}.
\section{Conclusion}
\label{sec:conclusion}
The $U(1)_{B-L}$ extension of the SM was presented which is then spontaneously broken to residual $\mathbb{Z}_4$ symmetry. The $\mathbb{Z}_4$ symmetry is both responsible for the Dirac nature of neutrinos as well as for the stability of DM, a unique feature for this type of construction. Neutrino masses are generated radiatively through scotogenic scenario. Since the neutrinos are of Dirac type the neutrinoless double beta decay is exactly absent, but the $\mathbb{Z}_4$ symmetry allows for non-zero neutrinoless quadruple beta decay, which is despite being an experimentally tiny effect is the dominant of the neutrinoless multipole beta decays. If future experiments on $0\nu 2n \beta$ see no positive results in $0\nu 2\beta$ but do observe non-zero $0\nu 4 \beta$, this will be a strong indication toward neutrinos of Dirac type while still violating lepton number by 4 units and will hint toward this type of model. $\mathbb{Z}_4$ allows for several WIMP like DM candidates in our model: best DM candidate is Majorana $N$ which allows for small neutrino masses of $O(0.1\text{eV})$ scale, enhanced $0\nu 4\beta$ decay, $U(1)_{B-L}$ breaking scale as low as $O(10 \text{TeV})$, and DM masses of $O(1 \text{TeV})$; other possible DM candidate is $S$ real scalar field which has a radiative decay to neutrinos and is suitable long-lived DM candidate, making $S$ long-lived also suppresses $0\nu 4\beta$ decay, so it predicts no observable $0\nu 4\beta$ in current or future $0\nu 2n \beta$ experiments without fine-tuning. In many models like this one, $0\nu 4\beta$ might be predicted to be non-zero but even in that case it is expected to be well below the sensitivity of current and future experiments looking for $0\nu 2n \beta$ decays. Model presented here allows for arbitrary enhanced $0\nu 4\beta$ decay which can be made as large as $10^{16-19}$. The prize we pay for this is the introduction of $S$ field which gives us a freedom of the enhancement of $0\nu 4\beta$ without effecting neutrino mass generation and DM related processes (for the $N$ DM case). We have also shown that the model can satisfy all required collider constraints. More detailed collider phenomenology will be presented elsewhere. Here we focused on demonstrating a way for Dirac scotogenic neutrinos with $\Delta L=4$ and dominant $0\nu 4\beta$ decay multipole where Baryon and Lepton number symmetries and violations are obtained from $U(1)_{B-L}$ gauge symmetry.
\acknowledgments
This work was supported by the National Research Foundation of Korea Grants No. 2009-0083526, No. 2017K1A3A7A09016430, and No. 2017R1A2B4006338.
\appendix
\section{Note on removing phases from Yukawa terms in $N_{L,R}$ sector}
\label{ap:mN}
\begin{align}
    \mathcal{L}=&-\left(\bar{N}_L^c,\bar{N}_R\right)\left(\begin{matrix}0 & Y_{ND} v_4 \\ Y_{ND} v_4 & Y_{NM}^{\dagger} v_4\end{matrix}\right)\left(\begin{matrix}N_L \\ N_R^c\end{matrix}\right)+\text{H.c.}\\
    &=v^{T}Mv+\text{H.c.}=\underbrace{v^{T}U^{T}}_{v^{\prime T}}\underbrace{U^{*}MU^{\dagger}}_{M_D}\underbrace{Uv}_{v^{\prime}}+\text{H.c.},
\end{align}
with 
\begin{align}
    U(\phi,\theta,\Delta)&=e^{\imath\phi/2}\left(\begin{matrix}c & s \\ -s & c\end{matrix}\right)\left(\begin{matrix}e^{\imath \Delta} & 0 \\ 0 & e^{-\imath\Delta} \end{matrix}\right),\\
    v^{\prime}&=Uv=e^{\imath\left(\phi+\alpha_N\right)/2}\left(\begin{matrix}c & s \\ -s & c\end{matrix}\right)\left(\begin{matrix}e^{\imath \Delta} & 0 \\ 0 & e^{-\imath\Delta} \end{matrix}\right)\left(\begin{matrix}1 & 0 \\ 0 & e^{\imath\Delta\alpha}\end{matrix}\right)\left(\begin{matrix}N_L \\ N_R^c\end{matrix}\right)\\
    &=e^{\imath\left(\phi+\alpha_N+2\Delta\right)/2}\left(\begin{matrix}c & s \\ -s & c\end{matrix}\right)\left(\begin{matrix}N_L \\ N_R^c\end{matrix}\right),
\end{align}
where $c=$cos$\theta$, $s=$sin$\theta$, and in the last equality we have set $\Delta\alpha=2\Delta$. As can be seen $\Delta\alpha$ and $\alpha_N$ Majorana phases can be used to freely adjust $\phi$ and $\Delta$ phases in the unitary transformation. Next,
\begin{align}
    M_D&=U^{*}MU^{\dagger}=v_4 e^{-\imath\phi}\left(\begin{matrix}c & s \\ -s & c\end{matrix}\right)\left(\begin{matrix}e^{-\imath \Delta} & 0 \\ 0 & e^{\imath\Delta} \end{matrix}\right)\left(\begin{matrix}0 & \left|Y_{ND}\right|e^{\imath \phi_D} \\ \left|Y_{ND}\right|e^{\imath \phi_D} & \left|Y_{NM}\right|e^{\imath \phi_M}\end{matrix}\right)\left(\begin{matrix}e^{-\imath \Delta} & 0 \\ 0 & e^{\imath\Delta} \end{matrix}\right)\left(\begin{matrix}c & -s \\ s & c\end{matrix}\right)\\
    &=v_4 e^{-\imath\left(\phi-\phi_D\right)}\left(\begin{matrix}c & s \\ -s & c\end{matrix}\right)\left(\begin{matrix}0 & \left|Y_{ND}\right| \\ \left|Y_{ND}\right| & \left|Y_{NM}\right|e^{\imath \left(\phi_M-\phi_D+2\Delta\right)}\end{matrix}\right)\left(\begin{matrix}c & -s \\ s & c\end{matrix}\right)\\
    &=\left(\begin{matrix}\lambda_1 & 0 \\ 0 & \lambda_2\end{matrix}\right),
\end{align}
where
\begin{align}
    \text{tan}\left(2\theta\right)&=-2\frac{\left|Y_{ND}\right|}{\left|Y_{NM}\right|},\\
    \lambda_{1,2}&=\frac{v_4}{2}\left(\left|Y_{NM}\right|\pm\sqrt{\left|Y_{NM}\right|^2+4\left|Y_{ND}\right|^2}\right),\\
    \phi_D-\phi_M&=2\Delta=\Delta\alpha,\\
    \phi&=\phi_D.
\end{align}
As can be seen, Majorana phases of $N_{L,R}$ fermion fields can be used to remove phases from the mass matrix.
\section{$C_0(s_1,s_{12},s_2;m_0,m_1,m_2)$}
\label{sec:c0}
\begin{align}
    \label{eq:c0}
   &m_s^2 C_0(0,0,m_s^2;m_j,m_k,m_i) = \nonumber \\
   &\underset{\varepsilon\rightarrow 0^+}{\lim}\text{Li}_2\left[\frac{2(m_i^2-m_k^2)}{m_i^2 + m_j^2 - 2 m_k^2 - m_s^2 - \lambda^{1/2}(m_i^2,m_j^2,m_s^2)}+ i (m_i^2-m_k^2) \varepsilon\right] \nonumber \\
   &+ \underset{\varepsilon\rightarrow 0^+}{\lim}\text{Li}_2\left[\frac{2(m_i^2-m_k^2)}{m_i^2 + m_j^2 - 2 m_k^2 - m_s^2 + \lambda^{1/2}(m_i^2,m_j^2,m_s^2)} - i (m_i^2-m_k^2) \varepsilon\right] \nonumber \\
   &- \underset{\varepsilon\rightarrow 0^+}{\lim}\text{Li}_2\left[\frac{2(m_i^2 - m_k^2 - m_s^2)}{m_i^2 + m_j^2 - 2 m_k^2 - m_s^2 - \lambda^{1/2}(m_i^2,m_j^2,m_s^2)} + i (m_i^2 - m_k^2 - m_s^2) \varepsilon\right] \nonumber \\ 
   &- \underset{\varepsilon\rightarrow 0^+}{\lim}\text{Li}_2\left[\frac{2(m_i^2 - m_k^2 - m_s^2)}{m_i^2 + m_j^2 - 2 m_k^2 - m_s^2 + \lambda^{1/2}(m_i^2,m_j^2,m_s^2)} - i (m_i^2 - m_k^2 - m_s^2) \varepsilon\right] \nonumber \\
   &- \text{PolyLog}\left[2,\frac{(m_i^2 - m_k^2)(m_j^2 - m_k^2)}{m_i^2(m_j^2 - m_k^2) - m_k^2(m_j^2 - m_k^2 - m_s^2)}\right] \nonumber \\ 
   &+ \text{PolyLog}\left[2,\frac{(m_j^2 - m_k^2)(m_i^2 - m_k^2 - m_s^2)}{m_i^2(m_j^2 - m_k^2) - m_k^2(m_j^2 - m_k^2 - m_s^2)}\right],
\end{align}
where Kallen $\lambda$ is defined as 
\begin{align}
    \label{eq:kallenL}
    \lambda(x,y,z) &= x^2 + y^2 + z^2 - 2 (xy + xz + yz).
\end{align}
\section[N annihilation diagrams]{$N$ annihilation diagrams}
\label{sec:ssigmaN}
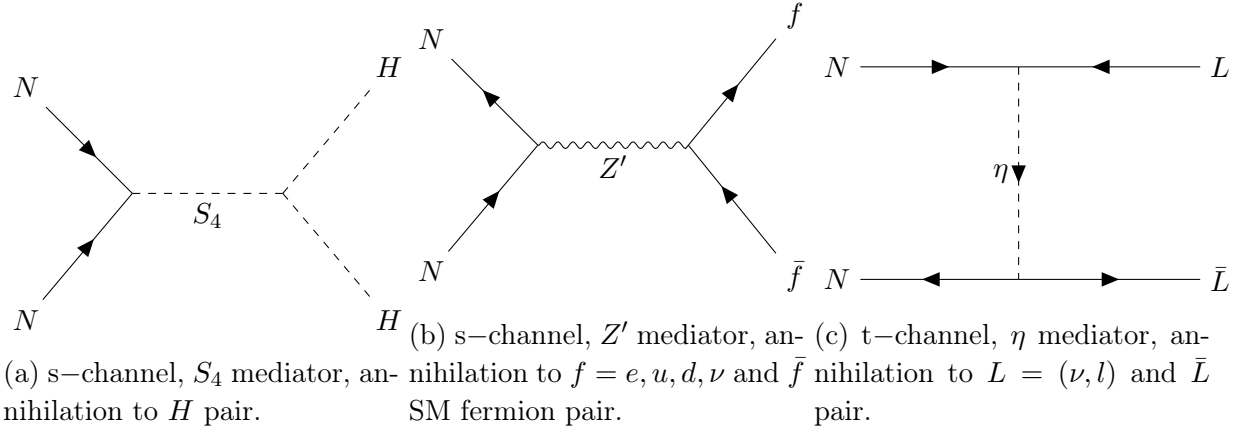
\begin{figure}[H]
    \centering
    \begin{subfigure}[b]{0.32\textwidth}
    \begin{tikzpicture}
\begin{feynman}
\vertex (i1) {$N$};
\vertex [below=1.414cm of i1] (im);
\vertex [below=1.414cm of im] (i2) {$N$};
\vertex [right=1.414cm of im] (a);
\vertex [right=2cm of a] (b);
\vertex [right=1.414cm of b] (fm);
\vertex [above=1.414cm of fm] (f1) {$H$};
\vertex [below=1.414cm of fm] (f2) {$H$};

\diagram* {
(i1) -- [fermion] (a) -- [scalar, edge label'=$S_4$] (b) -- [scalar] (f1),
(i2) -- [fermion] (a),
(b) -- [scalar] (f2),
};
\end{feynman}
\end{tikzpicture}
    \caption{s$-$channel, $S_4$ mediator, annihilation to $H$ pair.}
    \label{fig:Nsigma_ss_s}
    \end{subfigure}
    \begin{subfigure}[b]{0.32\textwidth}
    \begin{tikzpicture}
\begin{feynman}
\vertex (i1) {$N$};
\vertex [below=1.414cm of i1] (im);
\vertex [below=1.414cm of im] (i2) {$N$};
\vertex [right=1.414cm of im] (a);
\vertex [right=2cm of a] (b);
\vertex [right=1.414cm of b] (fm);
\vertex [above=1.414cm of fm] (f1) {$f$};
\vertex [below=1.414cm of fm] (f2) {$\bar{f}$};

\diagram* {
(i1) -- [anti fermion] (a) -- [boson, edge label'=$Z^{\prime}$] (b) -- [fermion] (f1),
(i2) -- [fermion] (a),
(b) -- [anti fermion] (f2),
};
\end{feynman}
\end{tikzpicture}
    \caption{s$-$channel, $Z^{\prime}$ mediator, annihilation to $f=e,u,d,\nu$ and $\Bar{f}$ SM fermion pair.}
    \label{fig:Nsigma_vf_s}
    \end{subfigure}
 \begin{subfigure}[b]{0.32\textwidth}
    \begin{tikzpicture}
\begin{feynman}
\vertex (i1) {$N$};
\vertex [below=2.828cm of i1] (i2) {$N$};
\vertex [right=2.414cm of i1] (a);
\vertex [right=2.414cm of i2] (b);
\vertex [right=2.414cm of a] (f1) {$L$};
\vertex [right=2.414cm of b] (f2) {$\bar{L}$};

\diagram* {
(i1) -- [fermion] (a) -- [anti fermion] (f1),
(i2) -- [anti fermion] (b) -- [fermion] (f2),
(a) -- [charged scalar, edge label'=$\eta$] (b) 
};
\end{feynman}
\end{tikzpicture}
    \caption{t$-$channel, $\eta$ mediator, annihilation to $L=(\nu,l)$ and $\Bar{L}$ pair.}
    \label{fig:Nsigma_ss_t}
    \end{subfigure}
  \caption[$N$ annihilation diagrams.]{$N$ annihilation diagrams.}
    \label{fig:Nsigma}
\end{figure}
\bibliography{references}

\begin{thebibliography}{65}%
\makeatletter
\providecommand \@ifxundefined [1]{%
 \@ifx{#1\undefined}
}%
\providecommand \@ifnum [1]{%
 \ifnum #1\expandafter \@firstoftwo
 \else \expandafter \@secondoftwo
 \fi
}%
\providecommand \@ifx [1]{%
 \ifx #1\expandafter \@firstoftwo
 \else \expandafter \@secondoftwo
 \fi
}%
\providecommand \natexlab [1]{#1}%
\providecommand \enquote  [1]{``#1''}%
\providecommand \bibnamefont  [1]{#1}%
\providecommand \bibfnamefont [1]{#1}%
\providecommand \citenamefont [1]{#1}%
\providecommand \href@noop [0]{\@secondoftwo}%
\providecommand \href [0]{\begingroup \@sanitize@url \@href}%
\providecommand \@href[1]{\@@startlink{#1}\@@href}%
\providecommand \@@href[1]{\endgroup#1\@@endlink}%
\providecommand \@sanitize@url [0]{\catcode `\\12\catcode `\$12\catcode
  `\&12\catcode `\#12\catcode `\^12\catcode `\_12\catcode `\%12\relax}%
\providecommand \@@startlink[1]{}%
\providecommand \@@endlink[0]{}%
\providecommand \url  [0]{\begingroup\@sanitize@url \@url }%
\providecommand \@url [1]{\endgroup\@href {#1}{\urlprefix }}%
\providecommand \urlprefix  [0]{URL }%
\providecommand \Eprint [0]{\href }%
\providecommand \doibase [0]{https://doi.org/}%
\providecommand \selectlanguage [0]{\@gobble}%
\providecommand \bibinfo  [0]{\@secondoftwo}%
\providecommand \bibfield  [0]{\@secondoftwo}%
\providecommand \translation [1]{[#1]}%
\providecommand \BibitemOpen [0]{}%
\providecommand \bibitemStop [0]{}%
\providecommand \bibitemNoStop [0]{.\EOS\space}%
\providecommand \EOS [0]{\spacefactor3000\relax}%
\providecommand \BibitemShut  [1]{\csname bibitem#1\endcsname}%
\let\auto@bib@innerbib\@empty
\bibitem [{\citenamefont {Aad}\ \emph {et~al.}(2012)\citenamefont {Aad} \emph
  {et~al.}}]{Aad:2012tfa}%
  \BibitemOpen
  \bibfield  {author} {\bibinfo {author} {\bibfnamefont {G.}~\bibnamefont
  {Aad}} \emph {et~al.} (\bibinfo {collaboration} {ATLAS}),\ }\href
  {https://doi.org/10.1016/j.physletb.2012.08.020} {\bibfield  {journal}
  {\bibinfo  {journal} {Phys. Lett.}\ }\textbf {\bibinfo {volume} {B716}},\
  \bibinfo {pages} {1} (\bibinfo {year} {2012})},\ \Eprint
  {https://arxiv.org/abs/1207.7214} {arXiv:1207.7214 [hep-ex]} \BibitemShut
  {NoStop}%
\bibitem [{\citenamefont {Chatrchyan}\ \emph {et~al.}(2012)\citenamefont
  {Chatrchyan} \emph {et~al.}}]{Chatrchyan:2012xdj}%
  \BibitemOpen
  \bibfield  {author} {\bibinfo {author} {\bibfnamefont {S.}~\bibnamefont
  {Chatrchyan}} \emph {et~al.} (\bibinfo {collaboration} {CMS}),\ }\href
  {https://doi.org/10.1016/j.physletb.2012.08.021} {\bibfield  {journal}
  {\bibinfo  {journal} {Phys. Lett.}\ }\textbf {\bibinfo {volume} {B716}},\
  \bibinfo {pages} {30} (\bibinfo {year} {2012})},\ \Eprint
  {https://arxiv.org/abs/1207.7235} {arXiv:1207.7235 [hep-ex]} \BibitemShut
  {NoStop}%
\bibitem [{\citenamefont {Fukuda}\ \emph {et~al.}(1996)\citenamefont {Fukuda}
  \emph {et~al.}}]{PhysRevLett.77.1683}%
  \BibitemOpen
  \bibfield  {author} {\bibinfo {author} {\bibfnamefont {Y.}~\bibnamefont
  {Fukuda}} \emph {et~al.},\ }\href
  {https://doi.org/10.1103/PhysRevLett.77.1683} {\bibfield  {journal} {\bibinfo
   {journal} {Phys. Rev. Lett.}\ }\textbf {\bibinfo {volume} {77}},\ \bibinfo
  {pages} {1683} (\bibinfo {year} {1996})}\BibitemShut {NoStop}%
\bibitem [{\citenamefont {Abdurashitov}\ \emph {et~al.}(2009)\citenamefont
  {Abdurashitov} \emph {et~al.}}]{PhysRevC.80.015807}%
  \BibitemOpen
  \bibfield  {author} {\bibinfo {author} {\bibfnamefont {J.~N.}\ \bibnamefont
  {Abdurashitov}} \emph {et~al.} (\bibinfo {collaboration} {SAGE
  Collaboration}),\ }\href {https://doi.org/10.1103/PhysRevC.80.015807}
  {\bibfield  {journal} {\bibinfo  {journal} {Phys. Rev. C}\ }\textbf {\bibinfo
  {volume} {80}},\ \bibinfo {pages} {015807} (\bibinfo {year}
  {2009})}\BibitemShut {NoStop}%
\bibitem [{\citenamefont {Fukuda}\ \emph {et~al.}(2002)\citenamefont {Fukuda}
  \emph {et~al.}}]{Fukuda:2002pe}%
  \BibitemOpen
  \bibfield  {author} {\bibinfo {author} {\bibfnamefont {S.}~\bibnamefont
  {Fukuda}} \emph {et~al.} (\bibinfo {collaboration} {Super-Kamiokande}),\
  }\href {https://doi.org/10.1016/S0370-2693(02)02090-7} {\bibfield  {journal}
  {\bibinfo  {journal} {Phys. Lett.}\ }\textbf {\bibinfo {volume} {B539}},\
  \bibinfo {pages} {179} (\bibinfo {year} {2002})},\ \Eprint
  {https://arxiv.org/abs/hep-ex/0205075} {arXiv:hep-ex/0205075 [hep-ex]}
  \BibitemShut {NoStop}%
\bibitem [{\citenamefont {Eguchi}\ \emph {et~al.}(2003)\citenamefont {Eguchi}
  \emph {et~al.}}]{PhysRevLett.90.021802}%
  \BibitemOpen
  \bibfield  {author} {\bibinfo {author} {\bibfnamefont {K.}~\bibnamefont
  {Eguchi}} \emph {et~al.} (\bibinfo {collaboration} {KamLAND Collaboration}),\
  }\href {https://doi.org/10.1103/PhysRevLett.90.021802} {\bibfield  {journal}
  {\bibinfo  {journal} {Phys. Rev. Lett.}\ }\textbf {\bibinfo {volume} {90}},\
  \bibinfo {pages} {021802} (\bibinfo {year} {2003})}\BibitemShut {NoStop}%
\bibitem [{\citenamefont {Fukuda}\ \emph {et~al.}(1998)\citenamefont {Fukuda}
  \emph {et~al.}}]{PhysRevLett.81.1562}%
  \BibitemOpen
  \bibfield  {author} {\bibinfo {author} {\bibfnamefont {Y.}~\bibnamefont
  {Fukuda}} \emph {et~al.} (\bibinfo {collaboration} {Super-Kamiokande
  Collaboration}),\ }\href {https://doi.org/10.1103/PhysRevLett.81.1562}
  {\bibfield  {journal} {\bibinfo  {journal} {Phys. Rev. Lett.}\ }\textbf
  {\bibinfo {volume} {81}},\ \bibinfo {pages} {1562} (\bibinfo {year}
  {1998})}\BibitemShut {NoStop}%
\bibitem [{\citenamefont {Ahn}\ \emph {et~al.}(2006)\citenamefont {Ahn} \emph
  {et~al.}}]{PhysRevD.74.072003}%
  \BibitemOpen
  \bibfield  {author} {\bibinfo {author} {\bibfnamefont {M.~H.}\ \bibnamefont
  {Ahn}} \emph {et~al.} (\bibinfo {collaboration} {K2K Collaboration}),\ }\href
  {https://doi.org/10.1103/PhysRevD.74.072003} {\bibfield  {journal} {\bibinfo
  {journal} {Phys. Rev. D}\ }\textbf {\bibinfo {volume} {74}},\ \bibinfo
  {pages} {072003} (\bibinfo {year} {2006})}\BibitemShut {NoStop}%
\bibitem [{\citenamefont {Abe}\ \emph {et~al.}(2013)\citenamefont {Abe} \emph
  {et~al.}}]{PhysRevLett.111.211803}%
  \BibitemOpen
  \bibfield  {author} {\bibinfo {author} {\bibfnamefont {K.}~\bibnamefont
  {Abe}} \emph {et~al.} (\bibinfo {collaboration} {T2K Collaboration}),\ }\href
  {https://doi.org/10.1103/PhysRevLett.111.211803} {\bibfield  {journal}
  {\bibinfo  {journal} {Phys. Rev. Lett.}\ }\textbf {\bibinfo {volume} {111}},\
  \bibinfo {pages} {211803} (\bibinfo {year} {2013})}\BibitemShut {NoStop}%
\bibitem [{\citenamefont {An}\ \emph {et~al.}(2012)\citenamefont {An} \emph
  {et~al.}}]{PhysRevLett.108.171803}%
  \BibitemOpen
  \bibfield  {author} {\bibinfo {author} {\bibfnamefont {F.~P.}\ \bibnamefont
  {An}} \emph {et~al.},\ }\href
  {https://doi.org/10.1103/PhysRevLett.108.171803} {\bibfield  {journal}
  {\bibinfo  {journal} {Phys. Rev. Lett.}\ }\textbf {\bibinfo {volume} {108}},\
  \bibinfo {pages} {171803} (\bibinfo {year} {2012})}\BibitemShut {NoStop}%
\bibitem [{\citenamefont {Ahn}\ \emph {et~al.}(2012)\citenamefont {Ahn} \emph
  {et~al.}}]{PhysRevLett.108.191802}%
  \BibitemOpen
  \bibfield  {author} {\bibinfo {author} {\bibfnamefont {J.~K.}\ \bibnamefont
  {Ahn}} \emph {et~al.} (\bibinfo {collaboration} {RENO Collaboration}),\
  }\href {https://doi.org/10.1103/PhysRevLett.108.191802} {\bibfield  {journal}
  {\bibinfo  {journal} {Phys. Rev. Lett.}\ }\textbf {\bibinfo {volume} {108}},\
  \bibinfo {pages} {191802} (\bibinfo {year} {2012})}\BibitemShut {NoStop}%
\bibitem [{\citenamefont {Abe}\ \emph {et~al.}(2012)\citenamefont {Abe} \emph
  {et~al.}}]{PhysRevD.86.052008}%
  \BibitemOpen
  \bibfield  {author} {\bibinfo {author} {\bibnamefont {Abe}} \emph {et~al.}
  (\bibinfo {collaboration} {Double Chooz Collaboration}),\ }\href
  {https://doi.org/10.1103/PhysRevD.86.052008} {\bibfield  {journal} {\bibinfo
  {journal} {Phys. Rev. D}\ }\textbf {\bibinfo {volume} {86}},\ \bibinfo
  {pages} {052008} (\bibinfo {year} {2012})}\BibitemShut {NoStop}%
\bibitem [{\citenamefont {Mohapatra}\ and\ \citenamefont
  {Senjanovic}(1980)}]{Mohapatra:1979ia}%
  \BibitemOpen
  \bibfield  {author} {\bibinfo {author} {\bibfnamefont {R.~N.}\ \bibnamefont
  {Mohapatra}}\ and\ \bibinfo {author} {\bibfnamefont {G.}~\bibnamefont
  {Senjanovic}},\ }\href {https://doi.org/10.1103/PhysRevLett.44.912}
  {\bibfield  {journal} {\bibinfo  {journal} {Phys. Rev. Lett.}\ }\textbf
  {\bibinfo {volume} {44}},\ \bibinfo {pages} {912} (\bibinfo {year} {1980})},\
  \bibinfo {note} {[,231(1979)]}\BibitemShut {NoStop}%
\bibitem [{\citenamefont {Minkowski}(1977)}]{Minkowski:1977sc}%
  \BibitemOpen
  \bibfield  {author} {\bibinfo {author} {\bibfnamefont {P.}~\bibnamefont
  {Minkowski}},\ }\href {https://doi.org/10.1016/0370-2693(77)90435-X}
  {\bibfield  {journal} {\bibinfo  {journal} {Phys. Lett.}\ }\textbf {\bibinfo
  {volume} {67B}},\ \bibinfo {pages} {421} (\bibinfo {year}
  {1977})}\BibitemShut {NoStop}%
\bibitem [{\citenamefont {Yanagida}(1979)}]{Yanagida:1979ab}%
  \BibitemOpen
  \bibfield  {author} {\bibinfo {author} {\bibfnamefont {T.}~\bibnamefont
  {Yanagida}}\ }(\bibinfo {year} {1979})\ p.~\bibinfo {pages} {95},\ \bibinfo
  {note} {(KEK., Tsukuba,), O. Sawada and A. Sugamoto (eds.)}\BibitemShut
  {NoStop}%
\bibitem [{\citenamefont {GellMann}\ \emph {et~al.}(1979)\citenamefont
  {GellMann}, \citenamefont {Ramond},\ and\ \citenamefont
  {Slansky}}]{GellMann:1979grs}%
  \BibitemOpen
  \bibfield  {author} {\bibinfo {author} {\bibfnamefont {M.}~\bibnamefont
  {GellMann}}, \bibinfo {author} {\bibfnamefont {P.}~\bibnamefont {Ramond}},\
  and\ \bibinfo {author} {\bibfnamefont {R.}~\bibnamefont {Slansky}},\
  }\href@noop {} {\emph {\bibinfo {title} {{Supergravity}}}}\ (\bibinfo {year}
  {1979})\ p.\ \bibinfo {pages} {315},\ \bibinfo {note} {(North-Holland.,
  Amsterdam), P. van Nieuwenhuizen and D. Z. Freedman (eds.)}\BibitemShut
  {NoStop}%
\bibitem [{\citenamefont {Schechter}\ and\ \citenamefont
  {Valle}(1980)}]{Schechter:1980gr}%
  \BibitemOpen
  \bibfield  {author} {\bibinfo {author} {\bibfnamefont {J.}~\bibnamefont
  {Schechter}}\ and\ \bibinfo {author} {\bibfnamefont {J.~W.~F.}\ \bibnamefont
  {Valle}},\ }\href {https://doi.org/10.1103/PhysRevD.22.2227} {\bibfield
  {journal} {\bibinfo  {journal} {Phys. Rev.}\ }\textbf {\bibinfo {volume}
  {D22}},\ \bibinfo {pages} {2227} (\bibinfo {year} {1980})}\BibitemShut
  {NoStop}%
\bibitem [{\citenamefont {Magg}\ and\ \citenamefont
  {Wetterich}(1980)}]{Magg:1980ut}%
  \BibitemOpen
  \bibfield  {author} {\bibinfo {author} {\bibfnamefont {M.}~\bibnamefont
  {Magg}}\ and\ \bibinfo {author} {\bibfnamefont {C.}~\bibnamefont
  {Wetterich}},\ }\href {https://doi.org/10.1016/0370-2693(80)90825-4}
  {\bibfield  {journal} {\bibinfo  {journal} {Phys. Lett.}\ }\textbf {\bibinfo
  {volume} {94B}},\ \bibinfo {pages} {61} (\bibinfo {year} {1980})}\BibitemShut
  {NoStop}%
\bibitem [{\citenamefont {Cheng}\ and\ \citenamefont
  {Li}(1980)}]{Cheng:1980qt}%
  \BibitemOpen
  \bibfield  {author} {\bibinfo {author} {\bibfnamefont {T.~P.}\ \bibnamefont
  {Cheng}}\ and\ \bibinfo {author} {\bibfnamefont {L.-F.}\ \bibnamefont {Li}},\
  }\href {https://doi.org/10.1103/PhysRevD.22.2860} {\bibfield  {journal}
  {\bibinfo  {journal} {Phys. Rev.}\ }\textbf {\bibinfo {volume} {D22}},\
  \bibinfo {pages} {2860} (\bibinfo {year} {1980})}\BibitemShut {NoStop}%
\bibitem [{\citenamefont {Mohapatra}\ and\ \citenamefont
  {Senjanovic}(1981)}]{Mohapatra:1980yp}%
  \BibitemOpen
  \bibfield  {author} {\bibinfo {author} {\bibfnamefont {R.~N.}\ \bibnamefont
  {Mohapatra}}\ and\ \bibinfo {author} {\bibfnamefont {G.}~\bibnamefont
  {Senjanovic}},\ }\href {https://doi.org/10.1103/PhysRevD.23.165} {\bibfield
  {journal} {\bibinfo  {journal} {Phys. Rev.}\ }\textbf {\bibinfo {volume}
  {D23}},\ \bibinfo {pages} {165} (\bibinfo {year} {1981})}\BibitemShut
  {NoStop}%
\bibitem [{\citenamefont {Weinberg}(1979)}]{Weinberg:1979sa}%
  \BibitemOpen
  \bibfield  {author} {\bibinfo {author} {\bibfnamefont {S.}~\bibnamefont
  {Weinberg}},\ }\href {https://doi.org/10.1103/PhysRevLett.43.1566} {\bibfield
   {journal} {\bibinfo  {journal} {Phys. Rev. Lett.}\ }\textbf {\bibinfo
  {volume} {43}},\ \bibinfo {pages} {1566} (\bibinfo {year}
  {1979})}\BibitemShut {NoStop}%
\bibitem [{\citenamefont {Zee}(1980)}]{Zee:1980ai}%
  \BibitemOpen
  \bibfield  {author} {\bibinfo {author} {\bibfnamefont {A.}~\bibnamefont
  {Zee}},\ }\href {https://doi.org/10.1016/0370-2693(80)90349-4,
  10.1016/0370-2693(80)90193-8} {\bibfield  {journal} {\bibinfo  {journal}
  {Phys. Lett.}\ }\textbf {\bibinfo {volume} {93B}},\ \bibinfo {pages} {389}
  (\bibinfo {year} {1980})},\ \bibinfo {note} {[Erratum: Phys.
  Lett.95B,461(1980)]}\BibitemShut {NoStop}%
\bibitem [{\citenamefont {Ma}(2006)}]{Ma:2006km}%
  \BibitemOpen
  \bibfield  {author} {\bibinfo {author} {\bibfnamefont {E.}~\bibnamefont
  {Ma}},\ }\href {https://doi.org/10.1103/PhysRevD.73.077301} {\bibfield
  {journal} {\bibinfo  {journal} {Phys. Rev.}\ }\textbf {\bibinfo {volume}
  {D73}},\ \bibinfo {pages} {077301} (\bibinfo {year} {2006})},\ \Eprint
  {https://arxiv.org/abs/hep-ph/0601225} {arXiv:hep-ph/0601225 [hep-ph]}
  \BibitemShut {NoStop}%
\bibitem [{\citenamefont {Fraser}\ \emph {et~al.}(2014)\citenamefont {Fraser},
  \citenamefont {Ma},\ and\ \citenamefont {Popov}}]{Fraser:2014yha}%
  \BibitemOpen
  \bibfield  {author} {\bibinfo {author} {\bibfnamefont {S.}~\bibnamefont
  {Fraser}}, \bibinfo {author} {\bibfnamefont {E.}~\bibnamefont {Ma}},\ and\
  \bibinfo {author} {\bibfnamefont {O.}~\bibnamefont {Popov}},\ }\href
  {https://doi.org/10.1016/j.physletb.2014.08.069} {\bibfield  {journal}
  {\bibinfo  {journal} {Phys. Lett.}\ }\textbf {\bibinfo {volume} {B737}},\
  \bibinfo {pages} {280} (\bibinfo {year} {2014})},\ \Eprint
  {https://arxiv.org/abs/1408.4785} {arXiv:1408.4785 [hep-ph]} \BibitemShut
  {NoStop}%
\bibitem [{\citenamefont {Ma}\ and\ \citenamefont {Popov}(2017)}]{Ma:2016mwh}%
  \BibitemOpen
  \bibfield  {author} {\bibinfo {author} {\bibfnamefont {E.}~\bibnamefont
  {Ma}}\ and\ \bibinfo {author} {\bibfnamefont {O.}~\bibnamefont {Popov}},\
  }\href {https://doi.org/10.1016/j.physletb.2016.11.027} {\bibfield  {journal}
  {\bibinfo  {journal} {Phys. Lett.}\ }\textbf {\bibinfo {volume} {B764}},\
  \bibinfo {pages} {142} (\bibinfo {year} {2017})},\ \Eprint
  {https://arxiv.org/abs/1609.02538} {arXiv:1609.02538 [hep-ph]} \BibitemShut
  {NoStop}%
\bibitem [{\citenamefont {Hirsch}\ \emph {et~al.}(2018)\citenamefont {Hirsch},
  \citenamefont {Srivastava},\ and\ \citenamefont {Valle}}]{Hirsch:2017col}%
  \BibitemOpen
  \bibfield  {author} {\bibinfo {author} {\bibfnamefont {M.}~\bibnamefont
  {Hirsch}}, \bibinfo {author} {\bibfnamefont {R.}~\bibnamefont {Srivastava}},\
  and\ \bibinfo {author} {\bibfnamefont {J.~W.~F.}\ \bibnamefont {Valle}},\
  }\href {https://doi.org/10.1016/j.physletb.2018.03.073} {\bibfield  {journal}
  {\bibinfo  {journal} {Phys. Lett.}\ }\textbf {\bibinfo {volume} {B781}},\
  \bibinfo {pages} {302} (\bibinfo {year} {2018})},\ \Eprint
  {https://arxiv.org/abs/1711.06181} {arXiv:1711.06181 [hep-ph]} \BibitemShut
  {NoStop}%
\bibitem [{\citenamefont {Benavides}\ \emph {et~al.}(2015)\citenamefont
  {Benavides}, \citenamefont {Epele}, \citenamefont {Fanchiotti}, \citenamefont
  {Canal},\ and\ \citenamefont {Ponce}}]{Benavides:2015afa}%
  \BibitemOpen
  \bibfield  {author} {\bibinfo {author} {\bibfnamefont {R.~H.}\ \bibnamefont
  {Benavides}}, \bibinfo {author} {\bibfnamefont {L.~N.}\ \bibnamefont
  {Epele}}, \bibinfo {author} {\bibfnamefont {H.}~\bibnamefont {Fanchiotti}},
  \bibinfo {author} {\bibfnamefont {C.~G.}\ \bibnamefont {Canal}},\ and\
  \bibinfo {author} {\bibfnamefont {W.~A.}\ \bibnamefont {Ponce}},\ }\href
  {https://doi.org/10.1155/2015/813129} {\bibfield  {journal} {\bibinfo
  {journal} {Adv. High Energy Phys.}\ }\textbf {\bibinfo {volume} {2015}},\
  \bibinfo {pages} {813129} (\bibinfo {year} {2015})},\ \Eprint
  {https://arxiv.org/abs/1503.01686} {arXiv:1503.01686 [hep-ph]} \BibitemShut
  {NoStop}%
\bibitem [{\citenamefont {Liu}\ \emph {et~al.}(2016)\citenamefont {Liu},
  \citenamefont {Zhang},\ and\ \citenamefont {Zhou}}]{Liu:2016oph}%
  \BibitemOpen
  \bibfield  {author} {\bibinfo {author} {\bibfnamefont {J.-H.}\ \bibnamefont
  {Liu}}, \bibinfo {author} {\bibfnamefont {J.}~\bibnamefont {Zhang}},\ and\
  \bibinfo {author} {\bibfnamefont {S.}~\bibnamefont {Zhou}},\ }\href
  {https://doi.org/10.1016/j.physletb.2016.07.043} {\bibfield  {journal}
  {\bibinfo  {journal} {Phys. Lett.}\ }\textbf {\bibinfo {volume} {B760}},\
  \bibinfo {pages} {571} (\bibinfo {year} {2016})},\ \Eprint
  {https://arxiv.org/abs/1606.04886} {arXiv:1606.04886 [hep-ph]} \BibitemShut
  {NoStop}%
\bibitem [{\citenamefont {Geng}\ \emph {et~al.}(2015)\citenamefont {Geng},
  \citenamefont {Huang}, \citenamefont {Tsai},\ and\ \citenamefont
  {Wang}}]{Geng:2015qha}%
  \BibitemOpen
  \bibfield  {author} {\bibinfo {author} {\bibfnamefont {C.-Q.}\ \bibnamefont
  {Geng}}, \bibinfo {author} {\bibfnamefont {D.}~\bibnamefont {Huang}},
  \bibinfo {author} {\bibfnamefont {L.-H.}\ \bibnamefont {Tsai}},\ and\
  \bibinfo {author} {\bibfnamefont {Q.}~\bibnamefont {Wang}},\ }\href
  {https://doi.org/10.1007/JHEP08(2015)141} {\bibfield  {journal} {\bibinfo
  {journal} {JHEP}\ }\textbf {\bibinfo {volume} {08}},\ \bibinfo {pages}
  {141}},\ \Eprint {https://arxiv.org/abs/1507.03455} {arXiv:1507.03455
  [hep-ph]} \BibitemShut {NoStop}%
\bibitem [{\citenamefont {Boucenna}(2014)}]{Boucenna:2014pga}%
  \BibitemOpen
  \bibfield  {author} {\bibinfo {author} {\bibfnamefont {S.~M.}\ \bibnamefont
  {Boucenna}},\ }\bibfield  {booktitle} {\emph {\bibinfo {booktitle}
  {{Proceedings, 4th Young Researchers Workshop: Physics Challenges in the LHC
  Era: Frascati, Rome, Italy, May 12-15, 2014}}},\ }\href@noop {} {\bibfield
  {journal} {\bibinfo  {journal} {Frascati Phys. Ser.}\ }\textbf {\bibinfo
  {volume} {59}},\ \bibinfo {pages} {19} (\bibinfo {year} {2014})}\BibitemShut
  {NoStop}%
\bibitem [{\citenamefont {Lim}\ \emph {et~al.}(2005)\citenamefont {Lim},
  \citenamefont {Takasugi},\ and\ \citenamefont {Yoshimura}}]{Lim:2005aa}%
  \BibitemOpen
  \bibfield  {author} {\bibinfo {author} {\bibfnamefont {C.~S.}\ \bibnamefont
  {Lim}}, \bibinfo {author} {\bibfnamefont {E.}~\bibnamefont {Takasugi}},\ and\
  \bibinfo {author} {\bibfnamefont {M.}~\bibnamefont {Yoshimura}},\ }\bibfield
  {booktitle} {\emph {\bibinfo {booktitle} {{NuFact04: Proceedings of the 6th
  International Workshop on Neutrino Factories and Superbeams Osaka, Japan,
  July 26-August 1, 2004}}},\ }\href
  {https://doi.org/10.1016/j.nuclphysbps.2005.05.066} {\bibfield  {journal}
  {\bibinfo  {journal} {Nucl. Phys. Proc. Suppl.}\ }\textbf {\bibinfo {volume}
  {149}},\ \bibinfo {pages} {354} (\bibinfo {year} {2005})},\ \bibinfo {note}
  {[,354(2005)]}\BibitemShut {NoStop}%
\bibitem [{\citenamefont {Witten}(2001)}]{Witten:2000dt}%
  \BibitemOpen
  \bibfield  {author} {\bibinfo {author} {\bibfnamefont {E.}~\bibnamefont
  {Witten}},\ }\bibfield  {booktitle} {\emph {\bibinfo {booktitle} {{Neutrino
  physics and astrophysics. Proceedings, 19th International Conference,
  Neutrino 2000, Sudbury, Canada, June 16-21, 2000}}},\ }\href
  {https://doi.org/10.1016/S0920-5632(00)00916-6} {\bibfield  {journal}
  {\bibinfo  {journal} {Nucl. Phys. Proc. Suppl.}\ }\textbf {\bibinfo {volume}
  {91}},\ \bibinfo {pages} {3} (\bibinfo {year} {2001})},\ \bibinfo {note}
  {[,3(2000)]},\ \Eprint {https://arxiv.org/abs/hep-ph/0006332}
  {arXiv:hep-ph/0006332 [hep-ph]} \BibitemShut {NoStop}%
\bibitem [{\citenamefont {Kim}\ and\ \citenamefont {Lee}(1999)}]{Kim:1999yu}%
  \BibitemOpen
  \bibfield  {author} {\bibinfo {author} {\bibfnamefont {J.~E.}\ \bibnamefont
  {Kim}}\ and\ \bibinfo {author} {\bibfnamefont {J.~S.}\ \bibnamefont {Lee}},\
  }\href@noop {} {\  (\bibinfo {year} {1999})},\ \Eprint
  {https://arxiv.org/abs/hep-ph/9907452} {arXiv:hep-ph/9907452 [hep-ph]}
  \BibitemShut {NoStop}%
\bibitem [{\citenamefont {Centelles~Chuliá}\ \emph {et~al.}(2019)\citenamefont
  {Centelles~Chuliá}, \citenamefont {Cepedello}, \citenamefont {Peinado},\
  and\ \citenamefont {Srivastava}}]{CentellesChulia:2019gic}%
  \BibitemOpen
  \bibfield  {author} {\bibinfo {author} {\bibfnamefont {S.}~\bibnamefont
  {Centelles~Chuliá}}, \bibinfo {author} {\bibfnamefont {R.}~\bibnamefont
  {Cepedello}}, \bibinfo {author} {\bibfnamefont {E.}~\bibnamefont {Peinado}},\
  and\ \bibinfo {author} {\bibfnamefont {R.}~\bibnamefont {Srivastava}},\
  }\href@noop {} {\  (\bibinfo {year} {2019})},\ \Eprint
  {https://arxiv.org/abs/1901.06402} {arXiv:1901.06402 [hep-ph]} \BibitemShut
  {NoStop}%
\bibitem [{\citenamefont {Ma}\ \emph {et~al.}(2016)\citenamefont {Ma},
  \citenamefont {Pollard}, \citenamefont {Popov},\ and\ \citenamefont
  {Zakeri}}]{Ma:2016nnn}%
  \BibitemOpen
  \bibfield  {author} {\bibinfo {author} {\bibfnamefont {E.}~\bibnamefont
  {Ma}}, \bibinfo {author} {\bibfnamefont {N.}~\bibnamefont {Pollard}},
  \bibinfo {author} {\bibfnamefont {O.}~\bibnamefont {Popov}},\ and\ \bibinfo
  {author} {\bibfnamefont {M.}~\bibnamefont {Zakeri}},\ }\href
  {https://doi.org/10.1142/S0217732316501637} {\bibfield  {journal} {\bibinfo
  {journal} {Mod. Phys. Lett.}\ }\textbf {\bibinfo {volume} {A31}},\ \bibinfo
  {pages} {1650163} (\bibinfo {year} {2016})},\ \Eprint
  {https://arxiv.org/abs/1605.00991} {arXiv:1605.00991 [hep-ph]} \BibitemShut
  {NoStop}%
\bibitem [{\citenamefont {Basso}\ \emph {et~al.}(2009)\citenamefont {Basso},
  \citenamefont {Belyaev}, \citenamefont {Moretti},\ and\ \citenamefont
  {Shepherd-Themistocleous}}]{Basso:2008iv}%
  \BibitemOpen
  \bibfield  {author} {\bibinfo {author} {\bibfnamefont {L.}~\bibnamefont
  {Basso}}, \bibinfo {author} {\bibfnamefont {A.}~\bibnamefont {Belyaev}},
  \bibinfo {author} {\bibfnamefont {S.}~\bibnamefont {Moretti}},\ and\ \bibinfo
  {author} {\bibfnamefont {C.~H.}\ \bibnamefont {Shepherd-Themistocleous}},\
  }\href {https://doi.org/10.1103/PhysRevD.80.055030} {\bibfield  {journal}
  {\bibinfo  {journal} {Phys. Rev.}\ }\textbf {\bibinfo {volume} {D80}},\
  \bibinfo {pages} {055030} (\bibinfo {year} {2009})},\ \Eprint
  {https://arxiv.org/abs/0812.4313} {arXiv:0812.4313 [hep-ph]} \BibitemShut
  {NoStop}%
\bibitem [{\citenamefont {Fonseca}(2012)}]{Fonseca:2011sy}%
  \BibitemOpen
  \bibfield  {author} {\bibinfo {author} {\bibfnamefont {R.~M.}\ \bibnamefont
  {Fonseca}},\ }\href {https://doi.org/10.1016/j.cpc.2012.05.017} {\bibfield
  {journal} {\bibinfo  {journal} {Comput. Phys. Commun.}\ }\textbf {\bibinfo
  {volume} {183}},\ \bibinfo {pages} {2298} (\bibinfo {year} {2012})},\ \Eprint
  {https://arxiv.org/abs/1106.5016} {arXiv:1106.5016 [hep-ph]} \BibitemShut
  {NoStop}%
\bibitem [{\citenamefont {Rizzo}(2018)}]{Rizzo:2018vlb}%
  \BibitemOpen
  \bibfield  {author} {\bibinfo {author} {\bibfnamefont {T.~G.}\ \bibnamefont
  {Rizzo}},\ }\href@noop {} {\  (\bibinfo {year} {2018})},\ \Eprint
  {https://arxiv.org/abs/1810.07531} {arXiv:1810.07531 [hep-ph]} \BibitemShut
  {NoStop}%
\bibitem [{\citenamefont {Holdom}(1986{\natexlab{a}})}]{Holdom:1985ag}%
  \BibitemOpen
  \bibfield  {author} {\bibinfo {author} {\bibfnamefont {B.}~\bibnamefont
  {Holdom}},\ }\href {https://doi.org/10.1016/0370-2693(86)91377-8} {\bibfield
  {journal} {\bibinfo  {journal} {Phys. Lett.}\ }\textbf {\bibinfo {volume}
  {166B}},\ \bibinfo {pages} {196} (\bibinfo {year}
  {1986}{\natexlab{a}})}\BibitemShut {NoStop}%
\bibitem [{\citenamefont {Holdom}(1986{\natexlab{b}})}]{Holdom:1986eq}%
  \BibitemOpen
  \bibfield  {author} {\bibinfo {author} {\bibfnamefont {B.}~\bibnamefont
  {Holdom}},\ }\href {https://doi.org/10.1016/0370-2693(86)90470-3} {\bibfield
  {journal} {\bibinfo  {journal} {Phys. Lett.}\ }\textbf {\bibinfo {volume}
  {B178}},\ \bibinfo {pages} {65} (\bibinfo {year}
  {1986}{\natexlab{b}})}\BibitemShut {NoStop}%
\bibitem [{\citenamefont {Akerib}\ \emph {et~al.}(2017)\citenamefont {Akerib}
  \emph {et~al.}}]{Akerib:2016vxi}%
  \BibitemOpen
  \bibfield  {author} {\bibinfo {author} {\bibfnamefont {D.~S.}\ \bibnamefont
  {Akerib}} \emph {et~al.} (\bibinfo {collaboration} {LUX}),\ }\href
  {https://doi.org/10.1103/PhysRevLett.118.021303} {\bibfield  {journal}
  {\bibinfo  {journal} {Phys. Rev. Lett.}\ }\textbf {\bibinfo {volume} {118}},\
  \bibinfo {pages} {021303} (\bibinfo {year} {2017})},\ \Eprint
  {https://arxiv.org/abs/1608.07648} {arXiv:1608.07648 [astro-ph.CO]}
  \BibitemShut {NoStop}%
\bibitem [{\citenamefont {Akerib}\ \emph {et~al.}(2018)\citenamefont {Akerib}
  \emph {et~al.}}]{Akerib:2018lyp}%
  \BibitemOpen
  \bibfield  {author} {\bibinfo {author} {\bibfnamefont {D.~S.}\ \bibnamefont
  {Akerib}} \emph {et~al.} (\bibinfo {collaboration} {LUX-ZEPLIN}),\
  }\href@noop {} {\  (\bibinfo {year} {2018})},\ \Eprint
  {https://arxiv.org/abs/1802.06039} {arXiv:1802.06039 [astro-ph.IM]}
  \BibitemShut {NoStop}%
\bibitem [{\citenamefont {Aprile}\ \emph {et~al.}(2018)\citenamefont {Aprile}
  \emph {et~al.}}]{Aprile:2018dbl}%
  \BibitemOpen
  \bibfield  {author} {\bibinfo {author} {\bibfnamefont {E.}~\bibnamefont
  {Aprile}} \emph {et~al.} (\bibinfo {collaboration} {XENON}),\ }\href
  {https://doi.org/10.1103/PhysRevLett.121.111302} {\bibfield  {journal}
  {\bibinfo  {journal} {Phys. Rev. Lett.}\ }\textbf {\bibinfo {volume} {121}},\
  \bibinfo {pages} {111302} (\bibinfo {year} {2018})},\ \Eprint
  {https://arxiv.org/abs/1805.12562} {arXiv:1805.12562 [astro-ph.CO]}
  \BibitemShut {NoStop}%
\bibitem [{\citenamefont {Staub}(2014)}]{Staub:2013tta}%
  \BibitemOpen
  \bibfield  {author} {\bibinfo {author} {\bibfnamefont {F.}~\bibnamefont
  {Staub}},\ }\href {https://doi.org/10.1016/j.cpc.2014.02.018} {\bibfield
  {journal} {\bibinfo  {journal} {Comput. Phys. Commun.}\ }\textbf {\bibinfo
  {volume} {185}},\ \bibinfo {pages} {1773} (\bibinfo {year} {2014})},\ \Eprint
  {https://arxiv.org/abs/1309.7223} {arXiv:1309.7223 [hep-ph]} \BibitemShut
  {NoStop}%
\bibitem [{\citenamefont {Porod}\ and\ \citenamefont
  {Staub}(2012)}]{Porod:2011nf}%
  \BibitemOpen
  \bibfield  {author} {\bibinfo {author} {\bibfnamefont {W.}~\bibnamefont
  {Porod}}\ and\ \bibinfo {author} {\bibfnamefont {F.}~\bibnamefont {Staub}},\
  }\href {https://doi.org/10.1016/j.cpc.2012.05.021} {\bibfield  {journal}
  {\bibinfo  {journal} {Comput. Phys. Commun.}\ }\textbf {\bibinfo {volume}
  {183}},\ \bibinfo {pages} {2458} (\bibinfo {year} {2012})},\ \Eprint
  {https://arxiv.org/abs/1104.1573} {arXiv:1104.1573 [hep-ph]} \BibitemShut
  {NoStop}%
\bibitem [{\citenamefont {Barducci}\ \emph {et~al.}(2018)\citenamefont
  {Barducci}, \citenamefont {Belanger}, \citenamefont {Bernon}, \citenamefont
  {Boudjema}, \citenamefont {Da~Silva}, \citenamefont {Kraml}, \citenamefont
  {Laa},\ and\ \citenamefont {Pukhov}}]{Barducci:2016pcb}%
  \BibitemOpen
  \bibfield  {author} {\bibinfo {author} {\bibfnamefont {D.}~\bibnamefont
  {Barducci}}, \bibinfo {author} {\bibfnamefont {G.}~\bibnamefont {Belanger}},
  \bibinfo {author} {\bibfnamefont {J.}~\bibnamefont {Bernon}}, \bibinfo
  {author} {\bibfnamefont {F.}~\bibnamefont {Boudjema}}, \bibinfo {author}
  {\bibfnamefont {J.}~\bibnamefont {Da~Silva}}, \bibinfo {author}
  {\bibfnamefont {S.}~\bibnamefont {Kraml}}, \bibinfo {author} {\bibfnamefont
  {U.}~\bibnamefont {Laa}},\ and\ \bibinfo {author} {\bibfnamefont
  {A.}~\bibnamefont {Pukhov}},\ }\href
  {https://doi.org/10.1016/j.cpc.2017.08.028} {\bibfield  {journal} {\bibinfo
  {journal} {Comput. Phys. Commun.}\ }\textbf {\bibinfo {volume} {222}},\
  \bibinfo {pages} {327} (\bibinfo {year} {2018})},\ \Eprint
  {https://arxiv.org/abs/1606.03834} {arXiv:1606.03834 [hep-ph]} \BibitemShut
  {NoStop}%
\bibitem [{\citenamefont {Arhrib}\ \emph {et~al.}(2016)\citenamefont {Arhrib},
  \citenamefont {Bœhm}, \citenamefont {Ma},\ and\ \citenamefont
  {Yuan}}]{Arhrib:2015dez}%
  \BibitemOpen
  \bibfield  {author} {\bibinfo {author} {\bibfnamefont {A.}~\bibnamefont
  {Arhrib}}, \bibinfo {author} {\bibfnamefont {C.}~\bibnamefont {Bœhm}},
  \bibinfo {author} {\bibfnamefont {E.}~\bibnamefont {Ma}},\ and\ \bibinfo
  {author} {\bibfnamefont {T.-C.}\ \bibnamefont {Yuan}},\ }\href
  {https://doi.org/10.1088/1475-7516/2016/04/049} {\bibfield  {journal}
  {\bibinfo  {journal} {JCAP}\ }\textbf {\bibinfo {volume} {1604}}\bibfield
  {number} {\bibinfo  {number} { (04)},\ \bibinfo {pages} {049}},\ }\Eprint
  {https://arxiv.org/abs/1512.08796} {arXiv:1512.08796 [hep-ph]} \BibitemShut
  {NoStop}%
\bibitem [{\citenamefont {Slatyer}\ and\ \citenamefont
  {Wu}(2017)}]{Slatyer:2016qyl}%
  \BibitemOpen
  \bibfield  {author} {\bibinfo {author} {\bibfnamefont {T.~R.}\ \bibnamefont
  {Slatyer}}\ and\ \bibinfo {author} {\bibfnamefont {C.-L.}\ \bibnamefont
  {Wu}},\ }\href {https://doi.org/10.1103/PhysRevD.95.023010} {\bibfield
  {journal} {\bibinfo  {journal} {Phys. Rev.}\ }\textbf {\bibinfo {volume}
  {D95}},\ \bibinfo {pages} {023010} (\bibinfo {year} {2017})},\ \Eprint
  {https://arxiv.org/abs/1610.06933} {arXiv:1610.06933 [astro-ph.CO]}
  \BibitemShut {NoStop}%
\bibitem [{\citenamefont {Aghanim}\ \emph {et~al.}(2018)\citenamefont {Aghanim}
  \emph {et~al.}}]{Aghanim:2018eyx}%
  \BibitemOpen
  \bibfield  {author} {\bibinfo {author} {\bibfnamefont {N.}~\bibnamefont
  {Aghanim}} \emph {et~al.} (\bibinfo {collaboration} {Planck}),\ }\href@noop
  {} {\  (\bibinfo {year} {2018})},\ \Eprint {https://arxiv.org/abs/1807.06209}
  {arXiv:1807.06209 [astro-ph.CO]} \BibitemShut {NoStop}%
\bibitem [{\citenamefont {Heeck}\ and\ \citenamefont
  {Rodejohann}(2013)}]{Heeck:2013rpa}%
  \BibitemOpen
  \bibfield  {author} {\bibinfo {author} {\bibfnamefont {J.}~\bibnamefont
  {Heeck}}\ and\ \bibinfo {author} {\bibfnamefont {W.}~\bibnamefont
  {Rodejohann}},\ }\href {https://doi.org/10.1209/0295-5075/103/32001}
  {\bibfield  {journal} {\bibinfo  {journal} {EPL}\ }\textbf {\bibinfo {volume}
  {103}},\ \bibinfo {pages} {32001} (\bibinfo {year} {2013})},\ \Eprint
  {https://arxiv.org/abs/1306.0580} {arXiv:1306.0580 [hep-ph]} \BibitemShut
  {NoStop}%
\bibitem [{\citenamefont {Arnold}\ \emph {et~al.}(2017)\citenamefont {Arnold}
  \emph {et~al.}}]{Arnold:2017bnh}%
  \BibitemOpen
  \bibfield  {author} {\bibinfo {author} {\bibfnamefont {R.}~\bibnamefont
  {Arnold}} \emph {et~al.} (\bibinfo {collaboration} {NEMO-3}),\ }\href
  {https://doi.org/10.1103/PhysRevLett.119.041801} {\bibfield  {journal}
  {\bibinfo  {journal} {Phys. Rev. Lett.}\ }\textbf {\bibinfo {volume} {119}},\
  \bibinfo {pages} {041801} (\bibinfo {year} {2017})},\ \Eprint
  {https://arxiv.org/abs/1705.08847} {arXiv:1705.08847 [hep-ex]} \BibitemShut
  {NoStop}%
\bibitem [{\citenamefont {Guzowski}(2018)}]{Guzowski:2018neg}%
  \BibitemOpen
  \bibfield  {author} {\bibinfo {author} {\bibfnamefont {P.}~\bibnamefont
  {Guzowski}} (\bibinfo {collaboration} {NEMO-3}),\ }in\ \href@noop {} {\emph
  {\bibinfo {booktitle} {{Prospects in Neutrino Physics (NuPhys2017) London,
  United Kingdom, December 20-22, 2017}}}}\ (\bibinfo {year} {2018})\ \Eprint
  {https://arxiv.org/abs/1804.00280} {arXiv:1804.00280 [hep-ex]} \BibitemShut
  {NoStop}%
\bibitem [{\citenamefont {Kidd}\ and\ \citenamefont
  {Tornow}(2018)}]{Kidd:2018fbb}%
  \BibitemOpen
  \bibfield  {author} {\bibinfo {author} {\bibfnamefont {M.~F.}\ \bibnamefont
  {Kidd}}\ and\ \bibinfo {author} {\bibfnamefont {W.}~\bibnamefont {Tornow}},\
  }\href {https://doi.org/10.1103/PhysRevC.98.055501} {\bibfield  {journal}
  {\bibinfo  {journal} {Phys. Rev.}\ }\textbf {\bibinfo {volume} {C98}},\
  \bibinfo {pages} {055501} (\bibinfo {year} {2018})}\BibitemShut {NoStop}%
\bibitem [{\citenamefont {Carter}\ and\ \citenamefont
  {Heinrich}(2011)}]{Carter:2010hi}%
  \BibitemOpen
  \bibfield  {author} {\bibinfo {author} {\bibfnamefont {J.}~\bibnamefont
  {Carter}}\ and\ \bibinfo {author} {\bibfnamefont {G.}~\bibnamefont
  {Heinrich}},\ }\href {https://doi.org/10.1016/j.cpc.2011.03.026} {\bibfield
  {journal} {\bibinfo  {journal} {Comput. Phys. Commun.}\ }\textbf {\bibinfo
  {volume} {182}},\ \bibinfo {pages} {1566} (\bibinfo {year} {2011})},\ \Eprint
  {https://arxiv.org/abs/1011.5493} {arXiv:1011.5493 [hep-ph]} \BibitemShut
  {NoStop}%
\bibitem [{\citenamefont {Aaboud}\ \emph {et~al.}(2019)\citenamefont {Aaboud}
  \emph {et~al.}}]{Aaboud:2019zxd}%
  \BibitemOpen
  \bibfield  {author} {\bibinfo {author} {\bibfnamefont {M.}~\bibnamefont
  {Aaboud}} \emph {et~al.} (\bibinfo {collaboration} {ATLAS}),\ }\href@noop {}
  {\  (\bibinfo {year} {2019})},\ \Eprint {https://arxiv.org/abs/1901.10917}
  {arXiv:1901.10917 [hep-ex]} \BibitemShut {NoStop}%
\bibitem [{\citenamefont {Aaboud}\ \emph {et~al.}(2017)\citenamefont {Aaboud}
  \emph {et~al.}}]{Aaboud:2017buh}%
  \BibitemOpen
  \bibfield  {author} {\bibinfo {author} {\bibfnamefont {M.}~\bibnamefont
  {Aaboud}} \emph {et~al.} (\bibinfo {collaboration} {ATLAS}),\ }\href
  {https://doi.org/10.1007/JHEP10(2017)182} {\bibfield  {journal} {\bibinfo
  {journal} {JHEP}\ }\textbf {\bibinfo {volume} {10}},\ \bibinfo {pages}
  {182}},\ \Eprint {https://arxiv.org/abs/1707.02424} {arXiv:1707.02424
  [hep-ex]} \BibitemShut {NoStop}%
\bibitem [{\citenamefont {Radburn-smith}(2018)}]{Radburn-smith:2649415}%
  \BibitemOpen
  \bibfield  {author} {\bibinfo {author} {\bibfnamefont {B.~C.}\ \bibnamefont
  {Radburn-smith}} (\bibinfo {collaboration} {CMS Collaboration}),\ }\href
  {https://cds.cern.ch/record/2649415} {\emph {\bibinfo {title} {{Searches for
  new heavy resonances in final states with leptons and photons}}}},\ \bibinfo
  {type} {Tech. Rep.}\ \bibinfo {number} {CMS-CR-2018-371}\ (\bibinfo
  {institution} {CERN},\ \bibinfo {address} {Geneva},\ \bibinfo {year}
  {2018})\BibitemShut {NoStop}%
\bibitem [{\citenamefont {Jaegle}(2015)}]{TheBelle:2015mwa}%
  \BibitemOpen
  \bibfield  {author} {\bibinfo {author} {\bibfnamefont {I.}~\bibnamefont
  {Jaegle}} (\bibinfo {collaboration} {Belle}),\ }\href
  {https://doi.org/10.1103/PhysRevLett.114.211801} {\bibfield  {journal}
  {\bibinfo  {journal} {Phys. Rev. Lett.}\ }\textbf {\bibinfo {volume} {114}},\
  \bibinfo {pages} {211801} (\bibinfo {year} {2015})},\ \Eprint
  {https://arxiv.org/abs/1502.00084} {arXiv:1502.00084 [hep-ex]} \BibitemShut
  {NoStop}%
\bibitem [{\citenamefont {Giovannella}\ \emph {et~al.}(2011)\citenamefont
  {Giovannella} \emph {et~al.}}]{Giovannella:2011nh}%
  \BibitemOpen
  \bibfield  {author} {\bibinfo {author} {\bibfnamefont {S.}~\bibnamefont
  {Giovannella}} \emph {et~al.},\ }\bibfield  {booktitle} {\emph {\bibinfo
  {booktitle} {{Proceedings, 2nd Symposium on Prospects in the Physics of
  Discrete Symmetries (DISCRETE 2010): Rome, Italy, December 6-11, 2010}}},\
  }\href {https://doi.org/10.1088/1742-6596/335/1/012067} {\bibfield  {journal}
  {\bibinfo  {journal} {J. Phys. Conf. Ser.}\ }\textbf {\bibinfo {volume}
  {335}},\ \bibinfo {pages} {012067} (\bibinfo {year} {2011})},\ \Eprint
  {https://arxiv.org/abs/1107.2531} {arXiv:1107.2531 [hep-ex]} \BibitemShut
  {NoStop}%
\bibitem [{\citenamefont {Echenard}(2016)}]{Echenard:2016tiu}%
  \BibitemOpen
  \bibfield  {author} {\bibinfo {author} {\bibfnamefont {B.}~\bibnamefont
  {Echenard}} (\bibinfo {collaboration} {BaBar}),\ }\bibfield  {booktitle}
  {\emph {\bibinfo {booktitle} {{Proceedings, 37th International Conference on
  High Energy Physics (ICHEP 2014): Valencia, Spain, July 2-9, 2014}}},\ }\href
  {https://doi.org/10.1016/j.nuclphysbps.2015.09.414} {\bibfield  {journal}
  {\bibinfo  {journal} {Nucl. Part. Phys. Proc.}\ }\textbf {\bibinfo {volume}
  {273-275}},\ \bibinfo {pages} {2427} (\bibinfo {year} {2016})}\BibitemShut
  {NoStop}%
\bibitem [{\citenamefont {Arkani-Hamed}\ \emph {et~al.}(2009)\citenamefont
  {Arkani-Hamed}, \citenamefont {Finkbeiner}, \citenamefont {Slatyer},\ and\
  \citenamefont {Weiner}}]{ArkaniHamed:2008qn}%
  \BibitemOpen
  \bibfield  {author} {\bibinfo {author} {\bibfnamefont {N.}~\bibnamefont
  {Arkani-Hamed}}, \bibinfo {author} {\bibfnamefont {D.~P.}\ \bibnamefont
  {Finkbeiner}}, \bibinfo {author} {\bibfnamefont {T.~R.}\ \bibnamefont
  {Slatyer}},\ and\ \bibinfo {author} {\bibfnamefont {N.}~\bibnamefont
  {Weiner}},\ }\href {https://doi.org/10.1103/PhysRevD.79.015014} {\bibfield
  {journal} {\bibinfo  {journal} {Phys. Rev.}\ }\textbf {\bibinfo {volume}
  {D79}},\ \bibinfo {pages} {015014} (\bibinfo {year} {2009})},\ \Eprint
  {https://arxiv.org/abs/0810.0713} {arXiv:0810.0713 [hep-ph]} \BibitemShut
  {NoStop}%
\bibitem [{\citenamefont {Lees}\ \emph {et~al.}(2017)\citenamefont {Lees} \emph
  {et~al.}}]{Lees:2017lec}%
  \BibitemOpen
  \bibfield  {author} {\bibinfo {author} {\bibfnamefont {J.~P.}\ \bibnamefont
  {Lees}} \emph {et~al.} (\bibinfo {collaboration} {BaBar}),\ }\href
  {https://doi.org/10.1103/PhysRevLett.119.131804} {\bibfield  {journal}
  {\bibinfo  {journal} {Phys. Rev. Lett.}\ }\textbf {\bibinfo {volume} {119}},\
  \bibinfo {pages} {131804} (\bibinfo {year} {2017})},\ \Eprint
  {https://arxiv.org/abs/1702.03327} {arXiv:1702.03327 [hep-ex]} \BibitemShut
  {NoStop}%
\bibitem [{\citenamefont {Banerjee}\ \emph {et~al.}(2017)\citenamefont
  {Banerjee} \emph {et~al.}}]{Banerjee:2016tad}%
  \BibitemOpen
  \bibfield  {author} {\bibinfo {author} {\bibfnamefont {D.}~\bibnamefont
  {Banerjee}} \emph {et~al.} (\bibinfo {collaboration} {NA64}),\ }\href
  {https://doi.org/10.1103/PhysRevLett.118.011802} {\bibfield  {journal}
  {\bibinfo  {journal} {Phys. Rev. Lett.}\ }\textbf {\bibinfo {volume} {118}},\
  \bibinfo {pages} {011802} (\bibinfo {year} {2017})},\ \Eprint
  {https://arxiv.org/abs/1610.02988} {arXiv:1610.02988 [hep-ex]} \BibitemShut
  {NoStop}%
\bibitem [{\citenamefont {Racker}(2014)}]{Racker:2013lua}%
  \BibitemOpen
  \bibfield  {author} {\bibinfo {author} {\bibfnamefont {J.}~\bibnamefont
  {Racker}},\ }\href {https://doi.org/10.1088/1475-7516/2014/03/025} {\bibfield
   {journal} {\bibinfo  {journal} {JCAP}\ }\textbf {\bibinfo {volume} {1403}},\
  \bibinfo {pages} {025}},\ \Eprint {https://arxiv.org/abs/1308.1840}
  {arXiv:1308.1840 [hep-ph]} \BibitemShut {NoStop}%
\bibitem [{\citenamefont {Hugle}\ \emph {et~al.}(2018)\citenamefont {Hugle},
  \citenamefont {Platscher},\ and\ \citenamefont {Schmitz}}]{Hugle:2018qbw}%
  \BibitemOpen
  \bibfield  {author} {\bibinfo {author} {\bibfnamefont {T.}~\bibnamefont
  {Hugle}}, \bibinfo {author} {\bibfnamefont {M.}~\bibnamefont {Platscher}},\
  and\ \bibinfo {author} {\bibfnamefont {K.}~\bibnamefont {Schmitz}},\ }\href
  {https://doi.org/10.1103/PhysRevD.98.023020} {\bibfield  {journal} {\bibinfo
  {journal} {Phys. Rev.}\ }\textbf {\bibinfo {volume} {D98}},\ \bibinfo {pages}
  {023020} (\bibinfo {year} {2018})},\ \Eprint
  {https://arxiv.org/abs/1804.09660} {arXiv:1804.09660 [hep-ph]} \BibitemShut
  {NoStop}%
\end{thebibliography}%
\end{document}